\documentclass[a4paper,11pt]{article}
\pdfoutput=1 

\usepackage{jheppub} 
                     
\usepackage[T1]{fontenc} 
\usepackage{xcolor}
\usepackage{subfigure}
\usepackage{braket}
\usepackage{hyperref}

\begin{document}

\title{\boldmath Entanglement and symmetry resolution in two dimensional free quantum field theories
}

\vspace{.5cm}

\author{Sara Murciano$^1$, Giuseppe Di Giulio$^1$, and Pasquale Calabrese$^{1,2}$}
\affiliation{$^1$SISSA and INFN Sezione di Trieste, via Bonomea 265, 34136 Trieste, Italy.}
\affiliation{$^{2}$International Centre for Theoretical Physics (ICTP), Strada Costiera 11, 34151 Trieste, Italy.}

\emailAdd{smurcian@sissa.it}

\vspace{.5cm}

\abstract{We present a thorough analysis of the entanglement entropies related to different symmetry sectors of free quantum field theories (QFT) with an internal U(1) symmetry. 
We provide explicit analytic computations for the charged moments of Dirac and complex scalar fields in two spacetime dimensions, both in the massive and massless cases, 
using two different approaches. 
The first one is based on the replica trick, the computation of the partition function on Riemann surfaces with the insertion of a flux $\alpha$, and the introduction of properly modified twist fields, whose two-point function directly gives the scaling limit of the charged moments. With the second method, the diagonalisation in replica space maps the problem to the computation of a partition function on a cut plane, that can be written exactly in terms of the solutions of non-linear differential equations of the Painlev\'e V type. 
Within this approach, we also derive an asymptotic expansion for the short and long distance behaviour of the charged moments. 
Finally, the Fourier transform provides the desired symmetry resolved entropies: 
at the leading order, they satisfy entanglement equipartition and we identify the subleading terms that break it.
Our analytical findings are tested against exact numerical calculations in lattice models.}

\maketitle

\section{Introduction}
Symmetries are a pillar of modern physics. 
They can concern spacetime, as rotational or relativistic invariance, or can be internal symmetries, which do not touch the spacetime coordinates. 
Their exploration turned out to be a central theme in several fields ranging from elementary particles to the theory of phase transitions, from string theory to solid-state physics.
One century ago, Emmy Noether proved that every symmetry of a physical system leads to a corresponding conservation law. 
For example, the conserved electric charge is the generator of the $U(1)$ gauge symmetry of electromagnetism. 
Consequently, an evergreen research topic is the characterisation of how the presence of a symmetry influences the properties of a physical system. 
In particular, this manuscript addresses the question of how the entanglement splits into the different sectors of an internal symmetry. 

The von Neumann entropy is the most successful way to characterise the bipartite entanglement of a subsystem $A$ in a pure quantum state  \cite{intro1,intro2,eisert-2010,intro3}. 
Given the reduced density matrix (RDM) $\rho_A$ of a subsystem $A$, the entanglement entropy is defined as
\begin{equation}
\label{eq:entr0}
S_1=-\mathrm{Tr} \rho_A\ln \rho_A.
\end{equation}
A related family of functions, known as R\'enyi entropies, is
\begin{equation}
\label{eq:entr}
S_n=\frac{1}{1-n} \ln\mathrm{Tr} \rho_A^{n}.
\end{equation}
The essence of the replica trick is that the von Neumann entropy may be obtained as the limit $n \to 1$ of Eq. (\ref{eq:entr}). 
The reason of this way of proceeding is that for integer $n$, in the path-integral formalism, $\mathrm{Tr} \rho_A^{n}$ is the partition function on an $n$-sheeted Riemann surface 
$\mathcal{R}_n$ obtained by joining cyclically the $n$ sheets along the region $A$ \cite{cc-04,cw-94}. 
This approach, when applied to critical systems whose low energy physics is described by a (1+1) dimensional conformal field theory (CFT), leads to the famous scaling results \cite{cc-04,cc-09,hlw-94,vidal,vidal1,vidal2,cw-94}
\begin{equation}
\label{eq:entr2}
S_1=\frac{c}{3} \ln \frac{\ell}{\epsilon}, \qquad S_n=\frac{c}{6}\frac{n+1}{n} \ln \frac{\ell}{\epsilon} ,
\end{equation}
when the subsystem $A$ is an interval of length $\ell$ embedded in an infinite one-dimensional system and $\epsilon\ll\ell$ is an ultraviolet cutoff.
 
The possibility of measuring in an experiment the internal symmetry structure of the entanglement \cite{fis} went together with a new theoretical framework developed 
to address the problem \cite{goldstein,goldstein1}. 
Indeed, in Refs. \cite{goldstein,goldstein1} a simple generalisation of the replica trick has been proposed to relate the symmetry resolved quantities to the moments of 
$\rho_A$ on a modified Riemann surface: we refer to them as {\it charged moments}. 
Such technique allowed for the derivation of interesting results about the different symmetry-resolved contributions not only in CFTs, but also in the context of
free gapped and gapless systems of bosons and fermions, integrable spin chains, disordered systems and many more (the interested readers can consult the 
comprehensive literature on the subject  
\cite{goldstein2,xavier,riccarda,crc-20,SREE2dG,MDC-19-CTM,lr-14,ccgm-20,SREE,cms-13,clss-19,tr-19,wv-03,bhd-18,bcd-19,kusf-20,kusf-20b,mrc-20,trac-20,s17,ms-20}). 

The main goal of this manuscript is to investigate the field theoretical techniques for the computation of the charged moments in relativistic free two-dimensional quantum field 
theories (QFTs). The paper is organised as follows. In Section \ref{sec:general} we provide all the definitions concerning the measures of symmetry resolved entanglement  
and we briefly recall two tools for the computation of the entanglement in QFT, i.e. the twist fields \cite{cc-09,ccd-08,cd-09,dixon,k-87} and the Green's function technique 
in the replica space \cite{CFH,casini,ch-rev}. 
Our main findings are reported in Sections \ref{sec:twistG}, \ref{sec:part},  and \ref{sec:partS}: in the first we employ the twist fields to compute the charged moments both 
in the massless and in the massive context (in the limit $m\ell \gg1$).  
These results are extended in sections  \ref{sec:part} and \ref{sec:partS}, where we write down the explicit form of the charged moments for arbitrary $m\ell$ 
and provide analytic asymptotic expansions valid for large and small $m\ell$. 
These outcomes are the starting point for the computation of the symmetry resolved entanglement entropies. 
Numerical checks for free fermions and bosons on the lattice are also provided as a benchmark of the analytical results. 
In Sec. \ref{app:nishioka} we give a result for the charged moments of a free massive scalar theory across a hyperplane in generic Euclidean dimensions $d$.
We draw our conclusions in Section \ref{sec:concl}. 
Three appendices are also included: they provide details about the analytical and numerical computations.

\section{Symmetry resolution and QFT techniques}\label{sec:general}
\label{sec:SR}
In this section, we provide an overview of notions about symmetry resolved entropies we use throughout the manuscript. 
We also recall some standard techniques used to compute the entanglement entropy in relativistic free QFT.

\subsection{Symmetry resolved entanglement}
We consider a system with an internal $U(1)$ symmetry and its bipartition into two subsystems, $A$ and $B$. The charge operator $Q$ is the generator of the symmetry and we assume it obeys $Q_A\oplus Q_B=Q$,  where $Q_i$ is the charge in the subsystem  $i$. If the system described by the density matrix $\rho$ is in an eigenstate of $Q$, then $[\rho,Q]=0$.  
Tracing out the degrees of freedom of $B$, we obtain the RDM of $A$, $\rho_A=\mathrm{Tr}_B \rho$. 
Hence, taking the trace over $B$ of $[\rho,Q] = 0$, we find that $[\rho_A,Q_A] = 0$. This means that $\rho_A$ has a block-diagonal structure where each block corresponds to an eigenvalue $q$ of $Q_A$. The density matrix $\rho_A(q)$ corresponding to an eigenvalue $q$ is obtained by projecting $\rho_A$ onto the eigenspace of $Q_A$ with fixed $q$, as induced by the projector $\Pi_q$.
Therefore we can write
\begin{equation}
\label{eq:sum}
\rho_A=\oplus_q p_{A}(q)\rho_{A}(q),
\end{equation}
where $p_{A}(q)$ is the probability of finding $q$ in a measurement of $Q_A$ in the RDM $\rho_A$, i.e. $p_A(q)= \mathrm{Tr} \Pi_q \rho_A$. 
Within this convention, the density matrices $\rho_A (q)$ of different blocks are normalised as ${\rm tr}\rho_A (q)=1$.  
The amount of entanglement shared by $A$ and $B$ in each symmetry sector can be computed through the {\it symmetry resolved R\'enyi entropies}, defined as
\begin{equation}
\label{eq:RSREE}
S_{n}(q) \equiv \dfrac{1}{1-n}\ln \mathrm{Tr} \rho^n_A(q).
\end{equation}
The limit $n \to 1$ gives the {\it symmetry resolved entanglement entropy}, i.e.
\begin{equation}
\label{eq:SvNEE}
S_1(q) \equiv -\mathrm{Tr} \rho_A(q) \ln \rho_A(q).
\end{equation}
The total von Neumann entanglement entropy associated to $\rho_A$ in Eq. (\ref{eq:sum}) splits into \cite{nc-10}
\begin{equation}
\label{eq:SvN}
S_{1}=\displaystyle \sum_q p(q) S_{1}(q)- \displaystyle \sum_q p(q) \ln p(q).
\end{equation}
The two contributions are known as configurational and fluctuation (or number) entanglement entropy, respectively \cite{fis}. 
The configurational entropy is also related to the operationally accessible entanglement entropy of Refs. \cite{wv-03,bhd-18,bcd-19}, 
while the number entropy is the subject of a substantial recent activity \cite{fis,kusf-20,kusf-20b,riccarda,ms-20}. 
The calculation of the symmetry resolved entropies by the definition (\ref{eq:RSREE}) requires the knowledge of the entanglement spectrum of $\rho_A$ and its resolution in the charge sectors. However, this is a difficult task, especially for an analytic derivation.
As first proposed in \cite{goldstein}, we can rather focus on the charged moments of $\rho_A$ 
\begin{equation}
\label{eq:firstdef}
Z_n(\alpha)\equiv\mathrm{Tr}\rho_A^ne^{iQ_A \alpha},
\end{equation}
with $Z_1(\alpha=0)=1$, being $\mathrm{Tr}\rho_A=1$.
Similar charged moments have been already considered in the context of free field theories \cite{d-16}, in holographic settings \cite{Belin-Myers-13-HolChargedEnt,cnn-16}, 
as well as in the study of entanglement in mixed states \cite{ssr-17,shapourian-19}. 
In this specific case, the charged moments are not the main goal of our computation, but they represent a fundamental tool, 
because their Fourier transforms are the moments of the RDM restricted to the sector of fixed charge $q$ \cite{goldstein}, i.e.
\begin{equation}
\label{eq:defF}
\mathcal{Z}_n(q)\equiv \mathrm{Tr} (\Pi_{q}\,\rho^n_A)=\displaystyle \int_{-\pi}^{\pi}\dfrac{d\alpha}{2\pi}e^{-iq\alpha}Z_n(\alpha).
\end{equation}
(Here we assumed $Q$ to be the generator of a $U(1)$ symmetry and $q\in \mathbb Z$.)
Finally the symmetry resolved entropies are obtained as
\begin{equation}
\label{eq:SREE1}
S_n(q)=\dfrac{1}{1-n}\ln \left[ \dfrac{\mathcal{Z}_n(q)}{\mathcal{Z}^n_1(q)}\right], \qquad S_{1}(q)=\lim_{n\rightarrow 1} S_n(q).
\end{equation}
\subsection{Replica method and QFT}\label{sec:replica}
In the following sections we will mainly deal with a free fermionic field theory and with a complex scalar one, whose Euclidean actions are given, respectively, by
\begin{equation}\label{eq:actions}
\begin{split}
S_D=&\frac{1}{4 \pi} \displaystyle \int dz d\bar{z} \left( \psi^*_R \partial_{z}\psi_R+\psi^*_L\partial_{\bar{z}}\psi_L  +m(\psi^*_L\psi_R+\psi^*_R\psi_L\right), \\
S_S=&\frac{1}{4\pi}\displaystyle \int dz d\bar{z} \left(\partial_z \varphi^*\partial_{\bar{z}}\varphi +\partial_{\bar{z}} \varphi^*\partial_{z}\varphi +m^2\varphi^*\varphi\right),
\end{split}
\end{equation}
where we employ complex coordinates $(z,\bar{z})$ for the 2D spacetime.
In $S_D$ the fields $\psi_{R/L}$ are the chiral (right-moving $R$) and anti-chiral (left-moving $L$) components of the Dirac fermion. 
In $S_S$ the field $\varphi$ is a complex scalar. 
The actions in (\ref{eq:actions}) exhibit a $U(1)$ symmetry, i.e. a symmetry under phase transformations of the fields given, respectively, by 
\begin{equation}
\psi_{R/L} \to e^{i\alpha}\psi_{R/L} , \quad \psi_{R/L}^{* } \to e^{-i\alpha}\psi^*_{R/L}, \qquad \varphi \to  e^{i\alpha}\varphi, \quad \varphi^{* }\to e^{-i\alpha}\varphi^*.
\end{equation}
By Noether's theorem, this continuous symmetry transformation leads to a conserved quantity, which is the charge $Q$ we introduced before. 

The actions \eqref{eq:actions} played the role of the simplest massive quantum field theories to study the properties of the entanglement entropy. 
In the same spirit, they also represent the natural starting point for the field theoretical investigation of the charged moments and, as a consequence, of the 
symmetry resolved entropies. 

 In what follows we describe two powerful methods to calculate the entropy for free fields.  Starting from the replica trick described in the introduction, the first approach is based on a particular type of twist fields in quantum field theory that are related to branch points in the Riemann surface $\mathcal{R}_n$. 
 We denote them by $\mathcal{T}_n$ and $\mathcal{\tilde{T}}_n$. Their action, in operator formalism, is defined by \cite{cc-09,ccd-08,cd-09} 
\begin{equation}
\begin{split}
\mathcal{T}_n(z_1)\ \phi_i(z')&=\phi_{i+1}(z') \mathcal{T}_n(z_1),\\
\mathcal{\tilde{T}}_n(z_2)\phi_i(z')&=\phi_{i-1}(z')\mathcal{\tilde{T}}_n(z_2),
\end{split}
\end{equation}
where $z_1$ and $z_2$ are the endpoints of $A$, $z' \in A$ and $i=1,\dots,n$ with $n+1 \equiv 1$.
The two-point function of the twist fields directly gives \cite{cc-09}
\begin{equation}
\mathrm{Tr}\rho_A^n \propto \braket{\mathcal{T}_n(z_1)\mathcal{\tilde{T}}_n(z_2)}.
\end{equation}
In conformal invariant theories (e.g. when the mass terms in the actions (\ref{eq:actions}) vanish) the two-point function of twist fields is fixed by their scaling dimension, leading to Eq.\,(\ref{eq:entr2}). 
In some instances, a simplification arises by the diagonalisation in the replica space: the $n$-sheeted problem can be mapped to an equivalent one in which one deals with $
n$ decoupled and multivalued free fields, generically referred as $\tilde{\phi}_k$. 
Thus, also the twist fields can be written as products of fields acting only on $ \tilde{\phi}_k$, denoted as $\mathcal{T}_{n,k}$ and $\mathcal{\tilde{T}}_{n,k}$. The total partition function is a product of $n$ partition functions, $\zeta_k$, each one given by (up to unimportant multiplicative constant)
\begin{equation} 
\zeta_k \propto \braket{\mathcal{T}_{n,k}(z_1)\mathcal{\tilde{T}}_{n,k}(z_2)}.  
\end{equation}

The second approach is the one used in Refs. \cite{CFH,casini} for a fermionic and a complex scalar theory, respectively: it also relies on mapping the problem from the 
determination of the partition function on $\mathcal{R}_n$, to the computation of $n$ partition functions of a free field on a cut plane. However, the difference with respect to 
the previous 
approach is that each $\zeta_k$ is not computed as a two point-function of twist fields, but using the relation between the free energy and the Green's function of each sector $k$.
Denoting by $G_D$ the Green's function for the Dirac field and by $G_S$ the one for the scalar (in each sector $k$ of the $n$ copies), 
they are related to the corresponding partition function $\zeta_k$ by, respectively,
\begin{equation}
\label{eq:Greenvspartition}
\begin{split}
\partial_m\ln \zeta_k=\mathrm{tr} \,G_D, \qquad \partial_{m^2} \ln \zeta_k=-\int\,dr^2\,G_{S}({\bf{r},r'}).
\end{split}
\end{equation}
The strategy of Refs. \cite{CFH,casini} was to exploit the rotational and translational symmetry of the Helmholtz equations 
satisfied by $G_D$ and $G_S$ and analyse their behaviour at the singular endpoints of the cut $A$ so to determine the right hand sides of the above equations. 
The final expressions for $\zeta_k$ can be expressed in terms of the solution of second order non linear differential equations of the Painlev\'e V type.
Here we only report the final results for the R\'enyi entropies of free fields in the limit  $ m \ell \to 0$ \cite{CFH,casini}
\begin{equation}
\label{eq:entr3}
\begin{split}
S_n^D=& \frac{n+1}{6n}\left( \ln \frac{\ell}{\epsilon}-\frac{(m\ell)^2}{2}\ln^2 m\ell \right)+O((m\ell)^2 \ln m\ell), \\
S_n^S=& \frac{n+1}{3n}\ln \frac{\ell}{\epsilon}+\ln \frac{\ln m \ell}{\ln m \epsilon}+O(m\ell).
\end{split}
\end{equation}
These formulas have been obtained in the scaling regime with $t=m\ell$ fixed, in the conformal limit $m\ell\to 0$ and after taking the limit 
of large $\ell$.  Eq. \eqref{eq:entr3} shows the leading mass corrections to Eq. (\ref{eq:entr2}) for the theories in Eq. (\ref{eq:actions}) strongly depend on the statistics of the particles. 
The leading  mass correction vanishes for a Dirac field, while it is singular (like $ \ln(-\ln( m))$) for a Klein-Gordon field (both real and complex).  
In the literature, this infrared divergence is ascribed to the zero mode of the massless scalar theory \cite{unruh,bc-16}. 

\section{Twist Field Approach}\label{sec:twistG}
In this section we consider 1D critical and close to critical systems.  
We obtain a general exact result for the conformal invariant charged moments by exploiting the properties of some local operators known as modified or fluxed twist 
fields \cite{Belin-Myers-13-HolChargedEnt,d-16}. 
This result includes and generalises the ones in Ref. \cite{goldstein}. 
The same approach also provides the leading asymptotic behaviour of the charged moments for (free) massive field theories. 

\subsection{Modified Twist Fields}\label{sec:twist}
In a generic QFT, the replica trick for computing $Z_n(\alpha)$ defined in Eq.\,(\ref{eq:firstdef}) can be implemented by inserting an Aharonov-Bohm flux through a multi-sheeted Riemann surface $\mathcal{R}_n$, such that the total phase accumulated by the field upon going through the entire surface is $\alpha$ \cite{goldstein}. The result is that $Z_n(\alpha)$ is the partition function on such modified surface, that, following Ref. \cite{goldstein}, we dub $\mathcal{R}_{n,\alpha}$.
In QFT language, the insertion of the flux corresponds to a twisted boundary condition.
This boundary condition fuses with the twist fields at the endpoints of the subsystem $A$ resulting into two local operators $\mathcal{T}_{n,\alpha}$ and $\mathcal{\tilde{T}}_{n,\alpha}$.
These are modified versions of the standard twist fields $\mathcal{T}_{n}$ and $\mathcal{\tilde{T}}_{n}$ which take into account not only the internal permutational symmetry 
among the replicas but also the presence of the flux. 
Thus, the partition function on $\mathcal{R}_{n,\alpha}$ is determined by their two-point correlation function, that is the main object of interest in this section. 

As already mentioned in section \,\ref{sec:replica}, rather than dealing with fields defined on a non trivial manifold $\mathcal{R}_{n,\alpha}$,  it is more convenient to work on a single plane with a $n$-component field
\begin{equation}\label{eq:matrix}
\Phi = 
\begin{pmatrix}
\phi_1 \\
\phi_2 \\
\vdots   \\
\phi_n
\end{pmatrix},
\end{equation}
where $\phi_{j}$ is the field on the $j$-th copy (here the field $\phi_j$ generically refers to either a scalar field $\varphi_j$ or a chiral Dirac one $\psi_j$; 
the same applies to $\tilde{\phi}_k$ that we are going to introduce soon).
Upon crossing the cut $A$, the vector field $\Phi$ transforms according to the transformation matrix $T_{\alpha}$  
\begin{equation}\label{eq:matrix1}
T_{\alpha} = 
\begin{pmatrix}
0 & e^{i\alpha/n} &  &   \\
 & 0 & e^{i\alpha/n} &    \\
  &  & \ddots & \ddots  \\
(-1)^{(n+1)f}e^{i\alpha/n}  & &  & 0
\end{pmatrix} ,
\end{equation}
where $f = 1$ for free Dirac fermions and $f = 0$ for free complex scalars. When $\alpha=0$ we recover the usual transformations for the fields across the different replicas \cite{ch-rev}.
The matrix $T_{\alpha}$ has eigenvalues
\begin{equation}
\begin{split}
f=0: \qquad &\lambda_k=e^{i\frac{\alpha}{n}}e^{2\pi i \frac{k}{n}}, \quad k=0, \dots , n-1,\\
f=1: \qquad &\lambda_k=e^{i\frac{\alpha}{n}}e^{2\pi i \frac{k}{n}}, \quad k=-\frac{n-1}{2}, \dots , \frac{n-1}{2}.
\end{split}
\end{equation}
By diagonalising $T_{\alpha}$ with a unitary transformation, the problem is reduced to $n$ decoupled fields $\tilde{\phi}_k$ in a two dimensional spacetime. 
Thus, the total partition function is a product of the partition functions for each $k$ and the twist fields can be written as products of fields each acting on a different $\tilde{\phi}_{k}$, i.e.
\begin{equation}
\label{eq:prod_twist}
\begin{split}
f=0: \qquad &\mathcal{T}_{n,\alpha}=\displaystyle \prod_{k=0}^{n-1}\mathcal{T}_{n,k,\alpha},\qquad\qquad
\mathcal{\tilde{T}}_{n,\alpha}=\displaystyle \prod_{k=0}^{n-1}\mathcal{\tilde{T}}_{n,k,\alpha},\\
f=1: \qquad &\mathcal{T}_{n,\alpha}=\displaystyle \prod_{k=-\frac{n-1}{2}}^{\frac{n-1}{2}}\mathcal{T}_{n,k,\alpha},\qquad\,\,\,
\mathcal{\tilde{T}}_{n,\alpha}=
\displaystyle \prod_{k=-\frac{n-1}{2}}^{\frac{n-1}{2}}\mathcal{\tilde{T}}_{n,k,\alpha},
\end{split}
\end{equation}
with $\mathcal{T}_{n,k,\alpha}\tilde{\phi}_{k'}=\delta_{k,k'} e^{i\alpha/n}e^{2\pi i k/n} \tilde{\phi}_{k}$ and $\mathcal{\tilde{T}}_{n,k,\alpha}\tilde{\phi}_{k'}=\delta_{k,k'} e^{-i\alpha/n}e^{-2\pi i k/n} \tilde{\phi}_{k}$. 
Since the partition function on $\mathcal{R}_{n,\alpha}$ can be written as the two-point function of the modified twist fields, from (\ref{eq:prod_twist}) we have
\begin{equation}
\label{eq:prod_twist2}
\begin{split}
f=0: \qquad &\ln Z_n(\alpha)=\sum_{k=0}^{n-1} \ln\braket{\mathcal{T}_{n,k,\alpha} \mathcal{\tilde{T}}_{n,k,\alpha}},\\
f=1: \qquad &\ln Z_n(\alpha)=\sum_{k=-(n-1)/2}^{(n-1)/2} \ln\braket{\mathcal{T}_{n,k,\alpha} \mathcal{\tilde{T}}_{n,k,\alpha}}.
\end{split}
\end{equation}
When dealing with a CFT (e.g. when $m=0$ in \eqref{eq:actions}) $\mathcal{T}_{n,k,\alpha}$ and $ \mathcal{\tilde{T}}_{n,k,\alpha}$ are primary operators 
and their two-point function is fixed by conformal invariance to be
\begin{equation}
\label{eq:twopointfunct}
\braket{\mathcal{T}_{n,k,\alpha} \mathcal{\tilde{T}}_{n,k,\alpha}} \propto \frac{1}{|u-v|^{4\Delta_{k}(\alpha)}},
\end{equation}
where (see the Appendix \ref{app:twist})
\begin{equation}
\label{eq:conformaldim}
\begin{split}
f=0: \qquad &\Delta_k(\alpha)=\frac{1}{2}\left( \frac{k}{n}+\frac{|\alpha|}{2\pi n}\right) \left(1-\frac{k}{n} -\frac{|\alpha|}{2\pi n}\right),\\
f=1: \qquad &  \Delta_{k}(\alpha)=\frac{1}{2}\left(\frac{k}{n}+\frac{\alpha}{2 \pi n } \right)^2.
\end{split}
\end{equation}
 Let us stress that, in order to have operators with positive conformal dimension, the phase that bosons pick up going around one of the entangling points must be $0< \frac{k}{n}+\frac{\alpha}{2\pi n}<1$. This can be achieved, since $\alpha \in [-\pi,\pi]$, by trading $\alpha$ with $|\alpha|$ when we deal with scalar field theories.
 
Using Eqs. (\ref{eq:prod_twist2}), (\ref{eq:twopointfunct}) and (\ref{eq:conformaldim}) the logarithm of the partition function on $\mathcal{R}_{n,\alpha}$ reads
\begin{equation}
\label{eq:moments}
\begin{split}
f=0: \quad &\ln Z_n(\alpha)=-4\ln \ell \displaystyle \sum_{k=0}^{n-1}\Delta_{k}(\alpha) = 
-\left[\frac{1}{3}\left(n-\frac{1}{n}\right)-\frac{\alpha^2}{2\pi^2 n}+\frac{|\alpha|}{\pi n}\right]\ln \ell, \\
f=1: \quad &   \ln Z_n(\alpha)=-4\ln \ell \displaystyle \sum_{k=-\frac{n-1}{2}}^{\frac{n-1}{2}}\Delta_{k}(\alpha) =-\left[\frac{1}{6}\left(n-\frac{1}{n} \right)+\frac{2}{n}\left( \frac{\alpha}{2\pi}\right)^2 \right]\ln \ell.
\end{split}
\end{equation}
The charged moments for the free massless Dirac field theory $(f=1)$ have been already worked out in the literature with different 
techniques \cite{Belin-Myers-13-HolChargedEnt,d-16,goldstein,xavier}. 
Instead, the charged moments for a free massless complex scalar field $(f=0)$ represent a new result (actually in 
Appendix A of \cite{Belin-Myers-13-HolChargedEnt} a result consistent with (\ref{eq:moments}) has been obtained using the heat kernel techniques). 

Let us stress that the presence of a flux in the Riemann surface changes some features of the twist fields in CFT: they remain primary operators (see Appendix \ref{app:twist} 
for details), but they do depend on the theory and are not anymore identified only by the central charge (see also \cite{goldstein}). 

\subsection{Massive field theory and flux insertion}\label{sec:ctheorem}

In this section we compute the charged moments $Z_n(\alpha)$ of a massive relativistic 2D QFT on the infinite line for a bipartition in two semi-infinite lines.
Thus, we follow the same logic as in \cite{cc-04} (i.e. the continuum version of Baxter corner transfer matrix approach \cite{baxter} for the 
reduced density matrix \cite{cc-04,ccp-10,eer-10,pkl-99}), 
which in turn parallels the proof of the c-theorem by Zamolodchikov \cite{z-86}. 
The results of this section are not limited to free theories but hold for {\it generic massive relativistic} QFT. 
Exploiting the rotational invariance about the origin of the Riemann surface $\mathcal{R}_{n,\alpha}$, the expectation values of the stress tensor of a massive euclidean QFT in complex coordinates, $T \equiv T_{zz}, \bar{T}\equiv T^*_{\bar{z}\bar{z}}$, and the trace, $\Theta \equiv 4T_{z\bar{z}}$, have the form
\begin{equation}\label{eq:defs}
\begin{split}
\braket{T(z,\bar{z})}&=F_{n,\alpha}(z\bar{z})/z^2, \\
\braket{\Theta(z,\bar{z})}-\braket{\Theta}_{1,\alpha=0}&=G_{n,\alpha}(z\bar{z})/z\bar{z}, \\
\braket{\bar{T}(z,\bar{z})}&=F_{n.\alpha}(z\bar{z})/\bar{z}^2,
\end{split}
\end{equation}
where $\braket{\Theta}_{1,\alpha=0}$ is a non-vanishing constant measuring the explicit breaking of scale invariance in the non-critical system, while $\braket{T}_{1,\alpha=0}$ and $\braket{\bar{T}}_{1,\alpha=0}$ both vanish.
These quantities are related by the conservation equations of the stress-energy tensor ($4\partial_{\bar z} T+\partial_z \Theta=0$) as
\begin{equation}
\label{eq:cons}
(z\bar{z})\left(F'_{n,\alpha}+\frac{1}{4}G'_{n,\alpha} \right)=\frac{1}{4}G_{n,\alpha}.
\end{equation}
The conservation equations as well as the rotational invariance are preserved in the presence of the flux $\alpha$ because the Riemann surface $\mathcal{R}_{n,\alpha}$ 
can be thought simply as a complex plane with the insertion of two modified twist fields, as discussed in the previous subsection.
Both $F_{n,\alpha}$ and $G_{n,\alpha}$ approach zero for $ |z| \gg \xi$, while when $|z| \ll \xi$, they approach the CFT values given by the conformal 
dimension of the modified twist field, $\Delta_n(\alpha)$ \cite{goldstein}. 
Hence we have
\begin{equation}\label{eq:bc}
\begin{split}
F^{CFT}_{n,\alpha} &\rightarrow \frac{c}{24}\left( 1-\frac{1}{n^2} \right)+\frac{\Delta_n(\alpha)}{n}, \\
G_{n,\alpha} &\rightarrow 0,
\end{split}
\end{equation}
and in particular for a massive Dirac field theory ($f=1$) and for a complex massive scalar theory ($f=0$), using the conformal weights (\ref{eq:conformaldim}), we have
\begin{equation}\label{eq:bc2}
\begin{split}
F^{f=1}_{n,\alpha} &\rightarrow \frac{1}{24}\left( 1-\frac{1}{n^2} \right)+\frac{1}{2n^2}\left(\frac{\alpha}{2\pi} \right)^2, \\ F^{f=0}_{n,\alpha} &\rightarrow \frac{1}{12}\left( 1-\frac{1}{n^2} \right)+\frac{1}{2n^2}\left[ \left(\frac{\alpha}{2\pi} \right)^2-\frac{|\alpha|}{2\pi}\right].  \\
\end{split}
\end{equation}
Changing variable to $R^2=z\bar z$, we can rewrite the Eq. (\ref{eq:cons}) as
\begin{equation}
\label{eq:cons1}
R^2 \frac{\partial}{\partial R^2}\left( F_{n,\alpha}(R^2)+\frac{1}{4}G_{n,\alpha}(R^2) \right)=\frac{1}{4}G_{n,\alpha}(R^2).
\end{equation}
The corresponding integrated form using the boundary conditions in Eq.\,(\ref{eq:bc}) is
\begin{equation}
\label{eq:int}
\int_0^{\infty} \frac{G_{n,\alpha}(R^2)}{R^2} dR^2=-\frac{c}{6}\left( 1-\frac{1}{n^2} \right)-\frac{4 \Delta_{n,\alpha}}{n}.
\end{equation}
Taking into account the normalisation of the stress tensor, the definition of $G_{n,\alpha}$ in Eq.\,(\ref{eq:defs}) and that $\displaystyle \int_{\mathcal{R}_n} \braket{\Theta_{n,\alpha}} \,dR^2$  corresponds to the variation of the free energy wrt a scale transformation (the mass $m$ in this case)  per each sheet of the whole $n$-sheeted surface, the left hand side of (\ref{eq:int}) is equal to
\begin{equation}
-\frac{2}{n}m\partial_m \ln Z_n(\alpha).
\label{der1}
\end{equation}
We can therefore integrate this equation at fixed $n$ and $\alpha$ to obtain $\ln Z_n(\alpha)$. 
The additive non-universal integration constant can be absorbed in a UV cutoff $\epsilon_{n,\alpha}$ that consequently depends both on 
the R\'enyi index $n$ and the parameter $\alpha$ (consistently with the lattice results in Ref. \cite{riccarda,MDC-19-CTM} for massless theories). 
Finally we get
\begin{equation}
\ln  Z_n(\alpha)=\left[ \frac{c}{12}\left( n-\frac{1}{n} \right)+2\Delta_n(\alpha)\right] \ln (m\epsilon_{n,\alpha}),
\label{eq:cthgen}
\end{equation}
or specialising to free Dirac  ($f=1$) or complex Klein-Gordon ($f=0$) fields
\begin{eqnarray}
\label{eq:cth}
f=1: &\quad&\ln  Z_n(\alpha)=\left[ \frac{1}{12}\left( n-\frac{1}{n} \right)+\frac{1}{n}\left(\frac{\alpha}{2\pi} \right)^2\right] \ln (m\epsilon_{n,\alpha}),\\
f=0: &&\ln  Z_n(\alpha)=\left[ \frac{1}{6}\left( n-\frac{1}{n} \right)-\frac{1}{n}\left(\frac{\alpha}{2\pi}\right)^2 +\frac{|\alpha|}{2\pi n} \right] \ln (m\epsilon_{n,\alpha}).
\label{eq:cth2}
\end{eqnarray}

We should mention that $\ln Z_n(\alpha)$ for the Klein-Gordon field matches the continuum limit of a chain  of complex oscillators 
obtained through the corner transfer matrix approach \cite{MDC-19-CTM}. 
We can specialise Eq. \eqref{eq:cthgen} 
to a Luttinger liquid with parameter $K$, whose underlying field theory is a $c=1$ CFT equivalent to a massless compact boson. In this case, one can use the results found in \cite{goldstein} for the conformal dimension of the modified twist field, obtaining
\begin{equation}
\ln  Z_n(\alpha)=\left[ \frac{1}{12}\left( n-\frac{1}{n} \right)+\frac{K}{n}\left(\frac{\alpha}{2\pi} \right)^2\right] \ln (m\epsilon_{n,\alpha}),
\label{ZnaK}
\end{equation}
which for $K=1$ gives the result found for fermions in Eq. (\ref{eq:cth}), as it should.

\subsection{From charged moments to symmetry resolved entropies}
\label{srl}
Performing the Fourier transforms of the charged moments above, one obtains symmetry resolved moments and entropies. 
For the Luttinger liquid, which includes Dirac Fermions at $K=1$, the $\alpha$ dependence of the leading term is the same as in the massless cases. 
Hence the analysis is identical to the one of Refs. \cite{goldstein,riccarda} and so we will just sketch the results here. 
The charged moments (ignoring for the time being the dependence on $n$ and $\alpha$ of
$\epsilon_{n,\alpha}$ in \eqref{ZnaK}),  are 
\begin{equation}
\label{Znq_FT}
\mathcal{Z}_n (q) \simeq (m\epsilon)^{\frac{1}{12} \left( n - \frac{1}{n} \right)} \sqrt{\frac{n \pi}{ K |\ln m\epsilon|}} e^{-\frac{n \pi^2 q^2 }{ K |\ln m\epsilon|}},
\end{equation}
and hence symmetry resolved entropies  
\begin{equation}
\label{Snq_FT}
S_n (q) = S_n - \frac12 \ln \left( \frac{K}{\pi} |\ln m \epsilon| \right)  + O(1), 
\end{equation}
with $S_n$ the total entropy. 
Exploiting the knowledge of ${\cal Z}_1(q)$ in \eqref{Znq_FT} we also easily get the number or  fluctuation entropy  
\begin{equation}
S_{\rm num}=-\int_{-\infty}^{\infty} {\cal Z}_1(q) \ln {\cal Z}_1(q)= \frac12 \ln \left( \frac{K}{\pi} |\ln m\epsilon| \right)  + O(1)\,,
\end{equation}
that in the sum for the total entropy cancels exactly the double logarithmic term in Eq. \eqref{Snq_FT}.

Although the massive complex boson has been already investigated in Ref. \cite{MDC-19-CTM}, there another critical limit has been taken. 
Here we are interested in the Fourier transform of Eq. \eqref{eq:cth2}.  
In the saddle point approximation, we can neglect the term $\propto \alpha^2$ in Eq. \eqref{eq:cth2} and the Fourier transform is 
\begin{equation}
\label{eq:FcritV}
\mathcal{Z}_n(q) \simeq Z_n(0)\frac{2n|\ln (m\epsilon)|}{4n^2\pi^2q^2 + \ln^2 (m \epsilon)} \simeq
Z_n(0) \frac{2n}{|\ln (m\epsilon)|} \left(1-\frac{4n^2\pi^2q^2}{\ln^2 (m \epsilon)} +\dots\right),
\end{equation}
and hence 
\begin{equation}
S_n(q)=S_n -\ln |\ln m \epsilon|+ O(1)\,,
\label{snqs}
\end{equation}
with $S_n$ the total entropy. 
Also in this case, one easily derives the number entropy from ${\cal Z}_1(q)$ obtaining again, at the leading order, the 
double logarithmic term in $S_n(q)$ in Eq. \eqref{snqs}, i.e. $S_{\rm num}=\ln (|\ln m\epsilon|)+ O(1)$. 

\section{The Green's function approach: The Dirac field}\label{sec:part}
In this section we derive the charged moments for a massive Dirac field for arbitrary mass and then consider the limits of 
small and large mass. 
In Sec. \ref{sec:twist} we showed that $Z_n(\alpha)$ can be written as product of partition functions $\zeta_a$ on the plane with proper boundary conditions along the cut $A$, 
explicitly given by
\begin{equation}\label{eq:bcsD}
\tilde{\psi}_k(e^{2\pi i} z,e^{-2\pi i} \bar{z})=e^{2\pi i a} \tilde{\psi}_k (z,\bar{z}), \qquad a=\frac{k}{n}+\frac{\alpha}{2\pi n},
\qquad k=-\frac{n-1}{2}, \cdots, \frac{n-1}{2}.
\end{equation}
Hence we have
\begin{equation}
\label{eq:sumpartition}
\ln Z_n(\alpha)=\sum_{k=-\frac{n-1}{2}}^{\frac{n-1}{2}} \ln\zeta_{\frac{k}{n}+\frac{\alpha}{2\pi n}}.
\end{equation}
Let us introduce the auxiliary {\it universal} quantities
\begin{equation}\label{eq:wa}
w_a\equiv\ell \partial_{\ell} \ln \zeta _a, \quad  \quad c_n(\alpha)\equiv\sum_k w_{\frac{k}{n}+\frac{\alpha}{2\pi n}},
\end{equation}
that, using (\ref{eq:sumpartition}), allow us to write the logarithmic derivative of the partition function in $\mathcal{R}_{n,\alpha}$ as
\begin{equation}\label{eq:c-funct}
c_n(\alpha)=\ell \frac{\partial \ln  Z_n(\alpha)}{\partial \ell}
\qquad\Rightarrow
\qquad
\ln  Z_n(\alpha)=\int_{\epsilon}^\ell \frac{c_n(\alpha)}{\ell'} d\ell' .
\end{equation}
For $n=1$, the function $c_n(\alpha)$ is the analogue of Zamolodchikov's $c$-function \cite{z-86} in the presence of the flux $\alpha$.
The cutoff $\epsilon$, in analogy to what discussed in Sec. \ref{sec:ctheorem} depends on both $\alpha$ and $n$, although we almost always 
omit such a dependence for conciseness. 

As already discussed in section \ref{sec:replica}, the key observation of this approach relies on the identity between the partition function $\zeta_a$ and the Green's function in
the same geometry (see Eq. (\ref{eq:Greenvspartition})). 
Through this relation, the function $w_a$ has been already obtained for generic values of $a$ for the massive Dirac fermion \cite{CFH,ch-rev}. 

\begin{figure}
\centering
\subfigure
{\includegraphics[width=0.48\textwidth]{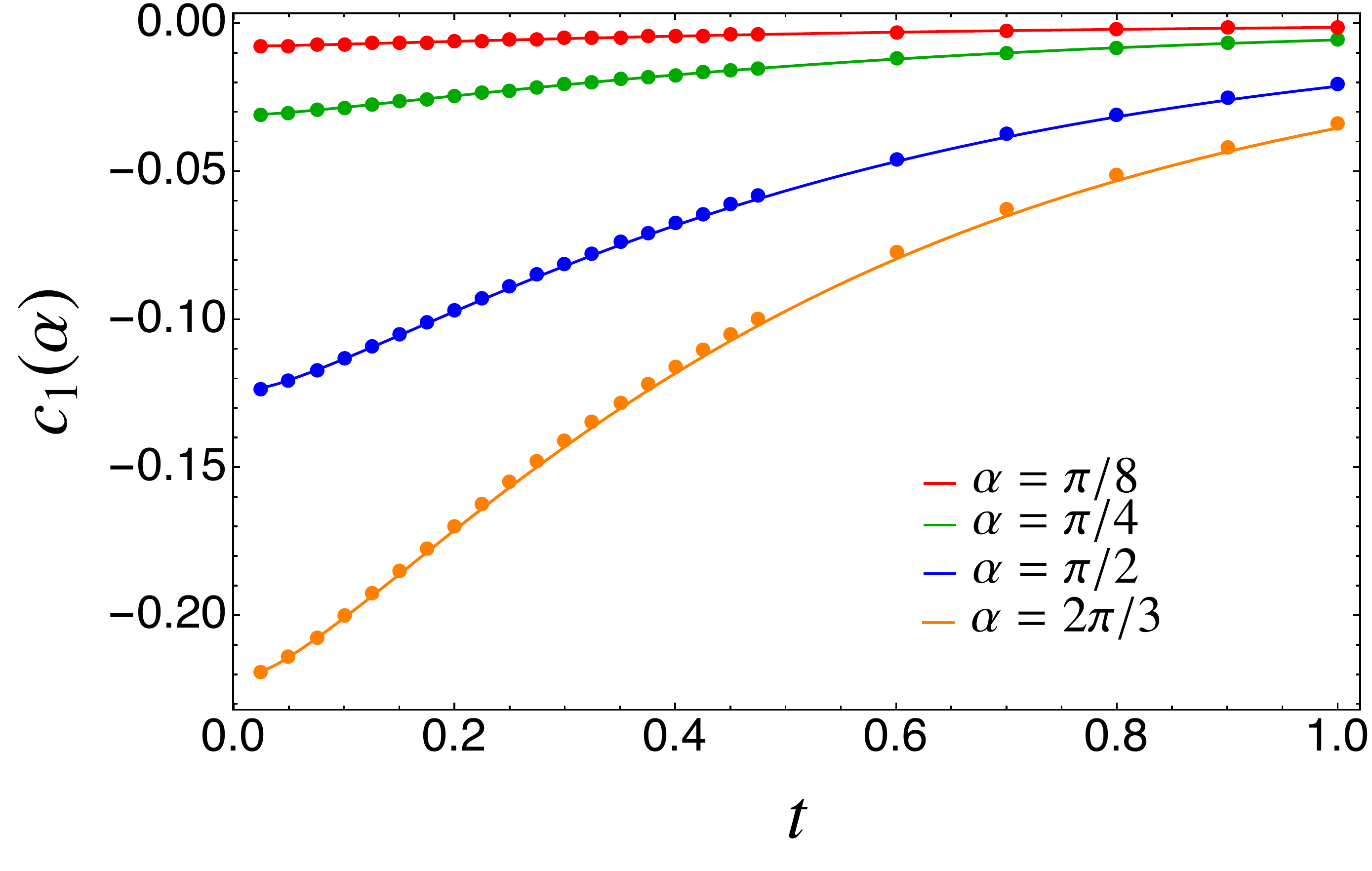}}
\subfigure
{\includegraphics[width=0.48\textwidth]{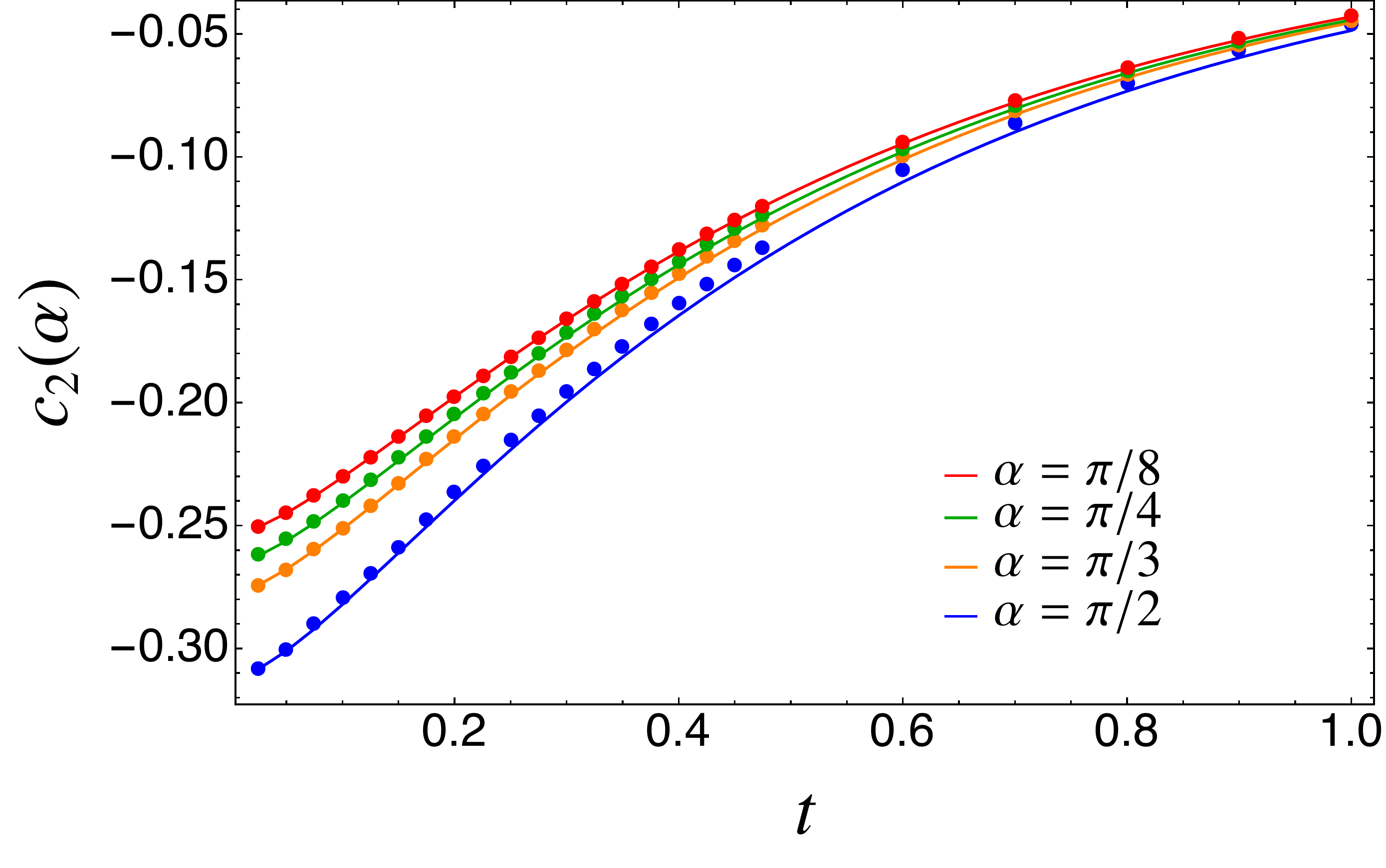}}
\caption{The universal constant $c_n(\alpha)$ extracted from the numerical solution of the Painlev\'e equation (\ref{eq:PainleveFermion}) for different values of $\alpha$ and $n$ 
as a function of $t=m\ell$ (full lines). 
The numerical data are obtained varying $\ell$ between $200$ and $400$ lattice points and properly choosing $m$ in such a way $t=m\ell \in (0,1)$. 
For larger $\alpha$ and $n$, we need larger subsystem size to have a good match between field theory and lattice calculation because lattice corrections become stronger.}
\label{fig:NumericalPainleveF}
\end{figure}

The method that we just reviewed provides exact results for the charged moments of a free Dirac field. 
Indeed, in Ref. \cite{CFH} it has been shown that the function $w_a$ defined in (\ref{eq:wa}) can be written as 
\begin{equation}\label{eq:wa1}
w_a(t)=-\int_{t}^{\infty} y v_a^2(y) dy,
\end{equation}
where $t=m \ell$ and $v_a$ is the solution of the Painlev\'e V equation
\begin{equation}
\label{eq:PainleveFermion}
v_a''+\frac{v_a'}{t}=-\frac{v_a}{1-v_a^2}(v_a')^2+v_a(1-v_a^2)+4\frac{\left(a-\frac{1}{2}\right)^2}{t^2(1-v_a^2)}v_a.
\end{equation}
This equation can be straightforwardly solved numerically with any standard algorithm for ordinary differential equations, once we impose the boundary condition as $t \to 0$
\cite{CFH} 
\begin{equation}\label{eq:bcF}
v_a(t)=-2a(\ln t +\kappa_D(a))+O(t^2),
\end{equation} 
where 
\begin{equation}
\kappa_D(a)=-\ln 2 +2\gamma_E+\frac{1}{2}(\psi(a)+\psi(-a)),
\end{equation}
with $\psi(z)\equiv \Gamma'(z)/\Gamma(z)$  the digamma function and 
$\gamma_E$ the Euler-Mascheroni  constant.
Plugging the numerical solution of the differential equation \eqref{eq:PainleveFermion} 
into Eq. \eqref{eq:wa}, we obtain the universal  constant $c_n(\alpha)$.  
Then, with the further integration \eqref{eq:c-funct}, the desired $\ln Z_n(\alpha)$ is found to the price of introducing the non-universal cutoff $\epsilon$.
As examples we report in Fig. \ref{fig:NumericalPainleveF} the plots of the resulting $c_n(\alpha)$ for few values of $\alpha$ and $n$ as functions of $t=m\ell$. 
In the figure we also compare our exact solution with the numerical results obtained from a lattice discretisation of the free Dirac theory 
(see Appendix \ref{app:lat} for details).  The agreement is excellent. We stress that in Fig. \ref{fig:NumericalPainleveF} 
there is no free parameter in matching analytical and numerical data for $c_n(\alpha)$ (as a difference compared to $Z_n(\alpha)$).

The method we just outlined provides exact results for the desired charged moments  and, by Fourier transform, the symmetry resolved entropies.
However, the procedure is completely numerical and we would appreciate an analytic handle on the subject. While in general this is not feasible, 
the limits of small and large $t$ are analytically treatable, as we are going to show.

\subsection{The expansion close to the conformal point $m\ell=0$}
\label{sec:CHfermionsCL}
Here we use the methods just introduced to derive an asymptotic expansion of the charged moments close to the conformal point, i.e.  for $t=m \ell \to 0$. 
In this limit, the expansion of the function $w_a(t)$ has been worked out in Ref. \cite{CFH}, obtaining
\begin{equation}
\label{eq:fermi}
w_a=-2a^2+a^2(1-2\kappa_D+2\kappa_D^2+(4 \kappa_D-2)\ln t +2\ln^2 t)t^2-2a^4t^4 \ln ^4t+{O}(t^4\ln ^3t),
\end{equation}
where we omitted the dependence on $a$ of $\kappa_D$. 
In order to compute $c_n(\alpha)$ through Eq. (\ref{eq:wa}), we again set $a=\frac{k}{n}+\frac{\alpha}{2\pi n}$ and we compute the following sums 
\begin{subequations}\label{eq:kappa0}
\begin{align}
\sum_k a^2 = &\frac{1}{12}\left(n-\frac{1}{n} \right)+\frac{\alpha^2}{4 n \pi^2} ,\label{eq:kappa0a} \\
\Omega_n(\alpha)\equiv & \sum_{k} a^2 (\psi(a)+\psi(-a)) \equiv \Omega_n(0)+\frac{\alpha^2}{2\pi^2 n} \omega_n +  \rho_n^\omega(\alpha), \\
\Lambda_n(\alpha)\equiv &\sum_{k} a^2(\psi(a)+\psi(-a))^2 \equiv  \Lambda_n(0) +\frac{\alpha^2}{2\pi^2 n} \lambda_n +  \rho^\lambda_n(\alpha) , 
\end{align}
\end{subequations}
where $\omega_n= \pi^2 n \Omega''_n(\alpha)$ and $\lambda_n= \pi^2 n \Lambda''_n(\alpha)$ so that the remainder functions 
$\rho_n^{\omega/\lambda}(\alpha)$ are $O(\alpha^4)$.
All the sums over $k$ run from $-\frac{n-1}{2}$ to $\frac{n-1}{2}$. 
The quantities $\Omega(n,\alpha)$ and $\Lambda(n,\alpha)$ (and their derivatives) can be rewritten using the integral representation 
$\psi(x)=-\gamma_E+\int_0^{1}dy \frac{1-y^{x-1}}{1-y}$ for the digamma function $\psi(x)$. 
This procedure allows for an analytic continuation in $n$, as detailed in appendix \ref{app:dirac}.

\begin{figure}
\centering
\subfigure
{\includegraphics[width=0.48\textwidth]{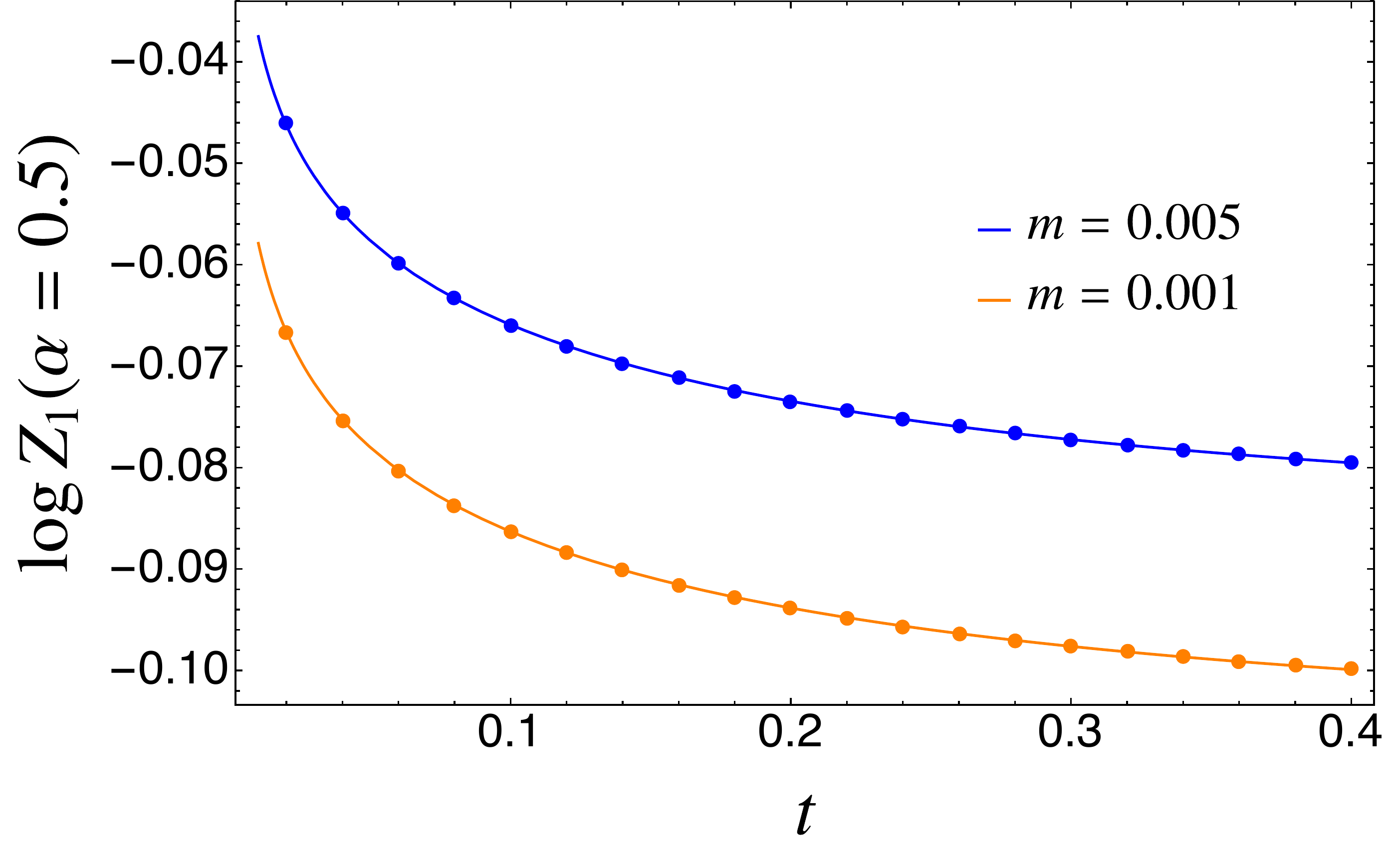}}
\subfigure
{\includegraphics[width=0.48\textwidth]{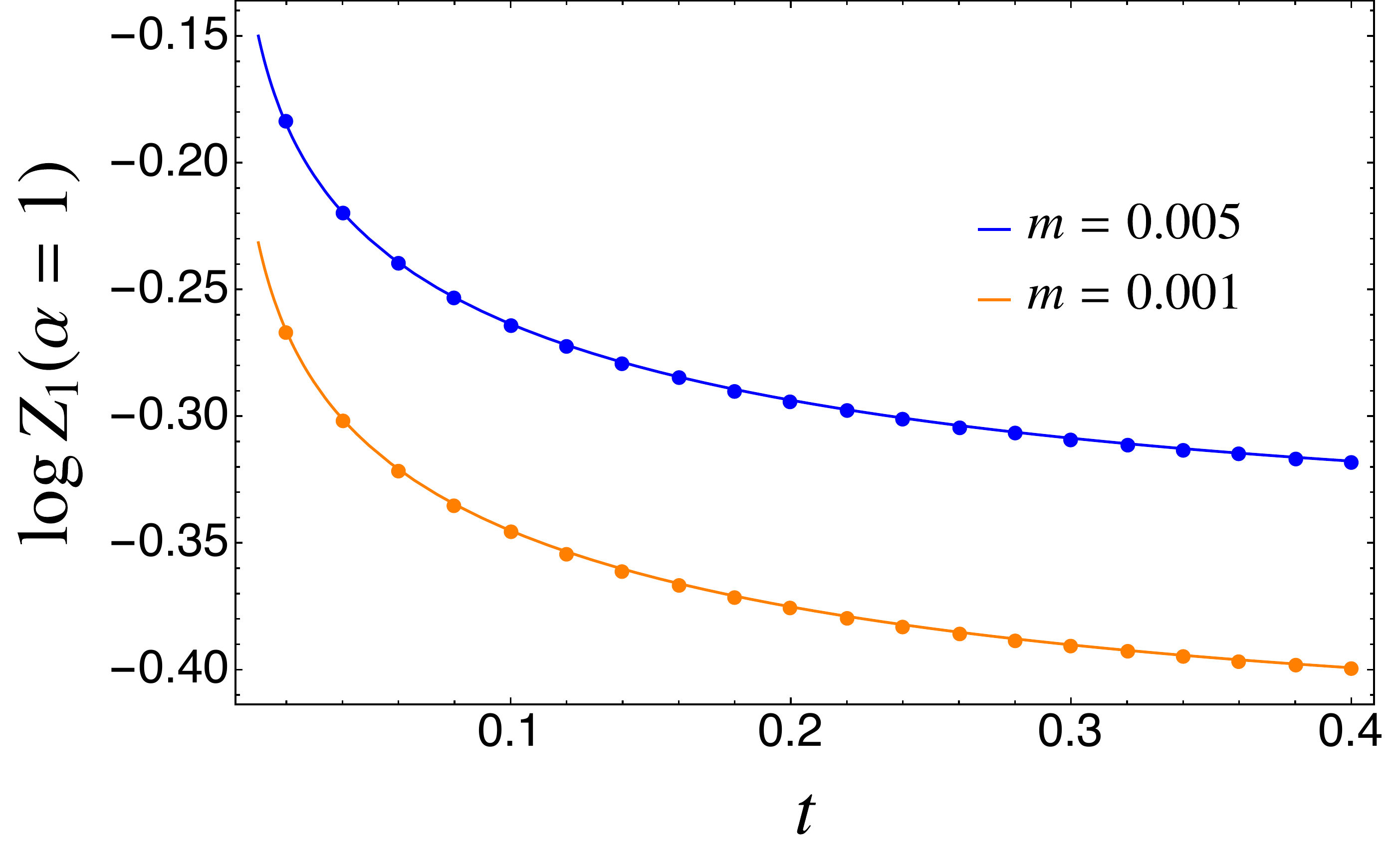}}
\caption{ Leading scaling behaviour of the charged R\'enyi entropies with the insertion of a flux $\alpha$. The numerical results (symbols) for two different values of $\alpha$ and masses $m$ are reported as functions of $t=m\ell$ when $n=1$. The data match well the prediction in Eq. (\ref{eq:totalsc}) (solid lines) which includes lattice corrections
as explained in the text.}
\label{fig:ZalphaF}
\end{figure}

From Eqs. (\ref{eq:fermi})-(\ref{eq:kappa0}) we obtain up to ${O}(t^2)$
\begin{equation}\label{eq:tointegrate1}
\begin{split}
c_n(\alpha)=&
\frac{\partial \ln Z_n(\alpha)}{\partial \ln \ell}
=
\displaystyle \sum_{k=-(n-1)/2}^{(n-1)/2}w_{\frac{k}{n}+\frac{\alpha}{2\pi n}}
=
\left( \frac{1-n^2}{6n}-\frac{\alpha^2}{2\pi^2 n} \right) (1-t^2\ln^2 t) +\\
&+\left[\left( \frac{1-n^2}{6n}  -\frac{\alpha^2}{2\pi^2 n} \right)(1+2\ln 2-4\gamma_E)
+2\Omega_n(\alpha)\right] t^2\ln t + \\
&- \left[\left( \frac{1-n^2}{12n}  -\frac{\alpha^2}{4\pi^2 n} \right)(1+2\ln 2-4\gamma_E+2(\ln 2-2\gamma_E)^2)+\right.  \\
&+   (1-4\gamma_E+2\ln 2)\Omega_n(\alpha)-\frac{\Lambda_n(\alpha)}{2} \Big] t^2 +{O}(t^4\ln ^3t). 
\end{split}
\end{equation} 
Eq. (\ref{eq:tointegrate1}) can be now integrated analytically, getting 
\begin{equation}\label{eq:totalsc}
\begin{split}
\ln Z_n(\alpha)=
-\left(\frac{1}{6}\left( n-\frac{1}{n}\right)+\frac{\alpha^2}{2\pi^2 n} \right)  \ln \frac{\ell}{\epsilon} + y_n(t)- \frac{\alpha^2}{2\pi^2 n} z_n(t)+ \rho^z_n(\alpha, t) +o(t^3), 
\end{split}
\end{equation}
where we defined 
\begin{eqnarray}
y_n(t)&=&{\frac{t^2}{6}\left ( n-\frac{1}{n}\right) \Big ( \frac{1}{2}\ln t^2 - \ln t(1-2\gamma_E+\ln 2)  +\frac{3}{4}+2\gamma_E^2 +\frac{\ln^2 2}{2} } \\
& & -2\gamma_E(1+\ln 2)+\ln 2 \Big )+  (\ln t -(\ln 2+1-2\gamma_E)) t^2\Omega_n(0)+ \frac{t^2}{4}\Lambda_n(0),  \nonumber \\
z_n(t)&=& t^2\Big[ -\frac{\ln t^2}2 + \ln t(1-2\gamma_E+\ln 2 -\omega_n ) + \\ & &
-\frac{3}{4} -2\gamma_E^2-\frac{\ln^2 2}2 +2\gamma_E(1+\ln 2)-\ln 2 + (\ln 2+1-2\gamma_E) \omega_n - \frac{\lambda_n}{4}  \Big], \nonumber\\
\rho^z_n(\alpha, t)&=&t^2[ (\ln t -(\ln 2+1-2\gamma_E)) \rho^\omega_n(\alpha)+\rho^\lambda_n(\alpha)] ,
\end{eqnarray}
and $\rho_n^z(\alpha)$ is defined so that $\rho_n^z(\alpha)=O(\alpha^4)$.
Notice, as we already stressed a few times, that in Eq. \eqref{eq:totalsc} the cutoff $\epsilon$ comes as an additive constant of integration and it generically 
depends on both $n$ and $\alpha$.

Eq. \eqref{eq:totalsc} represents our final field theoretical result for the charged entropies. 
We wish to test this prediction against exact lattice computations obtained with the methods in Appendix \ref{app:lat}.
However, in order to perform a direct comparison with lattice data, we have to take into account the additional non-universal contribution coming from the discretisation of 
the spatial coordinate, i.e. the explicit expression for the cutoff $\epsilon$ in Eq.~\eqref{eq:totalsc} that, as already mentioned, does depend on $\alpha$ and cannot be 
simply read off from the result at $\alpha=0$. 
We assume here (as Eq. \eqref{eq:totalsc} suggests at leading order) that the cutoff does not depend on the mass; 
consequently we can use the exact value for $m=0$ \cite{riccarda} obtained with the use of Fisher-Hartwig techniques. 
The final result of Ref. \cite{riccarda} may be written as 
\begin{multline}
\label{upsilon}
\left(\frac{1}{6}\left( n-\frac{1}{n}\right)+\frac{\alpha^2}{2\pi^2 n} \right)\ln (2\epsilon)= \\
\Upsilon_n{(\alpha)}= {n i}\int_{-\infty}^\infty  dw [\tanh( \pi w)-\tanh (\pi n w+i\alpha/2)]  \ln \frac{\Gamma(\frac12 +iw)}{\Gamma(\frac12 -iw)} \,, 
\end{multline}
and in particular we will use
\begin{equation}
\gamma(n)\equiv \frac12 \frac{\partial^2\Upsilon_n(\alpha)}{\partial \alpha^2}\Big|_{\alpha=0}
=\frac{ni}{4}\displaystyle \int_{-\infty}^{\infty} dw [\tanh^3(\pi n w )-\tanh(\pi n w)]\ln \dfrac{\Gamma(\frac{1}{2}+iw)}{\Gamma(\frac{1}{2}-iw)}.
\label{gamman}
\end{equation}
In Ref. \cite{riccarda} it has been shown that the cutoff in \eqref{upsilon} is very well described by the quadratic expansion in $\alpha$ and higher corrections 
$O(\alpha^4)$ are negligible for most practical purposes.

\begin{figure}
\centering
\subfigure
{\includegraphics[width=0.49\textwidth]{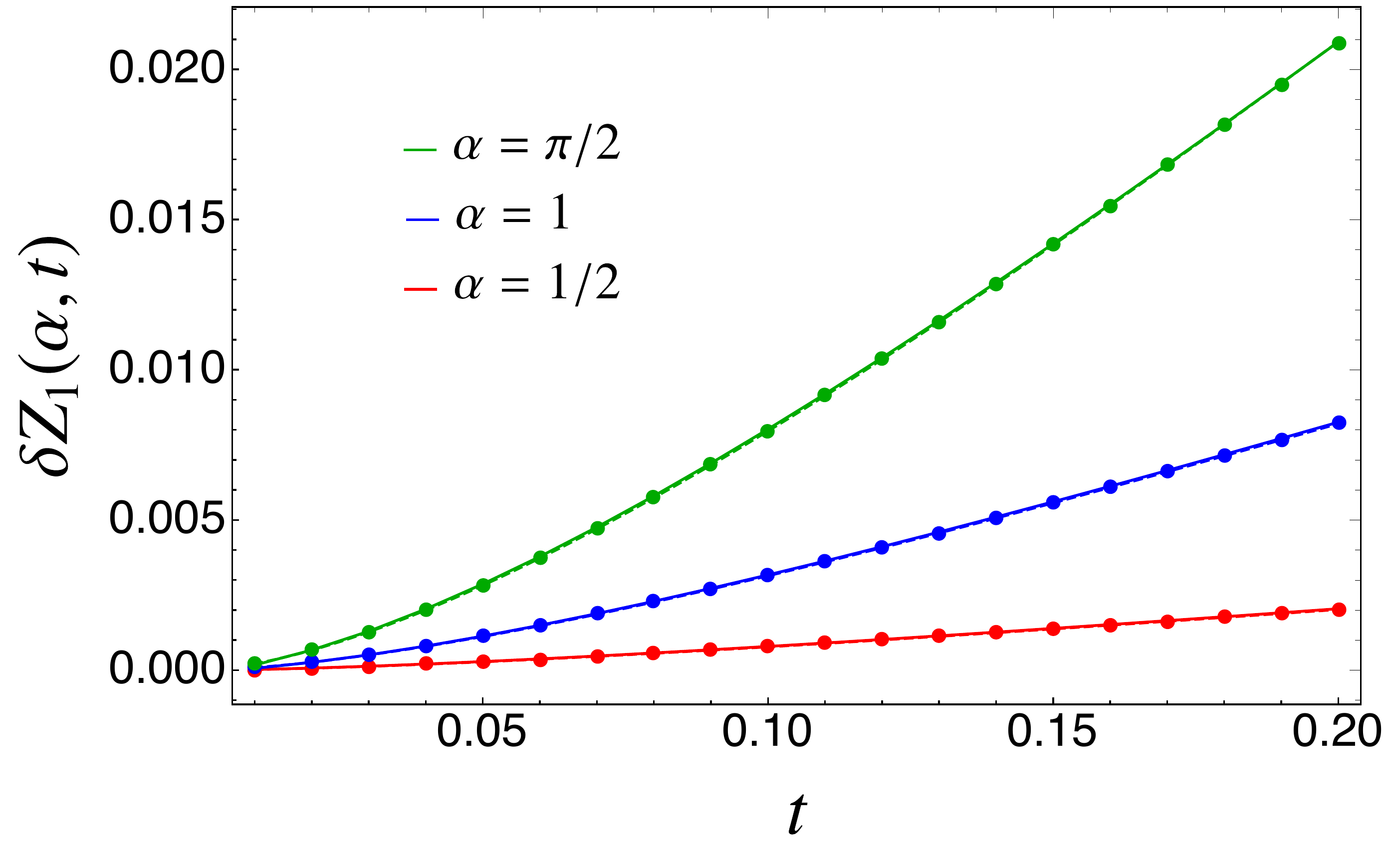}}
\subfigure
{\includegraphics[width=0.49\textwidth]{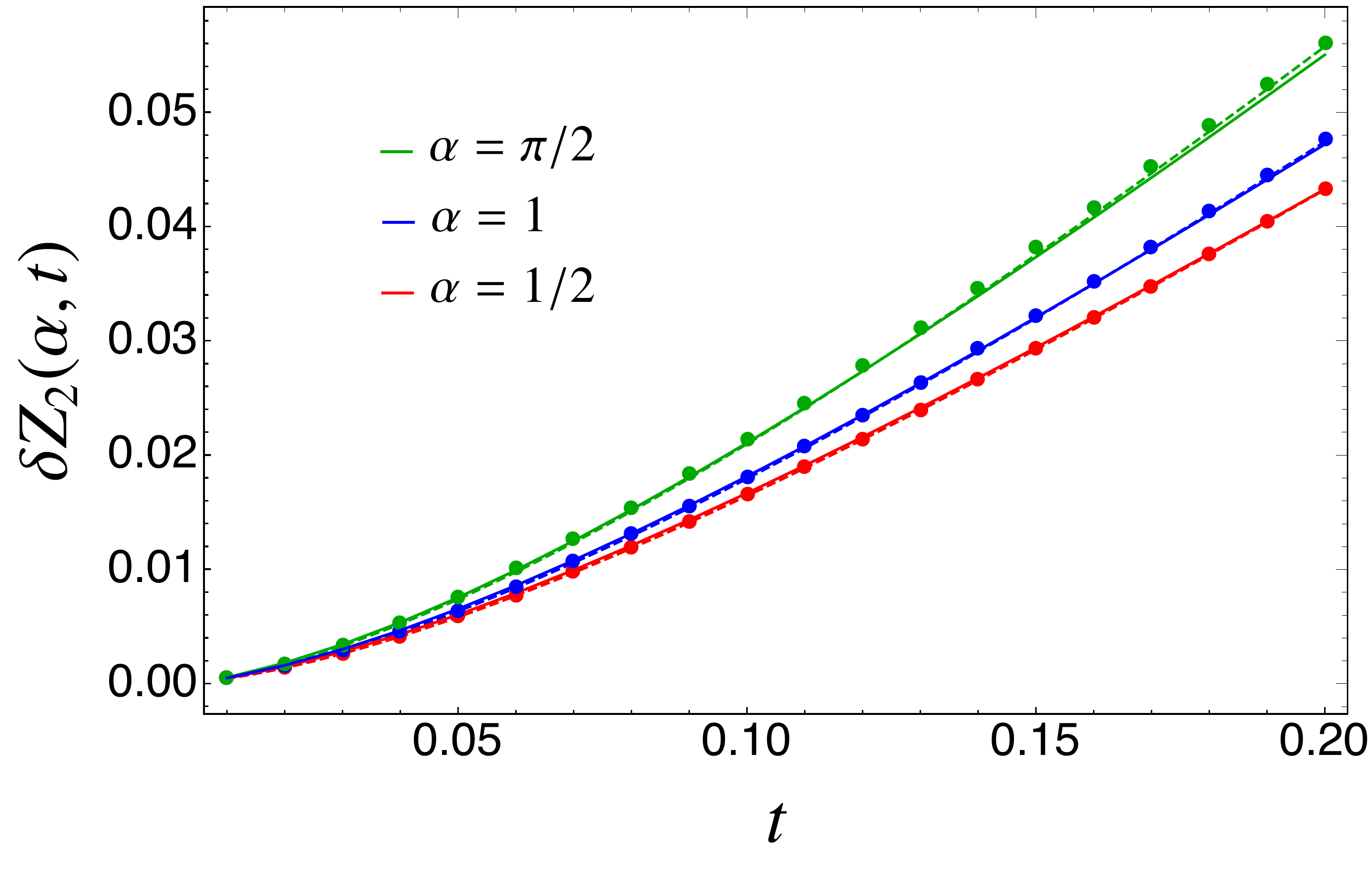}}
\caption{Subtracted universal charged entropy $\delta Z(\alpha,t)$ in Eq. \eqref{zeta}. 
Left (right) panel is for $n=1$ ($n=2$). 
The dashed lines are the small $t$ expansion in Eq. (\ref{eq:totalsc}) for $n=1$ while the solid lines are the Painlev\'e exact solution. 
The tiny discrepancies observed in some cases are finite $\ell$ corrections. 
 }
\label{fig:tcontribution}
\end{figure} 
 
In Fig. \ref{fig:ZalphaF} we report the numerical data for the charged moments with the insertion of a flux $\alpha$ for two values of $\alpha$ and $m$ with $n=1$. 
The data are well described by the theoretical prediction (\ref{eq:totalsc}) with the cutoff \eqref{upsilon}.  
Finally, in order to have a test of the prediction \eqref{eq:totalsc} that does not rely on an independent lattice calculation we can 
consider the difference between the charged entropy at finite $t$ (i.e. finite mass) and the massless one. 
Specifically we consider
\begin{equation}
\delta Z(\alpha,t)= \ln Z_n(\alpha,m) -\ln Z_n(\alpha,m=0),
\label{zeta}
\end{equation}
in which both the cutoff and $\ell$ dependences cancel and it becomes a {\it universal} function solely of $t$ (closely related to $c_n(\alpha)$).
The results for $\delta Z(\alpha,t)$ are reported in Fig. \ref{fig:tcontribution}. 
The agreement of the numerics with the prediction \eqref{eq:totalsc} is perfect for small $t$.
Furthermore, the differences emerging for larger $t$ are correctly captured by the numerical exact solution of the Painlev\'e equation \eqref{eq:PainleveFermion}. 
The small discrepancies visible in the figure are just finite size effects that are stronger for larger values of $n$ and $\alpha$.

\subsection{From the charged moments to symmetry resolution}

\begin{figure}
\centering
\subfigure
{\includegraphics[width=0.48\textwidth]{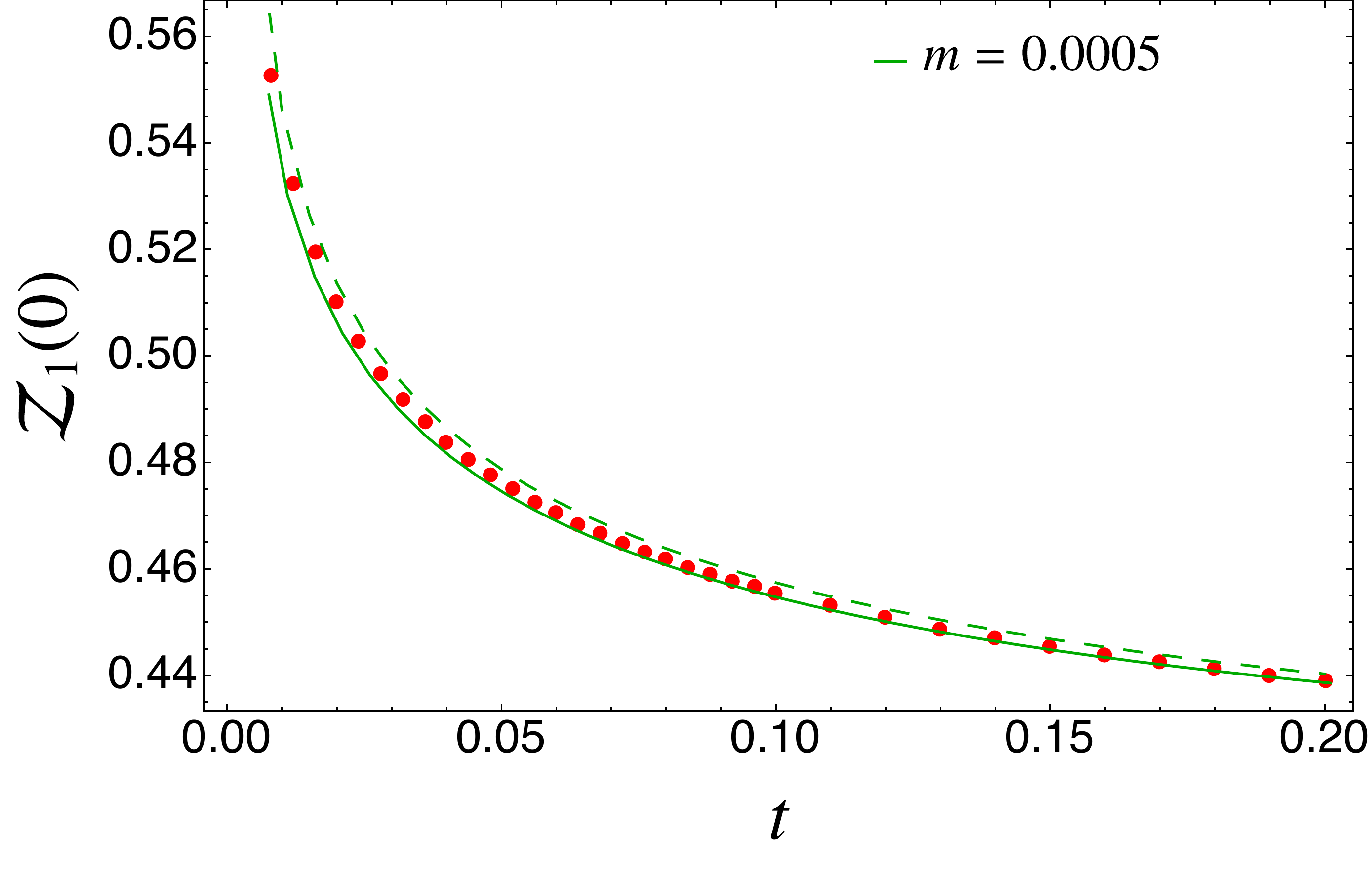}}
\subfigure
{\includegraphics[width=0.48\textwidth]{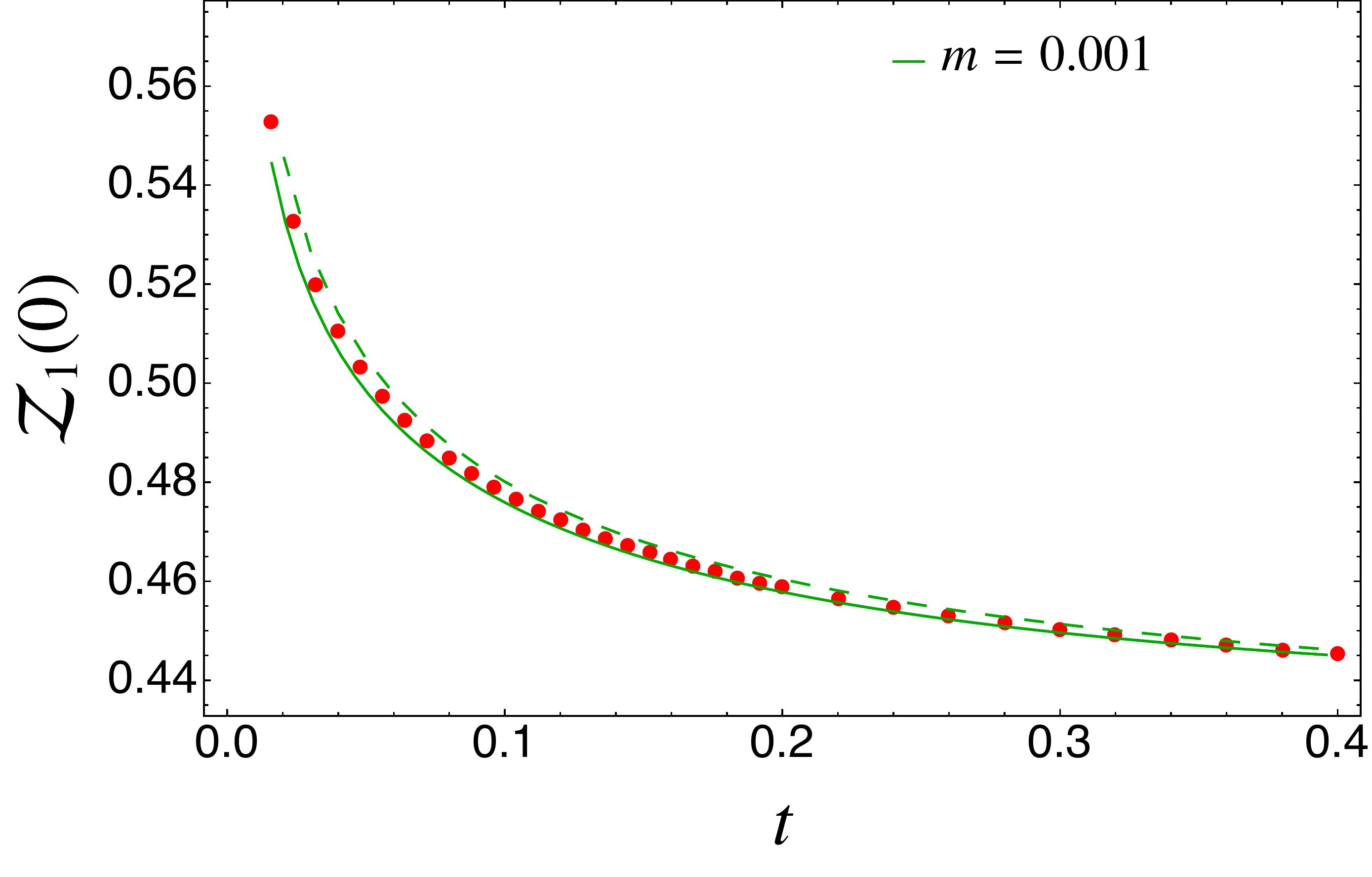}}
\subfigure
{\includegraphics[width=0.49\textwidth]{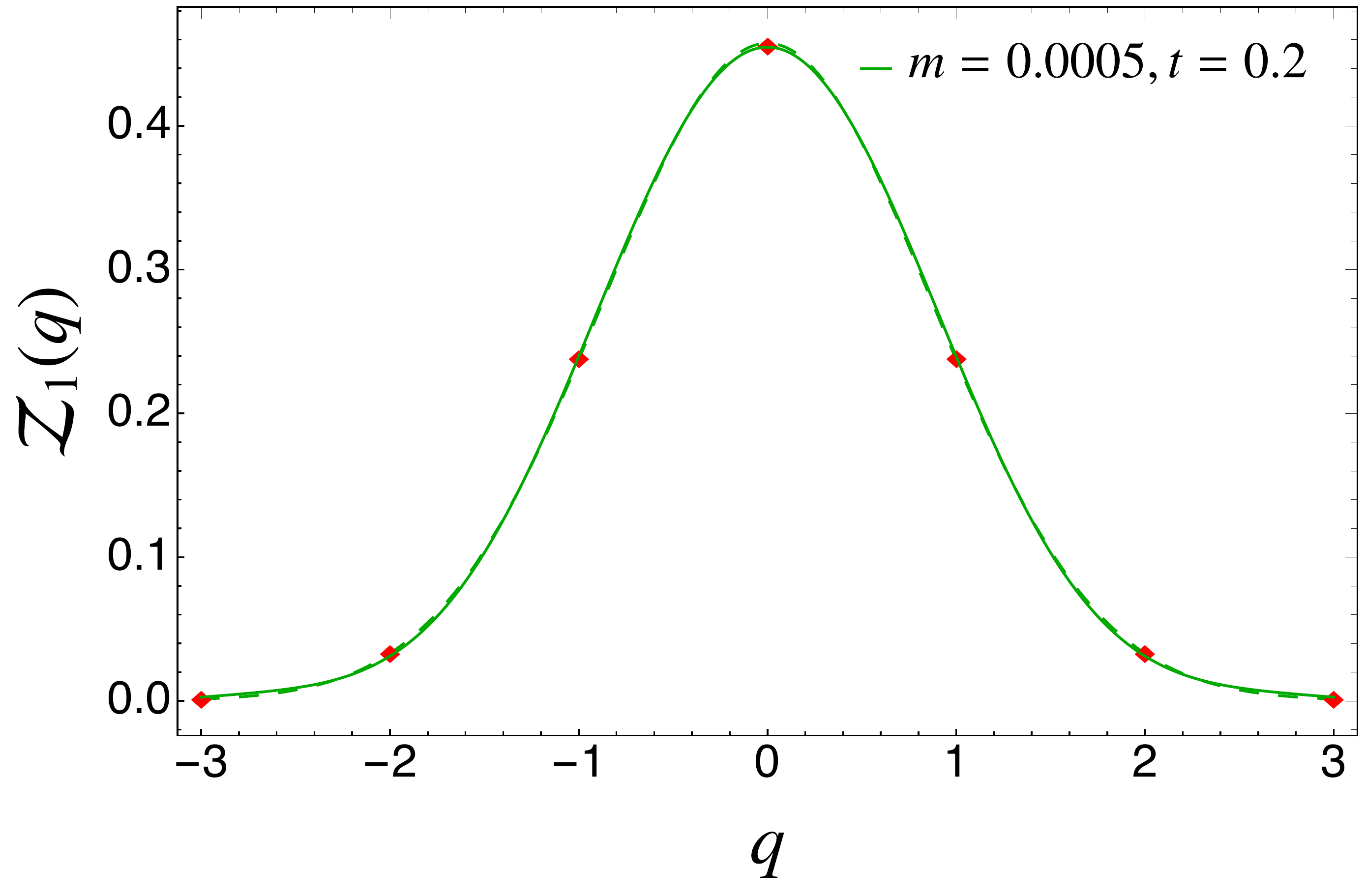}}
\subfigure
{\includegraphics[width=0.49\textwidth]{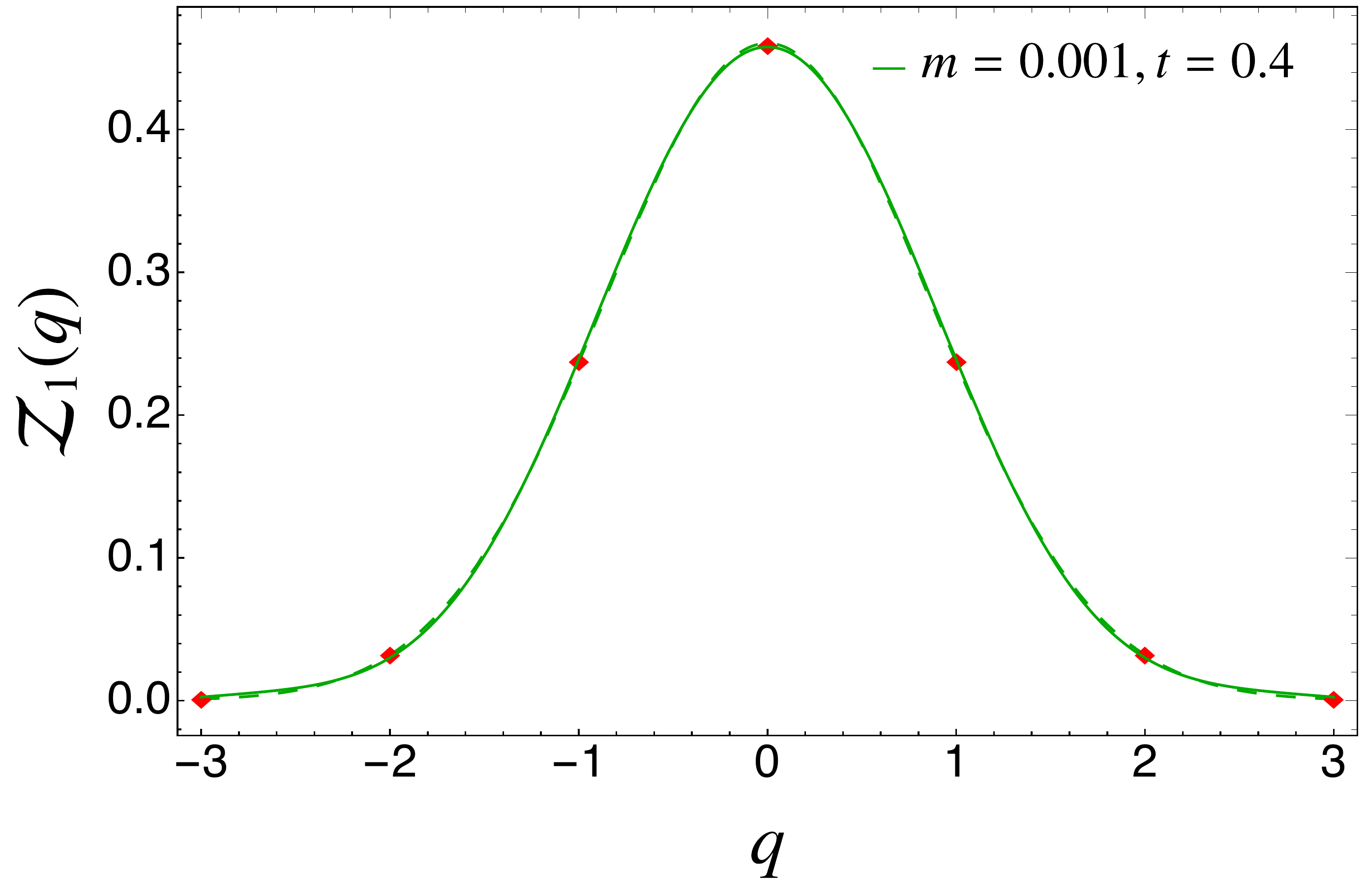}}
\caption{The probability $\mathcal{Z}_1(q)$. Top:  As a function of $t=m\ell$ at fixed $q=0$ for mass $m=0.0005$ (left) and $m=0.001$ (right).
The dashed green line is $\mathcal{Z}_1(q)$ obtained by the saddle point approximation, i.e. Eq.\,(\ref{eq:SP-FTrans-step3}). 
The solid green line is the exact Fourier transform without taking the quadratic approximation. 
For large $\ell$ (and $t$ as a consequence) the saddle-point approximation converges to the exact value, as expected.
Bottom: The same at fixed $t$ as function of $q$.
}
\label{fig:znq}
\end{figure}

We are now ready to study the true symmetry resolution by performing the Fourier transform of $Z_n(\alpha)$.
In this Fourier transform we ultimately use a saddle-point approximation in which $Z_n(\alpha)$ is Gaussian and hence 
we truncate hereafter Eq. (\ref{eq:totalsc}) at quadratic order in $\alpha$.
Consequently, the charged partition function can be well approximated as
\begin{equation}
Z_n(\alpha)= Z_n(0)  e^{- b_n \alpha^2/2}, 
\end{equation}
where 
\begin{equation}
b_n(\ell,t)=\frac{1}{\pi^2 n}(\ln\ell + z_n(t)) -2\gamma(n)+\frac{\ln 2}{\pi^2 n}\equiv \frac{1}{\pi^2 n}\ln \ell  -h_n,
\label{bn}
\end{equation}
where we consistently approximated the cutoff at quadratic level and used the lattice cutoff with $\gamma(n)$ given in Eq. \eqref{gamman}.
A different cutoff just leads to a different additive constant in $b_n$ (i.e., a different definition of $h_n$), but we will use its precise form only for the comparison with 
numerics and so all the following formulas are completely general.

Now we can compute the Fourier transform (\ref{eq:defF}) that reads
\begin{equation}
\label{eq:FTrans}
\mathcal{Z}_n(q)=Z_n(0)\displaystyle \int_{-\pi}^{\pi}\dfrac{d\alpha}{2\pi}e^{-iq\alpha }e^{-\alpha^2 b_n(\ell,t)/2}.
\end{equation}
When $\ell\to\infty$, we can perform the integral by saddle point approximation and the integration domain can be extended to the whole real line.
We end up in a simple Gaussian integral, obtaining
\begin{equation}
\label{eq:SP-FTrans-step3}
\mathcal{Z}_n(q)=\frac{Z_n(0)}{\sqrt{2\pi b_n(\ell,t)}}e^{-\frac{q^2}{2 b_n(\ell,t)}}.
\end{equation}
We check Eq. (\ref{eq:SP-FTrans-step3}) against numerical computations in Figure \ref{fig:znq} focusing on $n=1$ 
and the agreement is perfect. 
We test both the scaling with $t=m \ell$ for fixed $q$ and at fixed $t$ as a function of $q$.

\begin{figure}
\centering
\subfigure
{\includegraphics[width=0.45\textwidth]{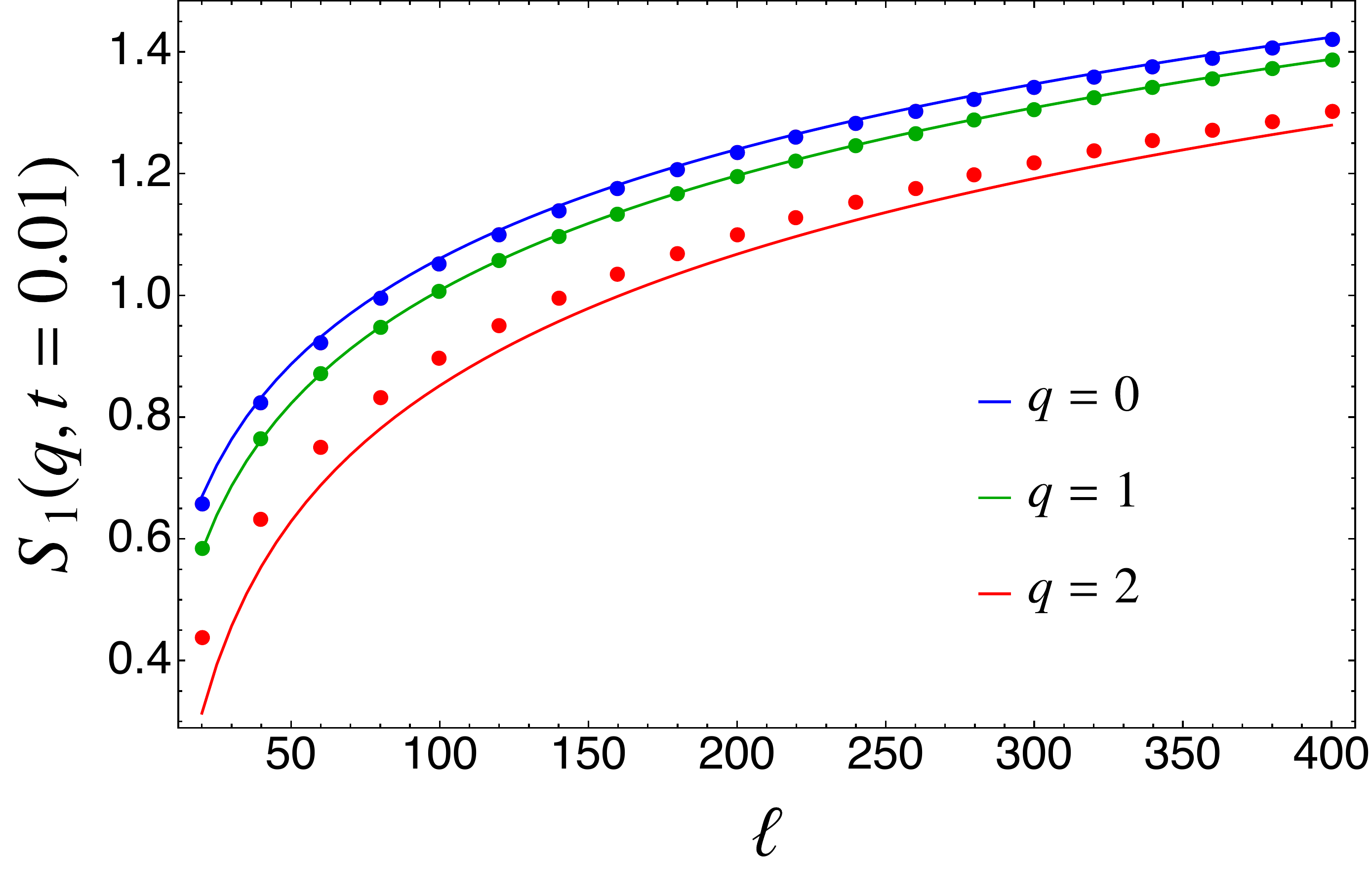}}
\subfigure
{\includegraphics[width=0.45\textwidth]{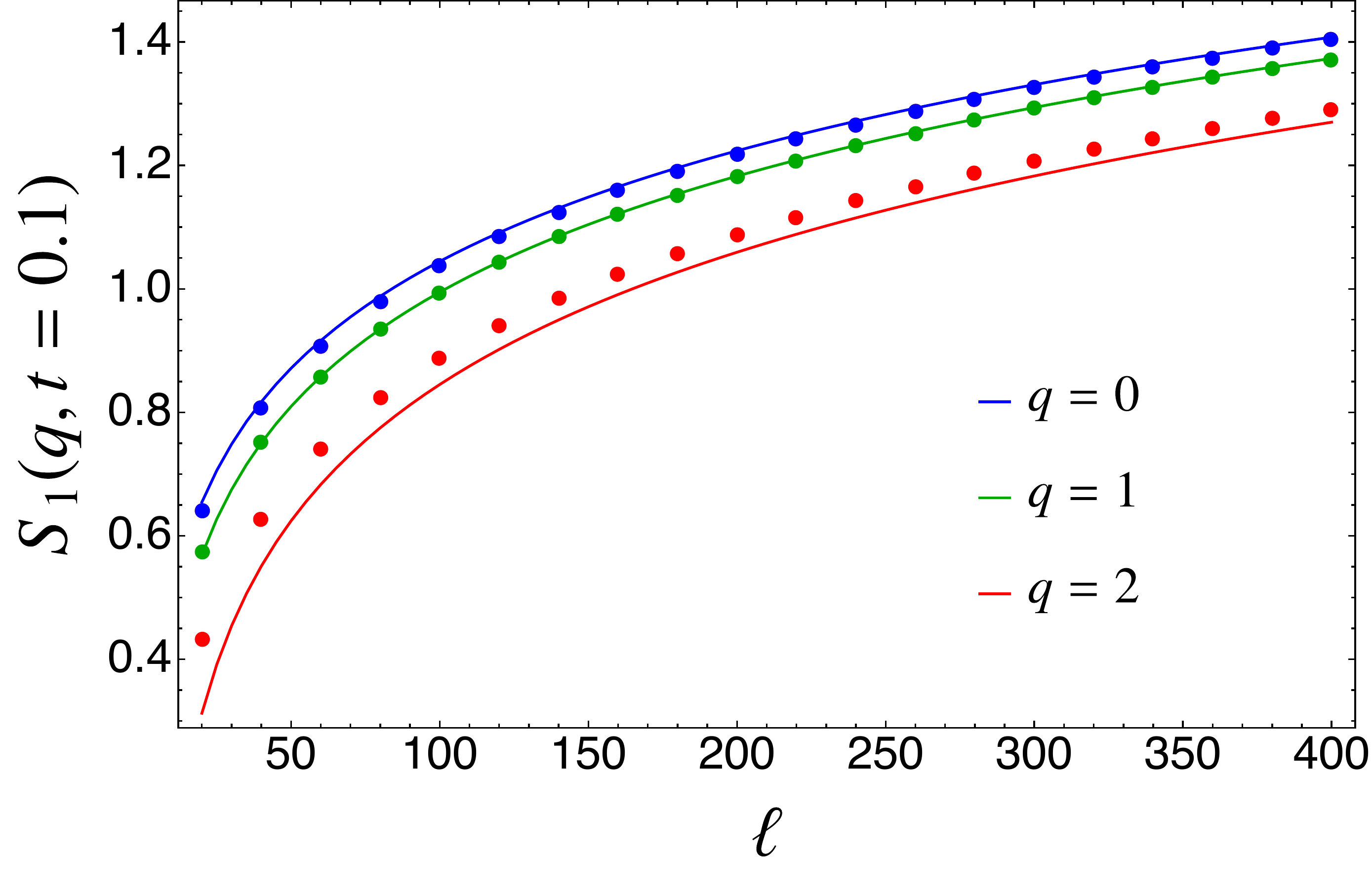}}
\subfigure
{\includegraphics[width=0.45\textwidth]{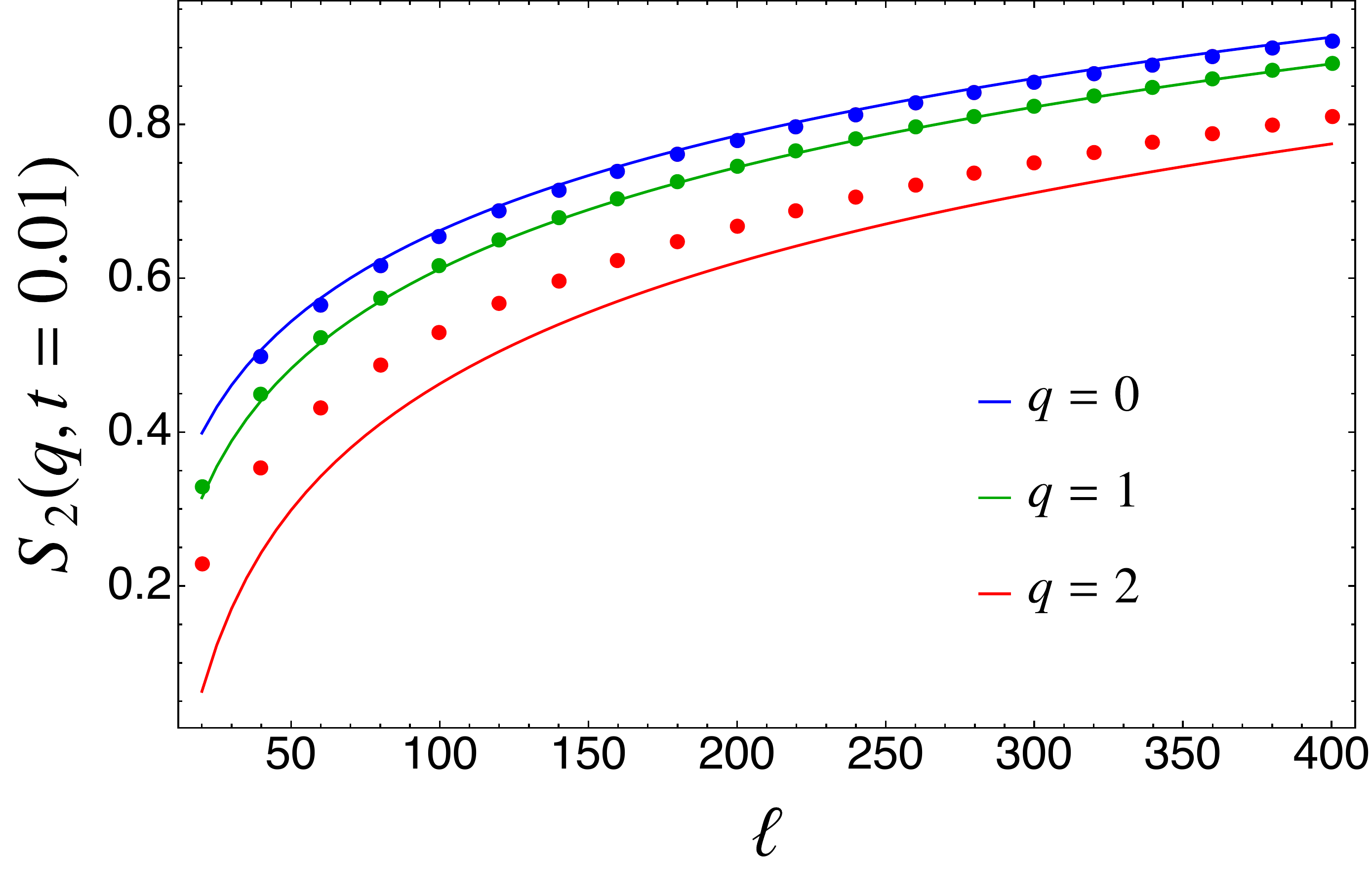}}
\subfigure
{\includegraphics[width=0.45\textwidth]{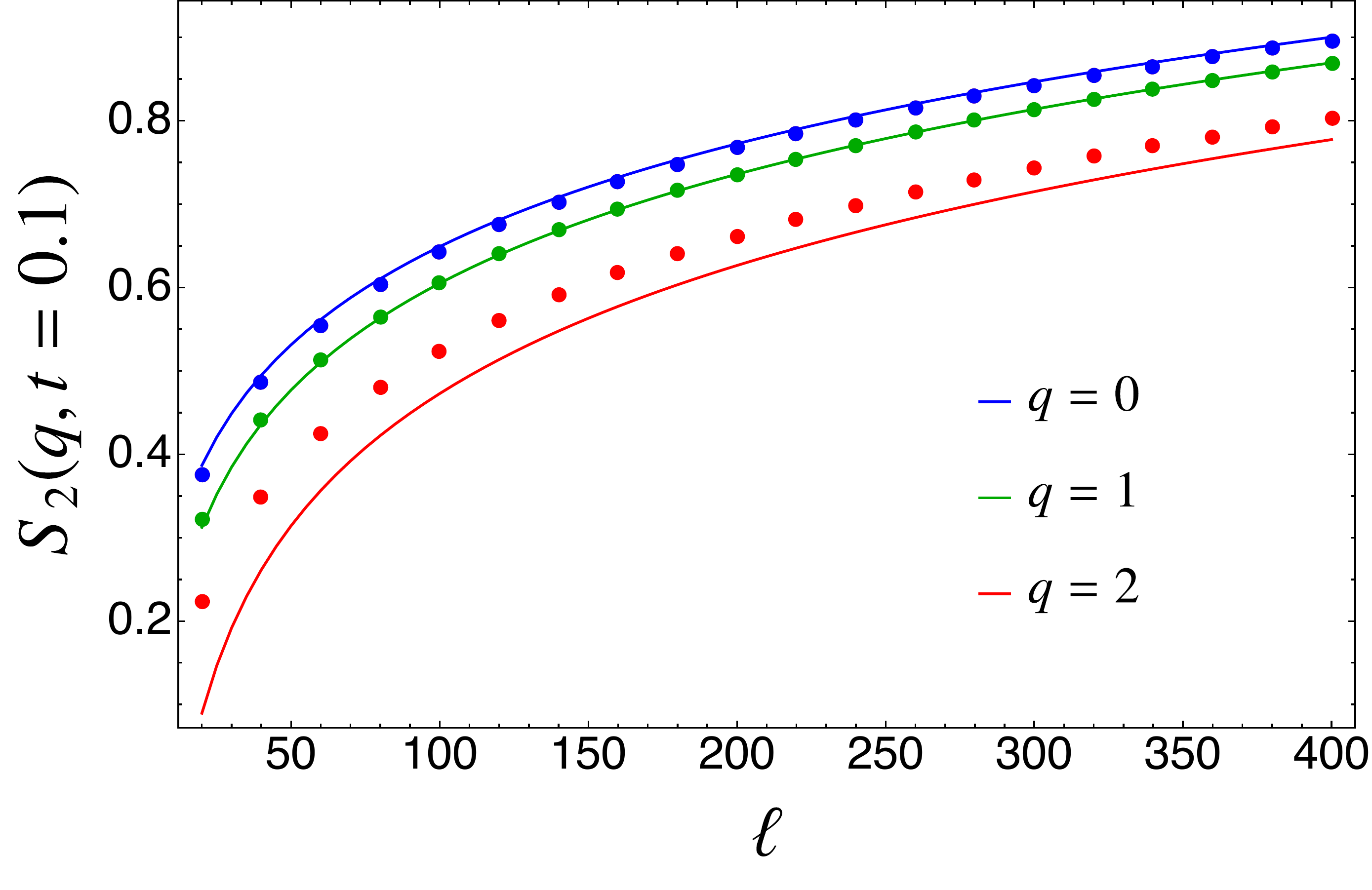}}
\caption{Symmetry resolved entanglement entropies for a few different values of $q$ and $n$ as functions of $\ell$.
The field theory prediction is tested against exact lattice computations. The agreement with Eq. (\ref{eq:SP-SRRE-v2Order}), that includes lattice effects, is remarkable.
For large $|q|$, the approximation at the order $q^2$ is no longer sufficient and neglected corrections to the scaling become important, as 
well known for the massless case \cite{riccarda}. 
}
\label{fig:snq}
\end{figure}

Now we are ready to compute the symmetry resolved R\'enyi entropies.  From the definition (\ref{eq:SREE1}) we have 
\begin{equation}
\label{eq:SP-SRRE}
S_n(q)=
S^D_n-\frac12\ln(2\pi)+\frac{1}{1-n}\ln\frac{b_1(\ell,t)^{n/2}}{b_n(\ell,t)^{1/2}}-\frac{q^2}{2(1-n)}\left( \frac{1}{b_n(\ell,t)}-\frac{n}{b_1(\ell,t)} \right),
\end{equation}
where $S^D_n$ is the total $n$-th R\'enyi entropy for the Dirac fields (cf. Eq. (\ref{eq:entr3}) up to ${O}(t^2\ln^2 t)$).
We can further expand the above equation for $\ell\to\infty$ since $b_n(\ell,t)$ diverges logarithmically, obtaining 
\begin{multline}
\label{eq:SP-SRRE-v2Order}
S_n(q)=
S^D_n-\frac{1}{2} \ln \Big(\frac{2}{\pi} \ln \delta_n \ell \Big)+ \frac{\ln n}{2(1-n)}-\frac{\pi^4n(h_1-nh_n)^2}{4(1-n)^2(\ln \ell)^2}+ 
\\ +
q^2n \pi^4\frac{h_1-nh_n}{2(1-n)(\ln \ell \kappa_n)^2} + o(\ln \ell^{-2}),
\end{multline}
where 
\begin{equation}
\label{eq:deltan}
\ln \delta_n=-\dfrac{\pi^2 n  (h_n-h_1)}{1-n},
\end{equation}
and 
\begin{equation}
\label{eq:kappan}
\ln \kappa_n=-\pi^2\frac{(h_1+n h_n)}{2}.
\end{equation}
The above formula is valid also for the symmetry resolved Von Neumann entropy taking properly the limits of the various pieces as $n\to 1$. 
By construction, the total entropy, $S^D_n$, coincides with the one obtained in \cite{CFH} for the massive fermions in the conformal limit up to ${O}(t^2)$.

\begin{figure}
\centering
\subfigure
{\includegraphics[width=0.45\textwidth]{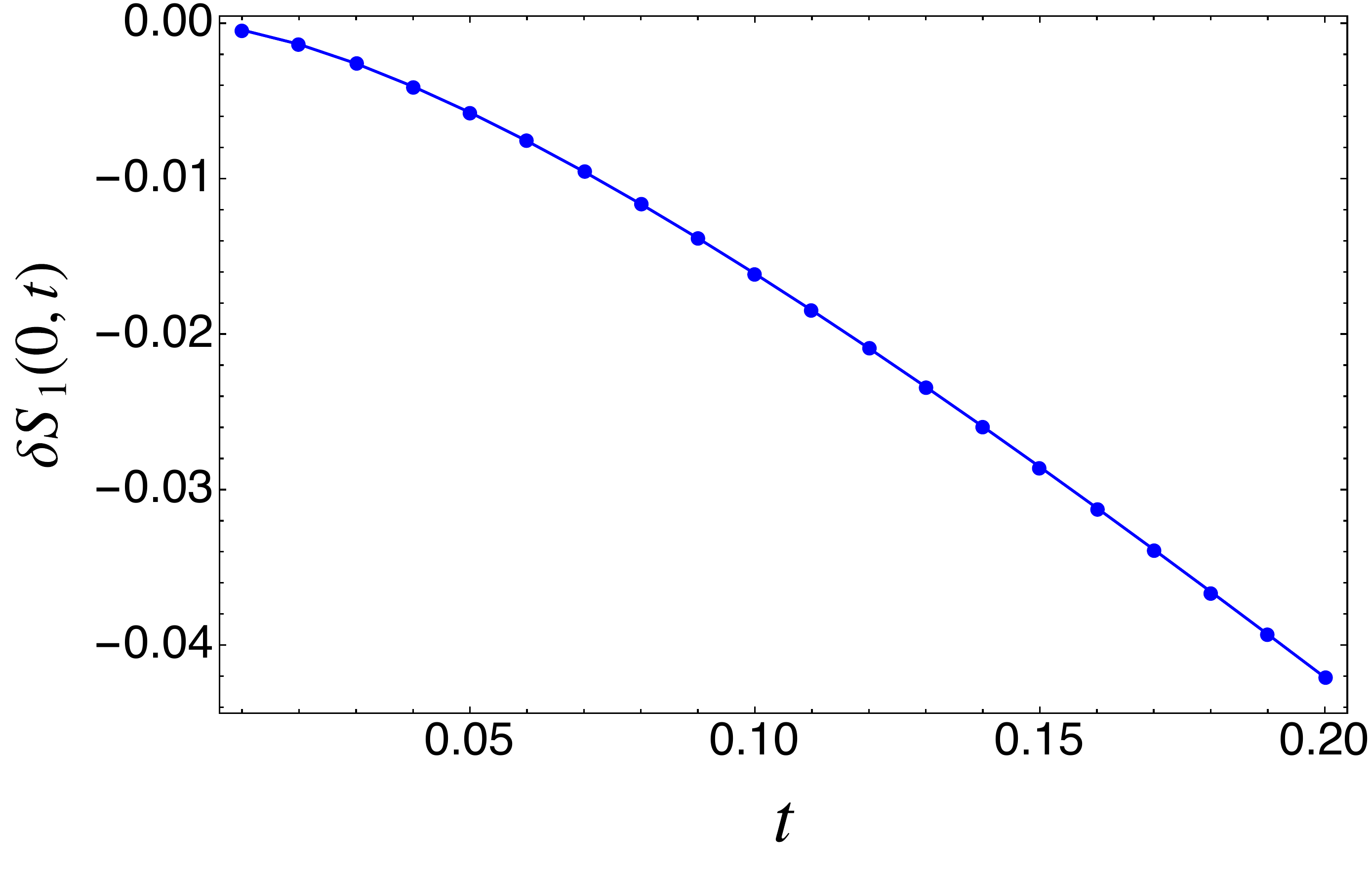}}
\subfigure
{\includegraphics[width=0.45\textwidth]{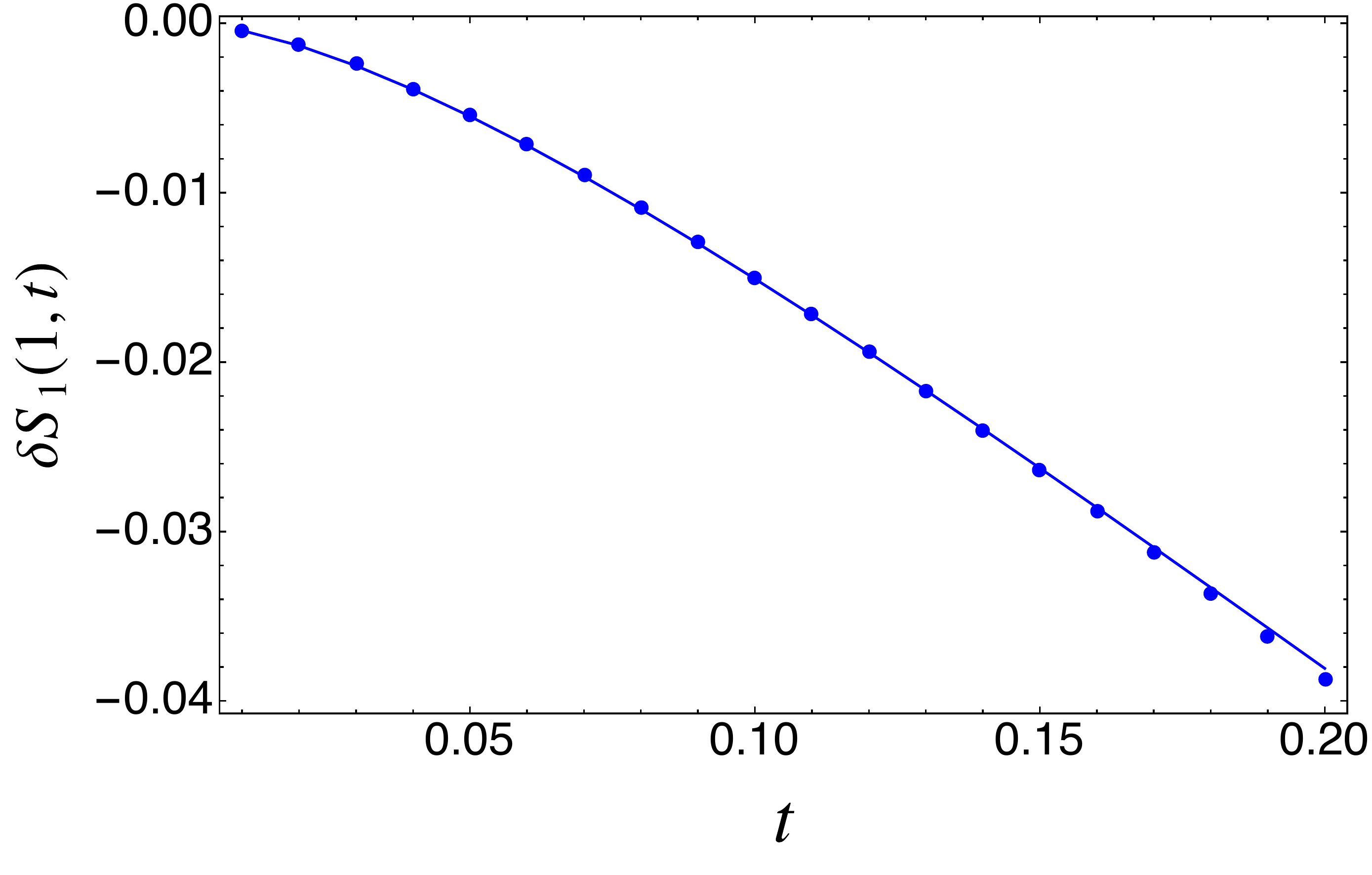}}
\caption{
Subtracted symmetry resolved von Neumann entropy  $\delta S_1(q,t)\equiv S_1(q,t)-S_1(q,t=0)$ for $q=0$ (left) and $q=1$ (right) as a function of $t$
(fixing $\ell =600$ and varying $m$). 
This subtracted quantity highlights the mass dependence of symmetry resolved entropies. 
The continuous lines are just the difference of the same subtracted entropies as obtained from the field theory expansion~\eqref{eq:SP-SRRE-v2Order}. 
}
\label{fig:snq2}
\end{figure}

Let us critically discuss the result in Eq. (\ref{eq:SP-SRRE-v2Order}).
The leading terms for large $\ell$ (up to $O((\ln\ell)^{-2})$) do not depend on $q$ and they are given by the total entropies $S_n^D$ in Eq. (\ref{eq:entr3}).
We then conclude that at this order, the presence of the mass does not break entanglement equipartition found in conformal field theory \cite{xavier}.
The first term breaking equipartition is at order $O((\ln\ell)^{-2})$ and its amplitude is governed by the constant $h_n$ defined in Eq. \eqref{bn}. 
This constant gets contributions both from the non-universal cutoff and from the mass; the two contributions have the same analytic features. 
In Fig. \ref{fig:snq} we test the accuracy of our total prediction against exact lattice numerical calculations. 
The agreement is remarkable for small values of $|q|$, but it worsens already at $q=2$; this does not come as a surprise since the same trend 
was already observed in the massless case \cite{riccarda}. 
Such discrepancies are entirely due to corrections of order $o(q^2)$ and are expected to reduce as $\ell$ gets larger. 
The drawback of the data reported in Fig. \ref{fig:snq} is that universal field theory mass contributions and the lattice non-universal terms are mixed up 
 and the latter are, by far, the largest one.   
It is then very difficult to observe the dependence on the mass in these plots.  
An effective and easy way to highlight the role of the mass is
to subtract from the symmetry resolved entropies their value for the massless case, i.e. considering the numerical evaluation of 
the  $\delta S_n(q,t)\equiv S_n(q,t) -S_n(q,0)$. 
Such subtracted entropies for $n=1$ and $q=0,1$ are reported in Fig. \ref{fig:snq2}, showing that the entropy is a monotonous decreasing function of $t$ 
(and hence of $m$ at fixed $\ell$).

\subsection{The long distance expansion.}\label{sec:LDF}
In this subsection we move to the analysis of the charged and symmetry resolved entropies in the limit of large $t$. 
The most effective way to proceed is, following Ref.  \cite{CFH}, to employ in Eq. (\ref{eq:PainleveFermion}) a boundary condition for $t\to\infty$, 
that takes the form \cite{CFH}
\begin{equation}\label{eq:bcL}
v_a(t)\sim \frac{2}{\pi} \sin (\pi a )K_{2a}(t),
\end{equation}
where $K_{2a}(t)$ is the modified Bessel function of the second kind. 
This is the starting point for a systematic expansion for large $t$ of the solution $v_a(t)$ of the differential equation  (\ref{eq:PainleveFermion}).
Plugging the resulting expansion into the integral (\ref{eq:wa1}) for $w_a(t)$, we get
\begin{equation}
w_a(t)=-e^{-2t}\frac{\sin^2(a\pi)}{\pi}\left(1+\frac{-1+16a^2}{4t}+O(t^{-2})\right).
\end{equation}
Summing over $a=\frac{k}{n}+\frac{\alpha}{2\pi n}$, we obtain the long distance asymptotic expansion for the universal factor $c_n(\alpha)$ 
\begin{equation}\label{eq:long}
c_n(\alpha)=\frac{e^{-2t}}{2\pi}\left( -n+\frac{(4-n^2)\pi^2-12\alpha^2}{12n t \pi^2}-\frac{2\csc \frac{\pi}{n}(\pi\cot \frac{\pi}{n} \cos \frac{\alpha}{n}+\alpha \sin \frac{\alpha}{n}) }{\pi n t }+O(t^{-2})\right),
\end{equation}
and for $n=1$ 
\begin{equation}
c_1(\alpha)=-\frac{e^{-2t}}{\pi}\sin^2\frac{\alpha}{2}\left(1+\frac{ \frac{4\alpha^2}{\pi^2}-1}{4t}+O(t^{-2})\right).
\label{eq:long2}
\end{equation}
This is consistent with the exact result $c_1(0)=0$ coming from the normalisation of the reduced density matrix. 
For $\alpha=0$, Eq. \eqref{eq:long} reproduces the known results \cite{CFH}.

\begin{figure}
\centering
\subfigure
{\includegraphics[width=0.48\textwidth]{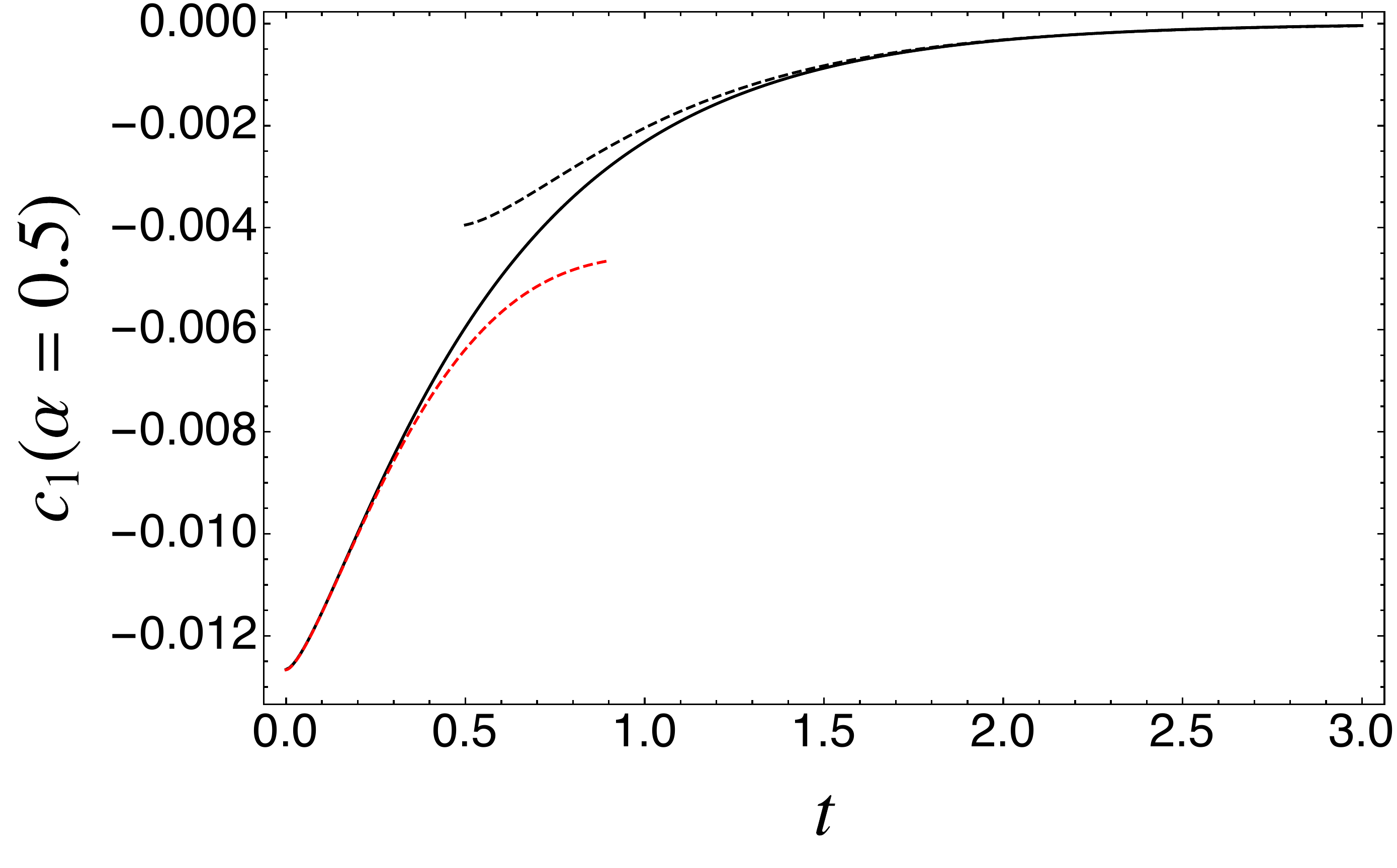}}
\subfigure
{\includegraphics[width=0.48\textwidth]{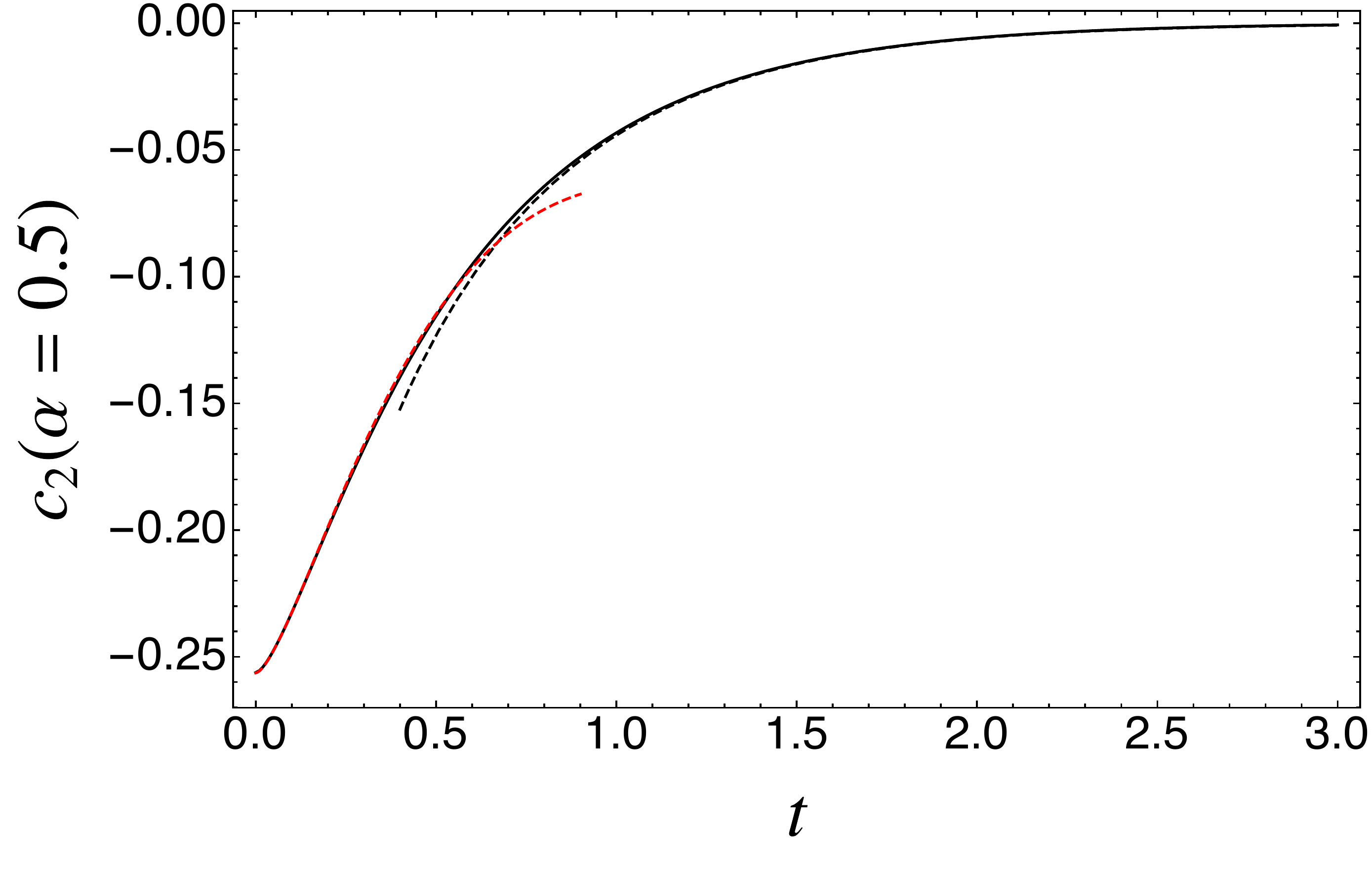}}
\caption{The solid lines are the functions $c_n(\alpha)$ obtained as exact numerical solutions of the differential equation (\ref{eq:PainleveFermion}). 
The dashed lines are the leading terms in the expansions for short (red) and long (black) distances, i.e. Eqs. (\ref{eq:tointegrate1}) and (\ref{eq:long}), respectively.}
\label{fig:c1}
\end{figure}

In Fig. \ref{fig:c1} we report the numerical exact solution of the Painlev\'e equation (\ref{eq:PainleveFermion}) for $c_n(\alpha)$; 
we focus on  $n=1,2$ and plot $c_n(\alpha)$ as a function of $t$. 
For large $t$, the solutions perfectly match the asymptotic expansions \eqref{eq:long} and \eqref{eq:long2}
(for completeness we also show the small $t$ expansion in Eq. (\ref{eq:tointegrate1})). 
Let us emphasise the presence of a discontinuity in $c_n(\alpha)$ for $n \to 1$ as a function of $n$:
it is due to the non-commutativity of the limits $n\to 1$ and $t \to \infty$, as well known and discussed at length in the literature for $\alpha=0$ \cite{CFH,cd-09}.
We show here that the presence of $\alpha$ does not cancel such a discontinuity, although for $\alpha\neq0$ the leading term is of the same order $e^{-2t}$.

The charged entropy is simply given by the integral
\begin{equation}
\label{eq:longmod0}
\ln Z_n(\alpha)= \int_{m\epsilon}^{m\ell} \frac{c_n(\alpha)}{t}dt.
\end{equation}
At large $t$, the function $c_n(\alpha)$ goes to zero exponentially in $t$ for any $n$; 
hence the charged entropies approach asymptotically a finite value for large $\ell$. 
This saturation value is determined entirely by the infrared physics, i.e. by the value of $c_n(\alpha)$ at small $t$, indeed
\begin{equation}
\label{eq:longmod}
\ln Z_n(\alpha) \simeq \ln Z_n^{(0)} (\alpha)\equiv   \int_{m\epsilon}^{\infty} \frac{c_n(\alpha)}{t}dt\simeq 
\left(\frac16 \Big(n-\frac1n\Big)+ \frac{\alpha^2}{2\pi^2 n}  \right)\ln m\epsilon.
\end{equation}
This dependence on $\ln(m\epsilon)$ coincides with the result in Sec. \ref{sec:ctheorem}
(following the analysis of the properties of the energy momentum tensor on $\mathcal{R}_{n,\alpha}$),  
up to a factor $2$ due to the number of the endpoints (cf. Eq. (\ref{eq:cth})).

The corrections in $m\ell$ to the $\ell$-independent result \eqref{eq:longmod} are obtained expanding the integral \eqref{eq:longmod0} in the ultraviolet.
Keeping for conciseness only the leading order in $t$ of Eqs. \eqref{eq:long} and \eqref{eq:long2} and
performing the integration, we get 
\begin{equation}\label{eq:longalpha}
\begin{split}
\ln Z_n(\alpha)&=\ln Z_n^{(0)} (\alpha)-\frac{n e^{-2t}}{4\pi t},\\
\ln Z_1(\alpha)&=\ln Z_1^{(0)} (\alpha)-\frac{e^{-2t}}{2\pi t}\sin^2\frac{\alpha}{2}.
\end{split}
\end{equation}
Once again, $Z_n(\alpha)$ are not continuous functions of $n$ close to $n=1$ (as it was already known for $\alpha=0$, see \cite{CFH}) and, above all, 
the correction of $\ln Z_n(\alpha)$ does not depend on $\alpha$ for $n \neq 1$ at this order. 
Subleading corrections to Eq. \eqref{eq:longalpha} can be straightforwardly and systematically worked out, but they are not illuminating, although they 
do depend on $\alpha$ also for $n\neq1$. 

For $n\neq1$, since the leading correction does not depend on $\alpha$, the Fourier transform is not affected and the symmetry resolved moments with $n\neq1$
just get a multiplicative correction to ${\cal Z}_n(q)$ in Eq. \eqref{Znq_FT} (so additive for the logarithm), given by 
\begin{equation}\label{eq:znq}
\delta \ln {\cal Z}_n(q)=  \frac{n e^{-2t}}{4\pi t}\,.
\end{equation}
For $n=1$ the net effect of the $\sin^2(\alpha/2)$ term in Eq. \eqref{eq:longalpha} is to renormalise the variance with an exponential additive correction, i.e. 
the desired probability is 
\begin{equation}\label{eq:z1q}
{\cal Z}_1(q)=  e^{-\frac{2q^2\pi^2}{4|\ln (m\epsilon)|+\pi e^{-2t}/t}}\sqrt{\frac{2\pi }{4|\ln (m\epsilon)|+\pi e^{-2t}/t}}
\end{equation}
The symmetry resolved R\'enyi entropies with $n\neq 1$ are straightforwardly obtained from Eq. \eqref{eq:SREE1}. 
Indeed, plugging Eqs. \eqref{eq:znq} and \eqref{eq:z1q} in \eqref{eq:SREE1}, we get
\begin{multline}
S_n(q)=-\frac{n+1}{6n}\ln (m \epsilon)-\frac{n e^{-2m\ell}}{4\pi m\ell(1-n)}+\frac{\ln n}{2(1-n)}-\frac{1}{2}\ln \left(\frac{2}{\pi}|\ln m \epsilon |-\frac{n e^{-2m\ell}}{(1-n)2m\ell} \right) +
\\+{O}((\ln m\epsilon)^{-1},e^{-3m\ell}).
\end{multline}
Such a result shows exact equipartition (at this order) which is a clear consequence of the simple form of \eqref{eq:znq}. 
This is reminiscent of the exact results for integrable models studied in Ref. \cite{MDC-19-CTM}.

The limit $n\to 1$ for the von Neumann entropy should be handled with care.  
We start rewriting Eq. \eqref{eq:SREE1} as
\begin{equation}\label{eq:nto1}
S_1(q)=\ln \mathcal{Z}_1(q)-\frac{1}{\mathcal{Z}_1(q)}\displaystyle \int_{-\pi}^{\pi}\frac{d\alpha}{2\pi}e^{-iq\alpha}Z_{1}(\alpha)\partial_n\ln Z_n(\alpha) |_{n= 1}.
\end{equation}
We use this equation to obtain the entire correction in $t$ due to the Bessel function and not only the leading exponential term (as done in Eq. \eqref{eq:longalpha}).
The crucial computation is
\begin{equation}\label{eq:goal}
\begin{split}
\partial_n c_n(\alpha)|_{n= 1}=-& \partial_n \int_t^{\infty} dy\, y \sum_{k=-(n-1)/2}^{(n-1)/2}v^2_a(y)\Bigg |_{n= 1}=\\
=&-\left(\frac{2}{\pi} \right)^2\partial_n\int_t^{\infty} dy\, y \sum_{k=-(n-1)/2}^{(n-1)/2}\sin^2(\pi a)K^2_{2a}(y)\Bigg |_{n= 1},
\end{split}
\end{equation}
where $a=\frac{k}{n}+\frac{\alpha}{2\pi n}$. 
We can use the integral representation for the Bessel function 
\begin{equation}
\label{eq:intBessel}
K_a(y)=\displaystyle \int_1^{\infty}du\, e^{-yu}\frac{(u+\sqrt{u^2-1})^a+(u+\sqrt{u^2-1})^{-a}}{2\sqrt{u^2-1}},
\end{equation}
to perform  the sum over $k$ in Eq. (\ref{eq:goal}). 
Once we plug Eq. \eqref{eq:intBessel} into Eq. (\ref{eq:goal}) , we get
\begin{equation}
\begin{split}
c_n(\alpha)=&-\frac{2}{\pi^2}\int_t^{\infty}dy\, y \int_1^{\infty} du \int_1^{\infty} dv \frac{e^{-y(u+v)}}{\sqrt{u^2-1}\sqrt{v^2-1}} \times \\
&\times \left(F_{n,\alpha}((u+\sqrt{u^2-1})(v+\sqrt{v^2-1}))+F_{n,\alpha}\left(\frac{(u+\sqrt{u^2-1})}{(v+\sqrt{v^2-1})}\right) \right),
\end{split}
\label{cF}
\end{equation}
where 
\begin{equation}
F_{n,\alpha}(z)=\frac{z^{\frac{\pi-\alpha}{n\pi}}}{4}\left(z-\frac{1}{z} \right)\left(\frac{1+z^{\frac{2\alpha}{n\pi}}}{z^{\frac{2}{n}}-1}+\frac{(z^{\frac{2(\pi+\alpha)}{n\pi}}-1)\cos \frac{\pi -\alpha}{n}+(z^{\frac{2}{n}}-z^{\frac{2\alpha}{n\pi}})\cos \frac{\pi +\alpha}{n}}{1+z^{\frac{4}{n}}-2z^{\frac{2}{n}}\cos (\frac{2\pi}{n})} \right).
\end{equation}
We now study the behaviour of $F_{n,\alpha}(z)$ when $n\to 1$. For $z=1$, the limit $n \to 1$ is singular.  
We can isolate this singularity using the polar variables $(n-1,z-1) \to (\rho \cos \theta , \rho \sin \theta)$ and expanding in the radial coordinate $\rho$. 
The result of this procedure is 
\begin{equation}
F_{n,\alpha}(z)=\frac{1}{2}-\frac{1}{2}\frac{(z-1)^2 \cos \alpha}{\pi^2(n-1)^2+(z-1)^2}+{O}(n-1,z-1),
\end{equation}
whose derivative with respect to $n$ is 
\begin{equation}
\partial_n F_{n,\alpha}(z)\Big |_{n\to 1}=\lim_{n\to 1} \frac{F_{n,\alpha}(z)-F_{1,\alpha}(z)}{n-1}=\pi^2\left(\frac{1}{2}-\sin^2\frac{\alpha}{2}\right)\delta(z-1).
\end{equation}
Plugging this result in Eq. \eqref{cF} and taking the derivative wrt $n$, we get
\begin{equation}
\partial_n c_n(\alpha) \big |_{n\to 1}=-\left(1-2\sin^2\frac{\alpha}{2} \right)\int_{t}^{\infty}dy y K_0(2y)=-\left(\frac{1}{2}-\sin^2\frac{\alpha}{2}\right)tK_1(2t),
\end{equation}
which, once integrated in $t$ according to Eq. \eqref{eq:longmod0}, gives the full ultraviolet behaviour of $\partial_n \ln Z_n(\alpha) \big |_{n\to 1}$, i.e.
\begin{equation}
\partial_n \ln Z_n(\alpha) \big |_{n\to 1}=\left(\frac{1}{3}-\frac{\alpha^2}{2\pi^2}\right)\ln (m\epsilon)+\left(\frac{1}{4}-\frac{1}{2}\sin^2\frac{\alpha}{2}\right)K_0(2t).
\end{equation}
Plugging the above derivative into Eq. \eqref{eq:nto1} finally yields 
\begin{multline}
\label{S1qlarge}
S_1(q)=-\frac{1}{3}\ln (m\epsilon)-\frac{1}{4}K_0(2m\ell)+\ln \mathcal{Z}_1(q)+\frac{\ln (m\epsilon)}{2\pi^2\mathcal{Z}_1(q)}\int_{-\pi}^{\pi}\frac{d\alpha}{2\pi}e^{-iq\alpha}Z_1(\alpha)\alpha^2+\\
+\frac{K_0 (2m\ell)}{2\mathcal{Z}_1(q)}\int_{-\pi}^{\pi}\frac{d\alpha}{2\pi}e^{-iq\alpha}Z_1(\alpha)\sin^2\frac{\alpha}{2} \\
\simeq -\frac{1}{3}\ln (m\epsilon)-\frac{1}{4}K_0(2m\ell)-\frac{1}{2}\ln \left( \frac{2 |\ln (m \epsilon)|}{\pi} 
\right)-\frac{1}{2}+{O}((\ln m\epsilon)^{-1}).
\end{multline}
The first two terms in \eqref{S1qlarge} are respectively the leading and the subleading terms in the total entanglement entropy of a massive Dirac field, 
in agreement with the known results in Refs. \cite{CFH,ccd-08,cd-09}.
The double logarithmic term appears only in the symmetry resolved result and, as already discussed in Eq. \eqref{Snq_FT}, it is related to the number entropy. 
The above derivation clearly highlight this correspondence. 
As for the R\'enyi entropy, at this order in $\ln m\epsilon$, there is perfect entanglement equipartition that will be broken by higher order terms.

\section{The Green's function approach: The complex scalar field}\label{sec:partS}
In this section we present a derivation of the charged moments for a complex massive scalar by generalising to $\alpha\neq 0$ the results obtained in \cite{casini,ch-rev}. 
In Sec. \ref{sec:twist} we showed that $Z_n(\alpha)$ can be written as product of partition functions on the plane with boundary conditions along the cut $A$
\begin{equation}\label{eq:bcsS}
\tilde{\varphi}_k(e^{2\pi i} z,e^{-2\pi i} \bar{z})=e^{2\pi i a} \tilde{\varphi}_k (z,\bar{z}), \qquad a=\frac{k}{n}+\frac{\alpha}{2\pi n}, \qquad k=0,\cdots n-1.
\end{equation}
Denoting, as usual, by $\zeta_a$ these partition functions we have
\begin{equation}
\label{eq:sumpartitionS}
\ln Z_n(\alpha)=\sum_{k=0}^{n-1} \ln\zeta_{\frac{k}{n}+\frac{\alpha}{2\pi n}}.
\end{equation}
%
As for the analogous case of fermions, cf. Eq. \eqref{eq:wa}, we define the auxiliary quantities 
\begin{equation}\label{eq:waS}
w_a\equiv\ell \partial_{\ell} \ln \zeta _a, \quad  \quad c_n(\alpha)\equiv\sum_{k=0}^{n-1} w_{\frac{k}{n}+\frac{\alpha}{2\pi n}},
\end{equation}
that, using (\ref{eq:sumpartitionS}), allow us to write the logarithmic derivative of the partition function in $\mathcal{R}_{n,\alpha}$ as
\begin{equation}\label{eq:c-functS}
c_n(\alpha)=\ell \frac{\partial \ln  Z_n(\alpha)}{\partial \ell}
\qquad\Rightarrow
\qquad
\ln  Z_n(\alpha)=\int_{\epsilon}^\ell \frac{c_n(\alpha)}{\ell'} d\ell' .
\end{equation}
Even here, for $n=1$, the function $c_n(\alpha)$ is the analogue of Zamolodchikov's $c$-function \cite{z-86} in the presence of the flux $\alpha$.

As already discussed in section \ref{sec:replica}, the key observation of this approach relies on the identity between the partition function $\zeta_a$ and the Green's function (see Eq.\,(\ref{eq:Greenvspartition})). 
Through this relation, the function $w_a$ has been obtained for generic values of $a$ also for bosonic free massive field theories \cite{ch-rev}. 
As already found in Sec.\,\ref{sec:twist} using twist fields, also this approach requires that $0< a <1$ for the scalar theory (see \cite{casini} for details). Thus, in order to compute $Z_n(\alpha)$,  we will use the results in \cite{ch-rev} setting $a=\frac{k}{n}+\frac{|\alpha|}{2\pi n}$ for the complex Klein Gordon theory.

\begin{figure}
\centering
\subfigure
{\includegraphics[width=0.48\textwidth]{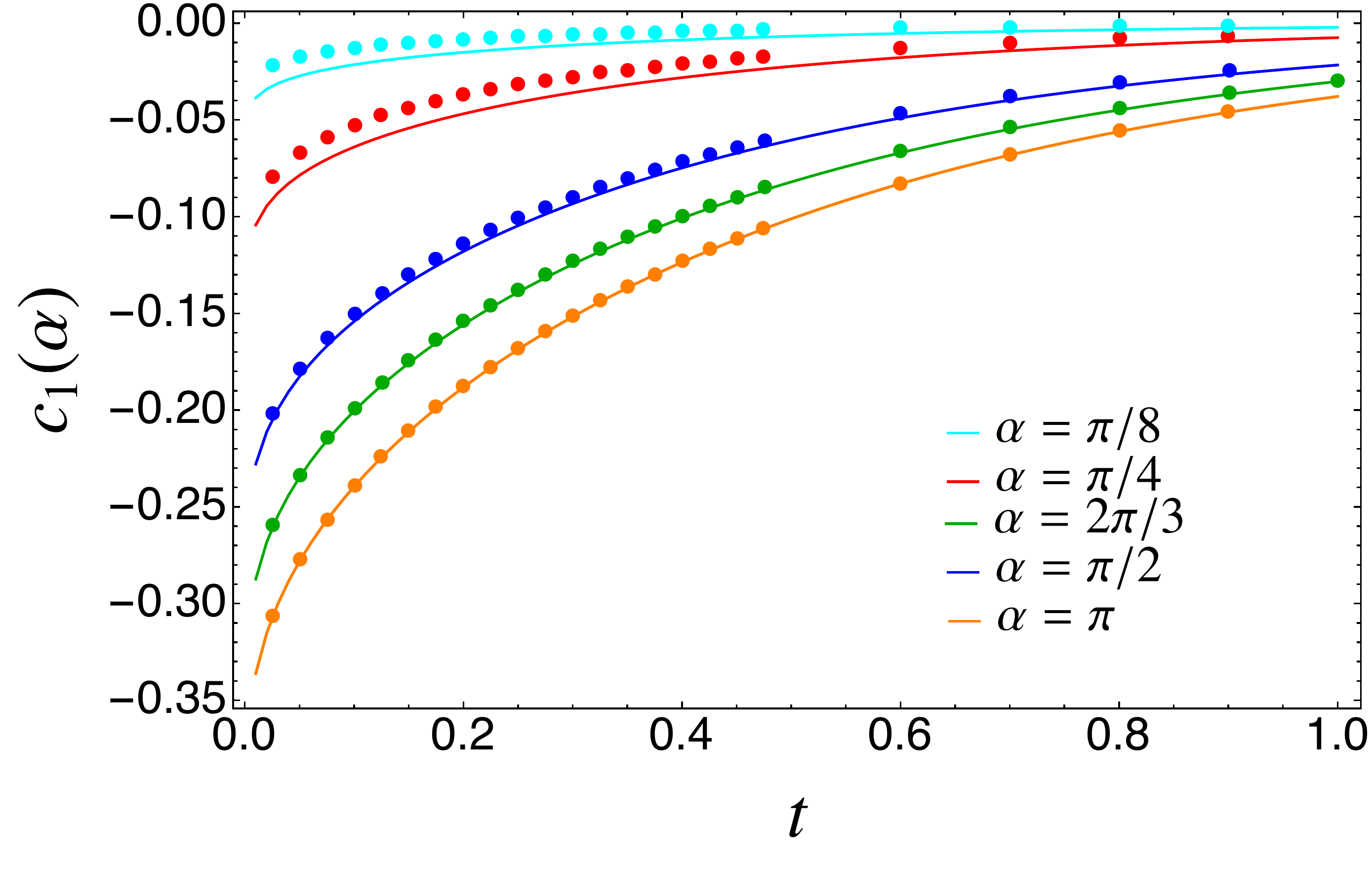}}
\subfigure
{\includegraphics[width=0.48\textwidth]{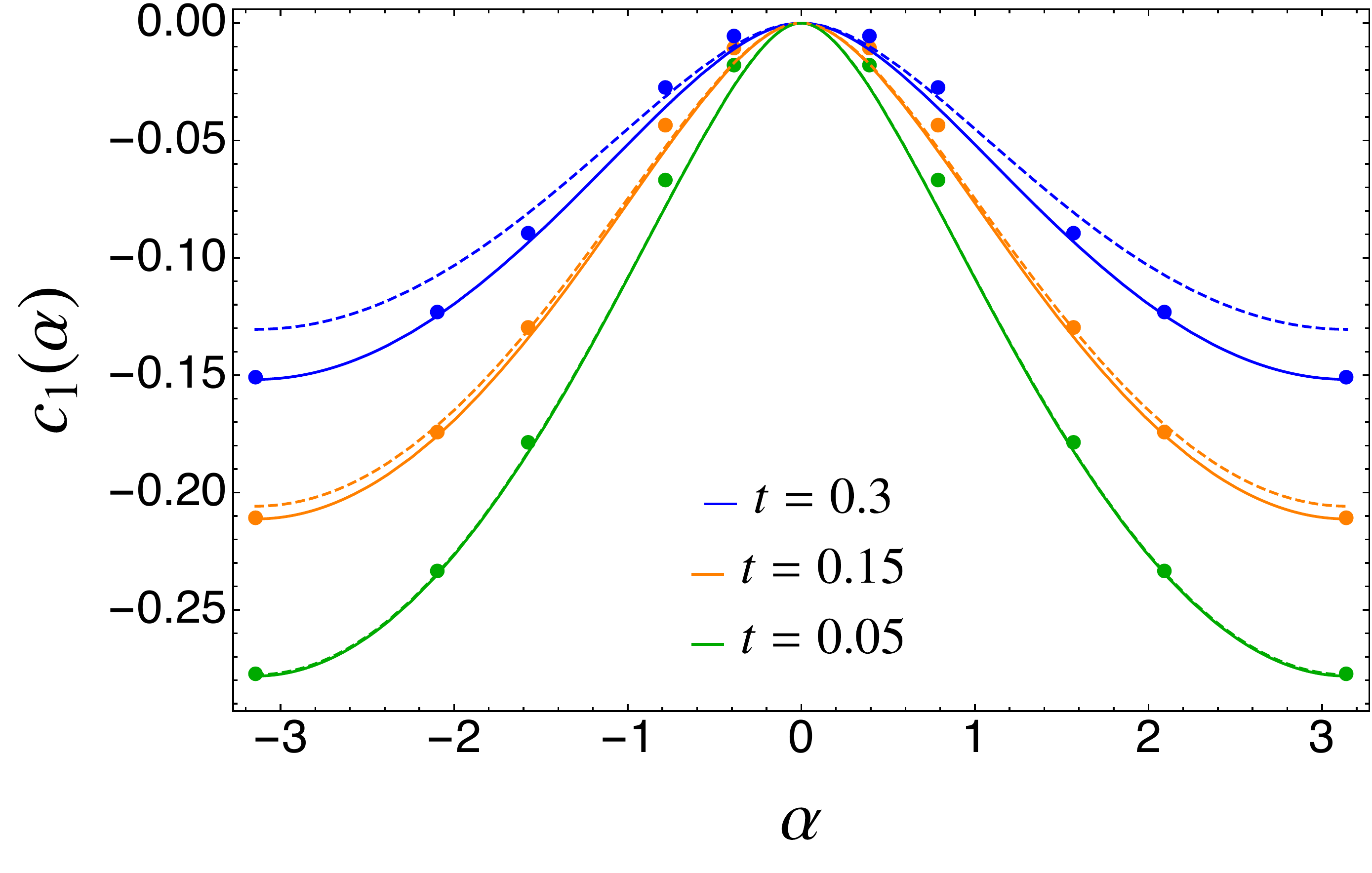}}
\caption{
Left panel: Logarithmic derivative of the charged moments, $c_1(\alpha)$, as a function of $t=m \ell$ obtained by solving numerically the Painl\'eve equation (\ref{eq:PainleveBoson}) (solid lines). The data (symbols) are obtained fixing (from top to bottom) $\ell=720,620,420,420,320$ and varying properly $m$ in such a way that 
$m\ell \in (0,1)$. As discussed in the main text, the agreement with the numerical data worsens as $\alpha$ decreases. Right panel: the same quantity is plotted as a function of $\alpha$ from the numerical solution of the Painlev\'e equation (solid lines), showing also in this case that the agreement as $\alpha \to 0$ is not excellent. 
The dashed lines represent the small $t$ expansion in Eq. (\ref{eq:tointegrate}): the smaller is $t$, the better the approximation works.}
\label{fig:NumericalPainlevealpha}
\end{figure}

Here we consider the complex massive non-compact bosonic field theory with action given by (\ref{eq:actions})  and mass $m$. 
The function $w_a$ with $0<a<1$ defined in (\ref{eq:waS}) can be written as \cite{casini}
\begin{equation}
\label{eq:wa1bosons}
w_a=-\int_{t}^{\infty} y u_a^2(y) dy,
\end{equation}
where $t=m \ell$ and $u_a$ is the solution of the Painlev\'e V equation
\begin{equation}
\label{eq:PainleveBoson}
u_a''+\frac{u_a'}{t}=\frac{u_a}{1+u_a^2}(u_a')^2+u_a(1+u_a^2)+4\frac{\left(a-\frac{1}{2}\right)^2}{t^2(1+u_a^2)}u_a.
\end{equation}
The solution of Eq. (\ref{eq:PainleveBoson}) is showed in Fig. \ref{fig:NumericalPainlevealpha}: 
the function $w_a$ for a generic value of $t$ can be obtained solving numerically Eq. (\ref{eq:PainleveBoson}) with the initial condition for $t\to 0$
\begin{equation}
u_a(t)=\frac{-1}{t\left(\ln t + \kappa_S(a) \right)} -a(a-1)t(\ln t +\kappa_S(a))+O(t),
\end{equation}
where \cite{ch-rev}
\begin{equation}
\kappa_S(a)= 2\gamma_E+\frac{\psi(1-a)+\psi(a)}{2}-\ln 2.
\end{equation}
In the figure we compare the exact result from field theory with numerical data for a chain of complex oscillators, obtained exploiting the techniques reviewed 
in Appendix \ref{app:lat}. 
We have a fairly good agreement between lattice and field theory, although for small values of $\alpha$ the agreement gets worse and one needs a larger and 
larger subsystem length $\ell$ on the lattice to match the continuum limit. 
This is not surprising, already in Ref. \cite{MDC-19-CTM} it was shown for the massless case that the lattice results approach the CFT ones in a non uniform way.
In the following we will further discuss this issue in the limits when we have an analytic handle on the problem. 

\subsection{The expansion close to the conformal point}\label{sec:CHboson}

In the conformal limit $t\to 0$ we have that the solution of the Painlev\'e equation admits the expansion \cite{casini}
\begin{equation}\label{eq:mainCH bosons}
w_a=-2a(1-a)-\dfrac{1}{\ln (t)+\kappa_S(a)}+O(t).
\end{equation}
Using Eq.\,(\ref{eq:waS}) we get
\begin{equation}\label{eq:tointegrate}
c_n(\alpha)=
\displaystyle \sum_{k=0}^{n-1}w_{\frac{k}{n}+\frac{|\alpha|}{2\pi n}}
=\frac{1-n^2}{3n}+\frac{\alpha^2}{2\pi^2 n}-\frac{|\alpha|}{\pi n}-\sum_{k=0}^{n-1}\frac{1}{ \ln t + \kappa_S(a(k))}+\dots.
\end{equation}
Let us discuss first at the level of the universal function $c_n(\alpha)$ the origin of the non uniform behaviour in $\alpha$. 
Eq. \eqref{eq:tointegrate} is an exact asymptotic expansion valid for any $\alpha\neq 0$.
For $\alpha\to0$, there is a clear problem with the constant $\kappa_{S}(a(0))$ (i.e. of the mode with $k=0$) which diverges as $\pi/|\alpha|$.
Hence, since $\ln t$ grows very slowly with $t$, the true asymptotic behaviour is attained only for $t\gtrsim e^{\pi/|\alpha|}$. 
For smaller values of $t$, the mode $k=0$ looks almost constant ($\sim |\alpha|/\pi$) and similar to the leading term.
Exactly for $\alpha=0$, the mode $k=0$ diverges, so its inverse is just zero and it does not affect the calculation. 
It is then clear that the approach to the asymptotic behaviour cannot be uniform in $\alpha$, as already observed numerically in Fig. \ref{fig:NumericalPainlevealpha}.

After having discussed this caveat with the small $\alpha$ behaviour, we are ready to integrate Eq. \eqref{eq:tointegrate} to get the charged moments,  
according to Eq. (\ref{eq:c-functS}), obtaining
\begin{equation}\label{eq:guesssbagliato}
\ln Z_n(\alpha)
=
\left(
\frac{1-n^2}{3n}+\frac{\alpha^2}{2\pi^2 n}-\frac{|\alpha|}{\pi n}
\right)\ln\left(\frac{\ell}{\epsilon}\right)
+
\sum_{k=0}^{n-1} \ln \left| \frac{\ln(m\epsilon) + \kappa_S(a(k))}{\ln(m \ell) + \kappa_S(a(k))} \right|.
\end{equation}
Let us remark that when $n=1$, the last sum reduces to the term with $k=0$. 
We recall that the cutoff $\epsilon$ depends both on $n$ and $\alpha$ making the analysis even more troubling.

\begin{figure}
\centering
\subfigure
{\includegraphics[width=0.43\textwidth]{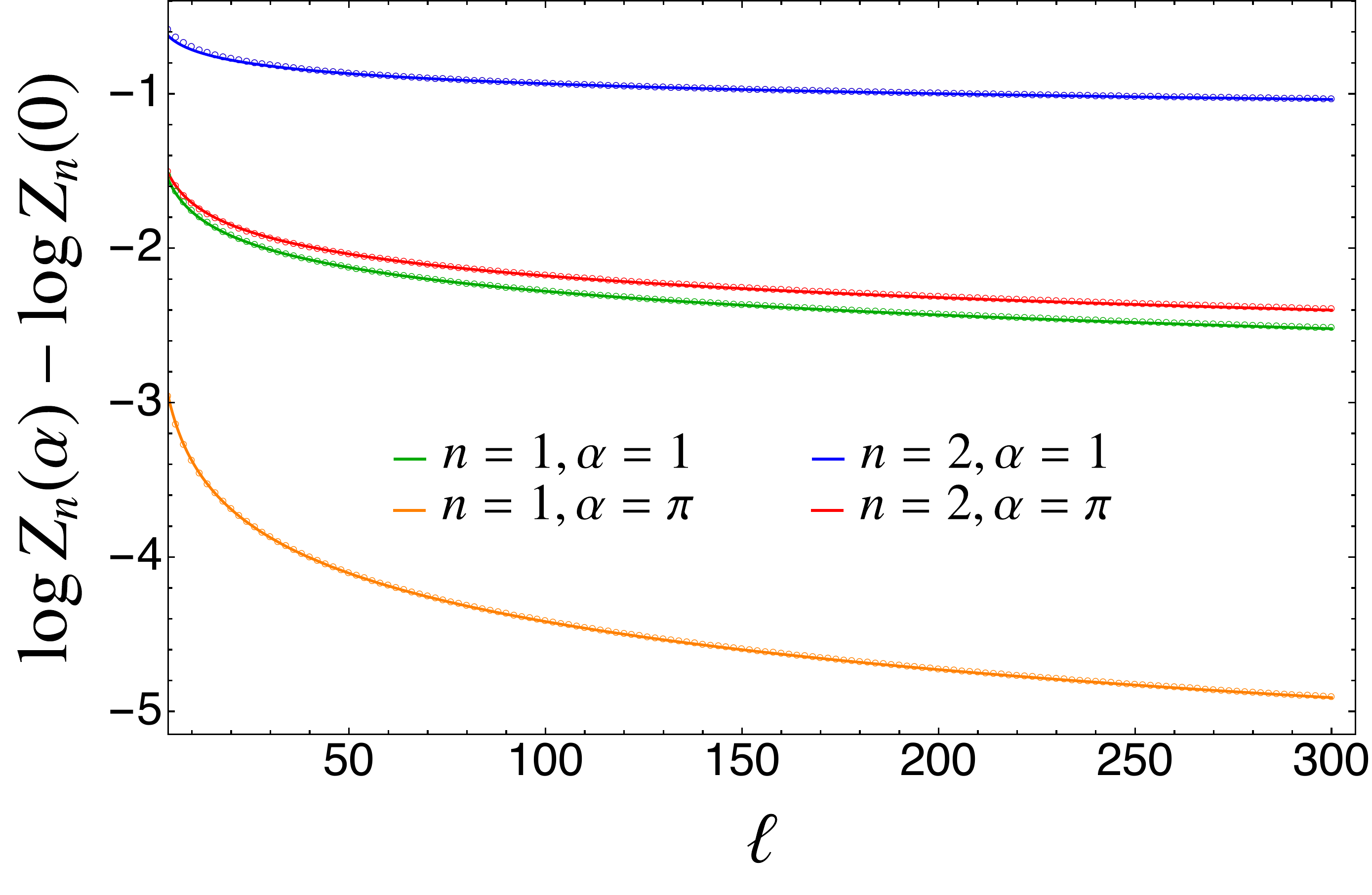}}
\subfigure
{\includegraphics[width=0.455\textwidth]{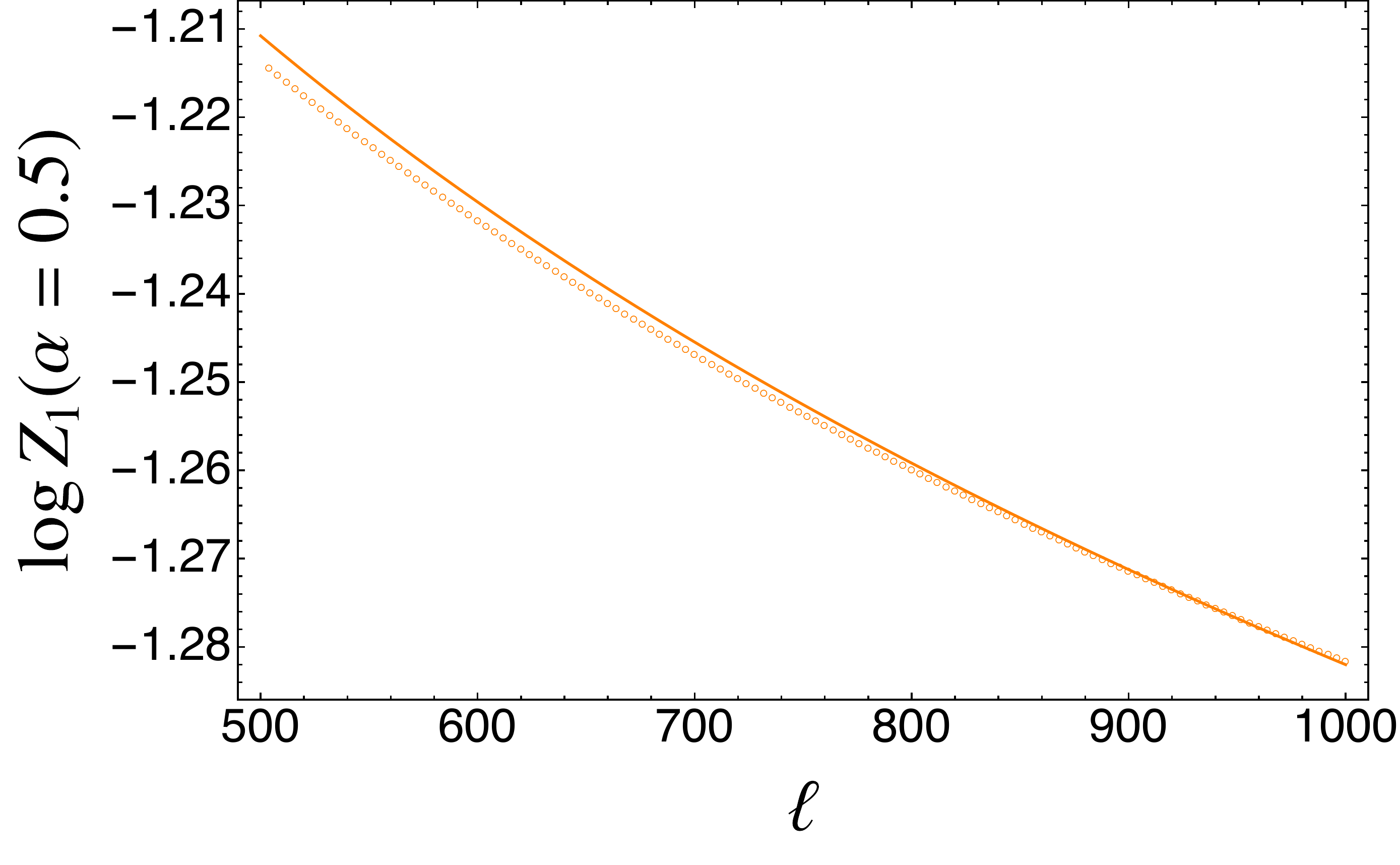}}
\caption{
Charged moments for the free massive boson close to the critical regime. 
We benchmark the analytical prediction (\ref{eq:guesssbagliato}) (solid lines) with the numerical data (symbols) for different values of $n$ and $\alpha$ at fixed $m=10^{-10}$: 
the analytical formula matches well the data for large values of $\alpha$ (left), but for smaller $\alpha$ (right) much larger values of $\ell$ are necessary to 
observe a fair match, as explained in the text.}\label{fig:massiveB}
\end{figure}

Even though we are in the conformal limit in which $m\ell \ll 1$, the additional constant $\kappa_S(a(0))$ cannot be neglected because of its divergent behaviour when 
$k=0$ and $\alpha=0$. 
The terms with $k>0$ do not present any problem and ${\kappa_S}(a(k))$ can be safely neglected. 
The mode with $k=0$ instead has three different regimes, depending on the value of $\kappa_S(k=0)$ which is governed by $\alpha$ as follows:
\begin{itemize}
\item 
for very small $\alpha$, i.e. such that $\alpha \lesssim n \pi/|\ln (m\epsilon/2) +\gamma_E|$, 
$\kappa_S(k=0)$ diverges faster than both $\ln m \ell$ and $\ln m \epsilon$. 
Hence, expanding the ratio in Eq. \eqref{eq:guesssbagliato}, this subleading term becomes of the same order of the leading one, i.e. 
\begin{equation}\label{eq:firstregime}
\ln \left| \frac{\ln(m\epsilon) + \kappa_S(a(0))}{\ln(m \ell) + \kappa_S(a(0))} \right|\xrightarrow[]{\alpha \to 0}\frac{|\alpha |}{\pi n} \ln \frac{\ell}{\epsilon}+\dots.,
\end{equation}
\item for intermediate values of $\alpha$, i.e. when $n \pi/|\ln (m\epsilon/2) +\gamma_E| \lesssim \alpha \lesssim  n \pi /|\ln (m\ell/2) +\gamma_E|$, 
we have 
\begin{equation}
\ln \left| \frac{\ln(m\epsilon) + \kappa_S(a(0))}{\ln(m \ell) + \kappa_S(a(0))} \right| \sim
\ln\left|  \frac{\ln(m\epsilon) }{\kappa_S(a(0))} \right|
+\dots,
\end{equation}
and hence this produces just an additive correction in $\ell$, but depending on $m\epsilon$;  
\item
for larger $\alpha$, i.e.  for $\alpha \gtrsim n \pi/|\ln (m\ell/2) +\gamma_E|$, the term $\kappa_S$ is a correction both for numerator and denominator 
and  so
\begin{equation}
\ln \left| \frac{\ln(m\epsilon) + \kappa_S(a(0))}{\ln(m \ell) + \kappa_S(a(0))} \right| \simeq  \ln \left| \frac{\ln(m\epsilon) }{\ln(m \ell) } \right|
+\dots,
\end{equation}
as for the terms with $k\neq 0$. We stress that this third regime is the true asymptotic one for large $\ell$ at fixed $\alpha$.
\end{itemize}

This competition among the three terms makes difficult the analytical treatment of the last sum in Eq.\,(\ref{eq:guesssbagliato}), 
and, at the same time, the non trivial dependence on the cutoff $\epsilon$ (that we recall also depends on $\alpha$ and $n$) complicates the comparison with the numerics. 
For this reason, we consider only the leading term in Eq. (\ref{eq:guesssbagliato}), which, strictly speaking, is valid in the massless case 
and in the third regime above. 
Such a leading term is the same provided as in the twist field approach (cf. Eq. (\ref{eq:moments})) and coincides with some equivalent ones in 
literature \cite{Belin-Myers-13-HolChargedEnt,MDC-19-CTM}. 
The main advantage of Eq. (\ref{eq:guesssbagliato}) is that it clearly shows what are the problems one faces when considering only the leading term. 
The comparisons with the numerics are shown in Figure \ref{fig:massiveB}:
we report the numerical data for different values of $n$ and $\alpha$; 
as expected from our previous discussion, the agreement with the predictions is very 
good for large $\alpha$, but it worsens as $\alpha$ gets smaller and $n$ gets larger.
Smaller is $\alpha$, larger is the value of $\ell$ on the lattice necessary to observe the true asymptotic behaviour.

\subsubsection{Symmetry resolution}

The symmetry resolved moments of the RDM can be computed through the Fourier transform of the leading term of the charged moments in Eq. (\ref{eq:guesssbagliato})
\begin{multline}\label{eq:erfi}
\mathcal{Z}_n(q)=\displaystyle \int_{-\pi}^{\pi}\dfrac{d\alpha}{2\pi}e^{-iq\alpha }Z_n(\alpha)=
Z_n(0)\displaystyle \int_{-\pi}^{\pi}\dfrac{d\alpha}{2\pi}e^{-iq\alpha }e^{\left( \frac{\alpha^2}{2\pi^2 n}-\frac{|\alpha|}{\pi n}\right) \ln \frac{\ell}{\epsilon}}=\\
Z_n(0)\left(\frac{\ell}{\epsilon}\right)^{-\frac1{2n}} \sqrt{\frac{n\pi}{8\ln (\ell/\epsilon)}}
(-1)^qe^{\frac{n\pi^2 q^2}{2\ln (\ell/\epsilon)}}\left[\mathrm{Erfi}\left(\frac{\ln (\ell/\epsilon) -n\pi i q}{\sqrt{2n \ln (\ell/\epsilon)}}\right)+\mathrm{Erfi}\left(\frac{\ln (\ell/\epsilon) +n\pi i q}{\sqrt{2n \ln (\ell/\epsilon)}}\right) \right],
\end{multline}
where $\mathrm{Erfi}(x)$ is the imaginary error function (the overall result is real and positive for $q\in \mathbb Z$)
\begin{equation}
\label{eq:erfi1}
\mathrm{Erfi}(x)=\frac{-2i}{\sqrt{\pi}}\displaystyle \int_0^{ix}dt\,e^{-t^2} \xrightarrow[]{x \rightarrow \infty}\frac{e^{x^{2}}}{\sqrt{\pi}x}.
\end{equation}
In the large $\ell$ limit, using the expansion in Eq. (\ref{eq:erfi1}), the charged moments in Eq. (\ref{eq:erfi}) can be can be approximated as
\begin{equation}\label{eq:erfiC}
\mathcal{Z}_n(q)=Z_n(0) \frac{n\ln \ell/\epsilon}{n^2\pi^2q^2+\ln^2 \ell/\epsilon}, 
\end{equation}
and hence the symmetry resolved entropies are given by
\begin{equation}\label{eq:symm}
S_n(q)=\frac{1}{1-n}\ln \frac{\mathcal{Z}_n(q)}{\mathcal{Z}_1(q)^n}\simeq S_n-\ln \ln \frac{\ell}\epsilon+\frac{\ln n}{1-n}, \quad S_1(q)\simeq S_1-\ln \ln \frac{\ell}\epsilon-1.
\end{equation}
The leading behaviour is described by the total R\'enyi entropies, with the usual correction $\ln \ln \ell$ that is independent on $q$, 
confirming the equipartition of the entanglement entropy for a complex massive scalar field theory,  in agreement with the result for massive harmonic chains \cite{MDC-19-CTM}
(although the critical limit considered there is different from the one here).
Let us mention that a further expansion of Eq. (\ref{eq:erfi}) leads to subleading corrections behaving as $q^2/(\ln \ell)^2$ which explicitly depend on $q$, 
breaking the equipartition of the entanglement. This kind of terms has been already found for bosonic systems in \cite{MDC-19-CTM}.

Let us now discuss the effect of the term that we disregarded in Eq. (\ref{eq:guesssbagliato}), namely the sum over $k$.
The mode with $k\neq 0$ would provide double logarithimic corrections encountered also in other contexts, like non unitary CFTs \cite{bcdr-14,cjs-17}.
These in principle are calculable and partially under control. 
We mention that such terms have a non-trivial dependence on $n$ in ${\cal Z}_n(q)$ and hence they are responsible of a further breaking of 
equipartition. 
Unfortunately, the determination of this correction is not easy because it is influenced by the precise dependence on $\alpha$ and $n$ of the  non-universal cutoff $\epsilon$
(as it should be clear from Eq. (\ref{eq:guesssbagliato})). 
Finally, as discussed for the charged moments, the effect of the mode $k=0$ is even more dramatic and too difficult to keep under control.


\subsection{The long distance expansion.}
\begin{figure}
\centering
\subfigure
{\includegraphics[width=0.48\textwidth]{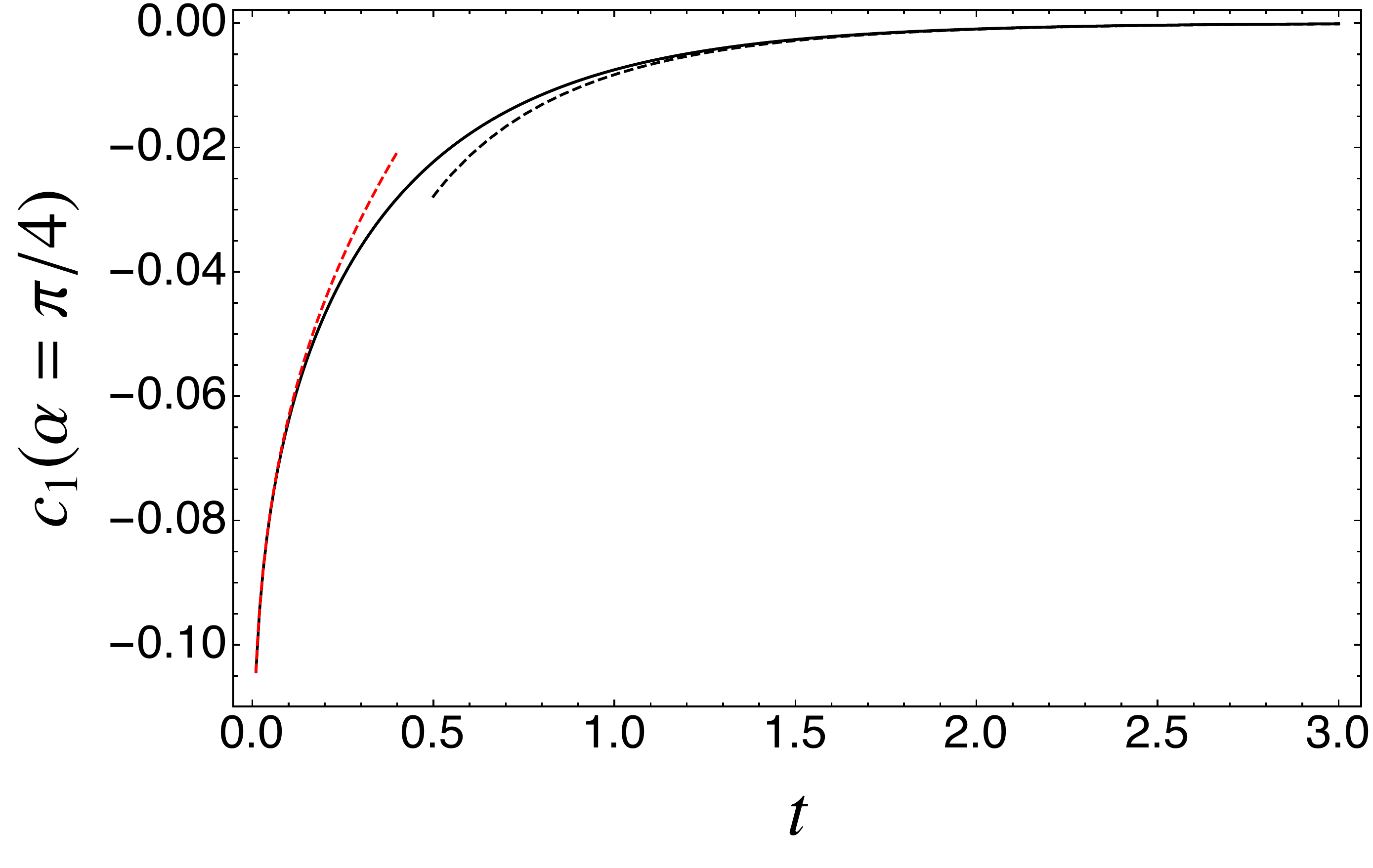}}
\subfigure
{\includegraphics[width=0.48\textwidth]{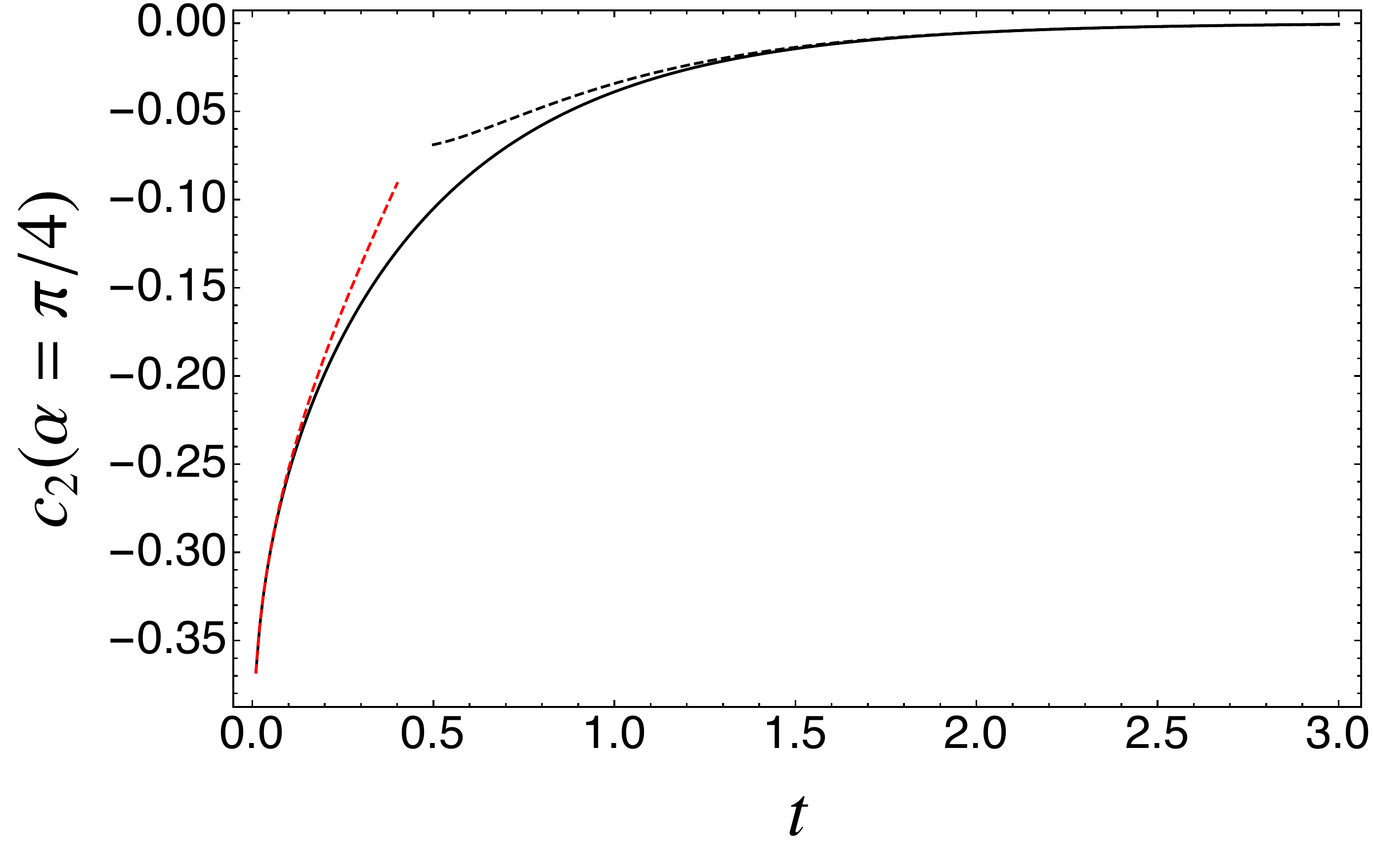}}
\caption{The solid lines are the functions $c_n(\alpha)$ as solutions of Eq. (\ref{eq:PainleveBoson}). 
The dashed lines are the short (red) and long (black) distance leading terms we evaluated analytically in Eq. (\ref{eq:tointegrate}) and Eq. (\ref{eq:longB}), respectively.}
\label{fig:LDB}
\end{figure}

The boundary condition for Eq. (\ref{eq:PainleveBoson}) in the limit in which $t \to \infty$ is \cite{casini}
\begin{equation}
u_a(t)\sim \frac{2}{\pi} \sin (\pi a )K_{1-2a}(t).
\end{equation}
The solution of Eq. (\ref{eq:PainleveBoson}) in the long distance regime together with Eq.\,(\ref{eq:wa1bosons}) gives
\begin{equation}\label{eq:longBosons}
w_a(t)=-e^{-2t}\frac{\sin^2(a\pi)}{\pi}\left(1+\frac{3-16a+16a^2}{4t}\right).
\end{equation}
Summing over $a=\frac{k}{n}+\frac{|\alpha|}{2\pi n}$, we get 
\begin{equation}\label{eq:longB}
\begin{split}
c_n(\alpha)&=\frac{e^{-2t}}{2\pi n t}\left(-n^2 t -\frac{8+n^2}{12}+\frac{2 |\alpha|}\pi -\frac{\alpha^2}{\pi^2}+{2}\Big( \csc^2\frac{\pi}{n}-\frac{|\alpha|}\pi \Big)\cos\frac{\alpha}{n} + \frac{2\alpha}{\pi} \cot\frac{\pi}{
   n} \sin\frac{\alpha}{n}\right),\\
c_1(\alpha)&=-\frac{e^{-2t}}{\pi}\sin^2\frac{\alpha}{2}\left(1+\frac{3+ \frac{4\alpha^2}{\pi^2}-\frac{8|\alpha|}{\pi}}{4t}\right).
\end{split}
\end{equation}
The long distance leading term in Eq.\,(\ref{eq:longB}) is showed in Fig.\,\ref{fig:LDB} for two different values of $n$: it approximates well the solution of the Painlev\'e equation (\ref{eq:PainleveBoson}) in the regime $t\gg 1$. The same feature was observed in Sec.\,\ref{sec:LDF} for the corresponding equations in fermionic systems as also the discontinuity for $n\to 1$, which can be ascribed to the non-commutativity of the limits $n \to 1$ and $t \to \infty$. 
Also for a complex scalar field, Eqs. (\ref{eq:longB}) show that the functions $c_n(\alpha)$ vanish for large $t$ and the charged moments stop growing. Hence,
\begin{equation}
\label{eq:longmodS}
\ln Z_n(\alpha)
 \simeq \ln Z_n^{(0)} (\alpha)\equiv   \int_{m\epsilon}^{\infty} \frac{c_n(\alpha)}{t}dt\simeq
  -\Big(\frac{1-n^2}{3n}+\frac{\alpha^2}{2\pi^2 n}-\frac{|\alpha|}{\pi n}\Big)\ln m\epsilon.
\end{equation}
As expected, the dependence on $\ln(m)$ coincides with the one reported in Eq. (\ref{eq:cth2}), up to a factor $2$ due to the number of endpoints.

Integrating $c_n(\alpha)$, we obtain up to order $O(e^{-2t}/t)$ 
\begin{equation}\label{eq:longalpha2B}
\begin{split}
\ln Z_n(\alpha)&=\ln Z_n^{(0)} (\alpha)-\frac{n e^{-2t}}{4\pi t},\\
\ln Z_1(\alpha)&=\ln Z_1^{(0)} (\alpha)-\frac{e^{-2t}}{2\pi t}\sin^2\frac{\alpha}{2},
\end{split}
\end{equation}
which are the same expressions found for fermions in Sec. \ref{sec:LDF}. The expression for $\delta \mathcal{Z}_n(q)$ is the same as in Eq. \eqref{eq:znq} for fermions, 
while $\mathcal{Z}_1(q)$ is given by
\begin{equation}
\mathcal{Z}_1(q)=\frac{|\ln m\epsilon|}{\pi^2q^2+\ln^2m\epsilon}+{O}(1/(\ln ^2 m\epsilon)),
\end{equation}
so that all contributions coming from the long-distance behaviour are negligible at order ${O}(1/(\ln m\epsilon))$. 
The resolved entropies are the ones given in \eqref{snqs}, where $S_n$ also takes into account the term $\frac{ne^{-2t}}{4\pi t}$. 
The limit $n\to 1$ can be solved through a technique similar to the one used in Sec. \ref{sec:LDF}.

\section{Charged moments across the hyperplane: massive scalar field}
\label{app:nishioka}

In this last section, we  provide a result for the charged moments of a free massive scalar theory across a hyperplane in $d$ Euclidean 
dimensions using a generalisation of the method reported in \cite{cc-04,Nishioka}. The action for a free complex massive scalar field is
\begin{equation}\label{eq:ham}
S=\int d^dx[\partial_{\mu}\varphi^{\dagger}(x)\partial^{\mu}\varphi(x)+m^2\varphi^{\dagger}(x)\varphi(x)].
\end{equation}
We denote the space coordinates by $x_i$, $i = 1,\cdots, d-1,$ and the Euclidean time by $x_0$. 
Let $A$ and $\bar{A}$ be regions with $x_1>0$ and $x_1\leq 0$, respectively. 
The entangling surface $\Sigma$ is chosen to be a $(d-2)$-dimensional hyperplane at $x_1 = 0$: $\Sigma = \{(x_0,x_i)|x_0 = x_1 = 0\}$. 
Let us introduce the metric of the spacetime
\begin{equation}
\label{metric}
\mathrm{d}s^2=\mathrm{d}r^2+r^2\mathrm{d}\tau^2+\displaystyle \sum_{i=2}^{d-1}\mathrm{d}x_i^2,
\end{equation}
where we have used the polar coordinates for the plane parametrised by $(x_0,x_1)$. 
The metric of the $n$-fold cover $\mathcal{M}_{n,\alpha}$ of the original spacetime pierced by a flux $\alpha$, that is constructed by gluing $n$ copies of the sheet with a cut along $A$, is still (\ref{metric}) with  $r\geq 0$ and $0\leq \tau \leq 2\pi n$. Thus, $\mathcal{M}_{n,\alpha}=\mathcal{C}_{n,\alpha} \times \mathbb{R}^{d-2}$, where $\mathcal{C}_{n,\alpha}$ is the two-dimensional cone parametrized by $(r,\tau)$.
  
In order to compute the charged moments, we can consider the theory with action in Eq. (\ref{eq:ham}) in terms of two real scalar fields $\varphi_1(x)$ and $\varphi_2(x)$ \cite{MDC-19-CTM}. The logarithm of the normalised charged moments $\ln Z^{\rm R}_n(\alpha)$ of {\it each real scalar field} on $\mathcal{M}_{n,\alpha}$ is given by
\begin{equation}
\label{eq:1loop}
\begin{split}
\ln Z^{\rm R}_n(\alpha)&=-\frac{1}{2} \ln \det (-\nabla^2_{\alpha}+m^2)=-\frac{1}{2} \mathrm{tr} \ln (-\nabla^2_{\alpha}+m^2)\\
&=\frac{1}{2} \displaystyle \int_{\epsilon^2}^{\infty} \frac{ds}{s}\mathrm{tr} \left[ e^{-s(-\nabla^2_{\alpha}+m^2)}-e^{-s}\right],
\end{split}
\end{equation}
where the parameter $\epsilon^2 \ll 1$ is introduced as a regulator for the UV divergences. Because of the direct product structure of $\mathcal{M}_{n,\alpha}$, the Laplacian decomposes into the sum of those on $\mathcal{C}_{n,\alpha}$ and $\mathbb{R}^{d-2}$: $\nabla^2_{\alpha}=\nabla^2_{\mathcal{C}_{n,\alpha}}+\nabla^2_{\mathbb{R}^{d-2}}$. The rotational symmetry of the cone $\mathcal{C}_{n,\alpha}$ allows the Fourier decomposition of the real fields by the modes $\exp(i \tau a)$, where $a=\frac{l}{n}+\frac{|\alpha|}{2\pi n}$, with integer $l$, such that, after a period $2\pi n$, they acquire a phase $e^{i\alpha}$, in agreement with the Aharonov-Bohm effect. Therefore, the eigenfunctions $\varphi_{k,a}(r,\tau)$ of the Laplacian are parametrized by $(k, a)$ satisfying
\begin{equation}
\label{eq:const1}
\begin{split}
\nabla^2_{\mathcal{C}_{n,\alpha}}\varphi_{k,a}(r,\tau)&=-k^2\varphi_{k,a}(r,\tau), k \in \mathbb{R}^{+}, \\
\varphi_{k,a}(r,\tau)&=\sqrt{\frac{k}{2\pi n}}e^{i\tau a}J_{|a|}(kr),
\end{split}
\end{equation}
where $J_a$ is the Bessel function of the first kind. The eigenfunctions form an orthonormal basis on the cone $\mathcal{C}_{n,\alpha}$, namely
\begin{equation}
\label{eq:norma}
\int_{\mathcal{C}_{n,\alpha}} \,\mathrm{d}^2x \varphi_{k,a}(x) \varphi^{*}_{k',a'}(x)=\delta_{na,na'}\delta(k-k').
\end{equation}
The orthonormal basis of the eigenfunctions of the Laplacian on $\mathbb{R}^{d-2}$ is spanned by the plane waves, $\varphi_{\mathbf{k}^{\bot}}(y)=\exp (i \mathbf{k}_{\bot} \cdot  \mathbf{y})/(2\pi)^{(d-2)/2}$, with eigenvalues $-k^2_{\bot}$. Exploiting these two sets of eigenfunctions, the trace of the kernel in Eq.\,(\ref{eq:1loop}) is
\begin{equation}
\label{eq:loop2}
\begin{split}
\mathrm{tr}[e^{-s(-\nabla^2+m^2)}] &= 
\int_{\mathcal{C}_{n,\alpha}}\mathrm{d}^2x\displaystyle \sum_{l=-\infty}^{\infty}\int_0^{\infty}\mathrm{d}k e^{-s(k^2+m^2)}\varphi_{k,a}(x)\varphi^*_{k,a}(x) \\
&\qquad \times \int_{\mathbb{R}^{d-2}}\mathrm{d}^{d-2}y \int \mathrm{d}^{d-2} k_{\bot}e^{-sk^2_{\bot}}\varphi_{\mathbf{k}^{\bot}}(y) \varphi^*_{\mathbf{k}^{\bot}}(y) \\
&= \dfrac{\mathrm{Vol}(\mathbb{R}^{d-2})}{n}\dfrac{e^{-sm^2}}{(4\pi s)^{(d-2)/2}}\left(- \zeta\left(-1,\frac{|\alpha|}{2\pi}\right)+n\displaystyle \int_0^{\infty} \mathrm{d}r\dfrac{1}{\sqrt{2\pi}}\right). 
\end{split}
\end{equation}
The IR divergence emerges from the following integrals
\begin{equation}
\label{eq:idbessel}
\begin{split}
\displaystyle \int_0^{\infty} \mathrm{d}k \, k e^{-sk^2}J_{a}(kr)^2&=\dfrac{e^{-r^2/(2s)}}{2s} I_{a} \left(\dfrac{r^2}{2s} \right), \\
\displaystyle \int_0^{\infty} \mathrm{d}r \, r e^{-r^2}I_a(r^2) &=-\dfrac{a}{2}+\displaystyle \int_0^{\infty} \mathrm{d}r\dfrac{1}{\sqrt{2\pi}},
\end{split}
\end{equation}
where $I_a$ is the modified Bessel function of the first kind. The second term on the right-hand side of Eq.\,(\ref{eq:idbessel}) is divergent, but it gives rise to a term proportional to $n$ and independent on $\alpha$, so it does not contribute to $\ln Z_n(\alpha)$. On the other hand, the UV divergence arises from the summation over the angular momentum $l$, which can be regularised, for example, by using the Hurwitz zeta function
\begin{equation}
\label{eq:hurwitz}
\displaystyle \sum_{l=-\infty}^{\infty} \bigg|l+\frac{|\alpha|}{2\pi}\bigg|=2\zeta\left(-1,\frac{|\alpha|}{2\pi}\right).
\end{equation}
The definition of the Hurwitz zeta function for complex arguments $s$ with Re$(s) > 1$ and $q$ with Re$(q) > 0$ is
\begin{equation}
\zeta(s,q)=\displaystyle 	\sum_{j=0}^{\infty}\frac{1}{(q+j)^s},
\end{equation}
whose analytic continuation satisfies the remarkable identity
\begin{equation}
\zeta\left(-1,\frac{|\alpha|}{2\pi}\right)= -\dfrac{\alpha^2}{8\pi^2n}+\dfrac{|\alpha|}{4\pi n}-\frac{1}{12}.
\end{equation}
The other kernel $\mathrm{tr}(e^{-s})$ is still divergent because it is proportional to the volume of $\mathcal{M}_{n,\alpha}$ but it is also proportional to $n$ coming from the volume of the cone, so, being independent on $\alpha$, it does not contribute to $\ln Z_n(\alpha)$. 

Thus, using Eqs. (\ref{eq:1loop}), (\ref{eq:loop2}),  and  \eqref{eq:hurwitz} we obtain the logarithm of the charged moments across a hyperplane $\mathbb{R}^{d-2}$ 
for a free complex massive scalar field \cite{MDC-19-CTM}
\begin{equation}
\label{eq:finalresult}
\begin{split}
\ln Z_n(\alpha)&=\ln Z^{\rm R}_n(\alpha)+\ln Z^{\rm R}_n(-\alpha)=\\
&=4\pi \mathrm{Vol}(\mathbb{R}^{d-2})\left(-\frac{1}{n} \zeta\left(-1,\frac{|\alpha|}{2\pi}\right)-\frac{n}{12 }\right)\int_{\epsilon^2}^{\infty}ds \dfrac{e^{-sm^2}}{(4\pi s)^{d/2}}\\
&=
4\pi \mathrm{Vol}(\mathbb{R}^{d-2})
\left( \dfrac{\alpha^2}{8\pi^2n}-\dfrac{|\alpha|}{4\pi n}-\frac{1}{12}\left(n-\frac{1}{n} \right)\right)
\int_{\epsilon^2}^{\infty}ds \dfrac{e^{-sm^2}}{(4\pi s)^{d/2}}.
\end{split}
\end{equation}
Very remarkably,  the dependence on $\alpha$ in Eq. (\ref{eq:finalresult}) is the same in any dimension and hence the symmetry resolved entropies are also the 
same in any dimension. 

As a consistency check, let us focus our attention on $d=2$: for a semi-infinite line, we obtain
\begin{equation}
\label{eq:finalresultd2}
\ln Z_n(\alpha)
=\left( \dfrac{\alpha^2}{8\pi^2n}-\dfrac{|\alpha|}{4\pi n}-\frac{1}{12}\left(n-\frac{1}{n} \right) \right)(-\mathrm{Ei}(-m^2\epsilon^2)),
\end{equation}
where we have used the definition of the exponential integral function
\begin{equation}
\mathrm{Ei}(x)=-\int_{-x}^{\infty} dt \frac{e^{-t}}{t}.
\end{equation}
By expanding around $\epsilon=0$, $-\mathrm{Ei}(-m^2\epsilon^2)\simeq \gamma_E+2\ln(m\epsilon)$ and considering $n=1$ we end up into
\begin{equation}
\label{eq:confirm}
\ln Z_1(\alpha)=\left( \dfrac{\alpha^2}{4\pi^2}-\dfrac{|\alpha|}{2\pi} \right)(-\ln(m\epsilon)-\frac{\gamma_E}{2}),
\end{equation}
where $\gamma_E$ is the Euler-Mascheroni constant. 
This coincides with what found previously in Eq. \eqref{eq:cth2} up to $O(1)$.
In $d=3$, Eq. (\ref{eq:finalresult}) coincides with the leading term obtained in \cite{mrc-20} for the critical limit of a two-dimensional harmonic lattice.

\section{Conclusions}\label{sec:concl}

In this manuscript we characterised the symmetry resolved entanglement for free massive fields in two dimensions, presenting the results for a Dirac field and a 
complex scalar theory.
We showed that two well known techniques in the framework of the replica trick can be adapted --by modifying the $n$-sheeted Riemann surface and the corresponding 
partition function-- to the calculation of charged moments. 
Both computations (via modified twist fields and the Green's function approach of Ref. \cite{ch-rev}) 
mainly rely on the boundary conditions of the fields at the endpoints of the entangling region. 
In the first framework, the conformal dimensions of the twist fields get modified as in Eq. (\ref{eq:moments}). 
In the second setting the change induced by the flux $\alpha$ lies in the precise form of the Painlev\'e V equations (\ref{eq:PainleveFermion}) and (\ref{eq:PainleveBoson})
providing the generalised partition function. 
These Painlev\'e equations are easily solved numerically for arbitrary values of the mass, but they can be also handled analytically 
in the limit of small masses, leading to the charged moments (\ref{eq:totalsc}) for the Dirac field and (\ref{eq:guesssbagliato}) for the scalar theory.
The opposite limit of mass much larger that the interval length can also be treated analytically.
For the free complex scalar, we also obtain general results for the charged moments in arbitrary dimension when the entangling surface is an hyperplane. 

From the Fourier transform of these charged moments, we extract the symmetry resolved R\'enyi entropies, stressing the relevant universal aspects.
At leading order for small $m$, the symmetry resolved entropies for both theories satisfy equipartition of entanglement \cite{xavier}. 
We also show that the entanglement equipartition is broken by the mass at order $(\ln \ell)^{-2}$, which is the same one found in other circumstances
\cite{riccarda,crc-20,MDC-19-CTM,mrc-20}.

There are two main aspects that our manuscript leave open for further study. 
The first one concerns the calculation of charged and symmetry resolved entropies in free scalars and fermions in arbitrary dimension and for 
entangling surfaces that are more complex than the simple hyperplane of Sec \ref{app:nishioka}. 
To this aim, we expect that some of the existing techniques in the literature, as e.g. in Refs. \cite{s-08,s-09,s-10,chm-11,chr-14,h-14,bchm-13,chl-09,ch-07,hn-13,s-11},  should be readily adapted to our problem. 
Furthermore, important insights could also come from the holographic correspondence for the entanglement entropy \cite{rt-06,nrt-09,Nishioka}.  
The other point is whether interacting QFTs can be handled in two dimensions, e.g. exploiting integrability techniques as those of Refs. \cite{ccd-08,cd-09,bc-16,c-17,clsv-19}. 


\section*{Acknowledgments}
We thank Luca Capizzi, D\'avid Horv\'ath, Pierluigi Niro and Paola Ruggiero for useful discussions. PC and SM acknowledge support from ERC under Consolidator grant number 771536 (NEMO).

\appendix
\section*{Appendices}

\section{Conformal dimensions of twist fields}\label{app:twist}
The goal of this section is to find the conformal dimension of the twist field $\mathcal{T}_{n,k,\alpha}$ defined in Eq. (\ref{eq:prod_twist}). 
We will call it generically $\mathcal{T}_{a}$, where $a=\frac{k}{n}+\frac{\alpha}{2\pi n }$, with $a \in [0,1]$. As already discussed in section \ref{sec:twist}, in the 
neighbourhood of a twist field the $k$-th component of $\phi$ undergoes a phase rotation 
\begin{equation}
\label{eq:rotation}
\tilde{\phi}_k(e^{2\pi i} z,e^{-2\pi i} \bar{z})=e^{2\pi i a} \tilde{\phi}_k (z,\bar{z}).
\end{equation}
Let us start from the case of the free complex scalar CFT with fields $(\varphi_k,\varphi_k^{*})$ and, following \cite{dixon}, consider the correlation function in the presence of four $\mathbb{Z}_n$ twist-fields
\begin{equation}
\label{eq:4twist}
g(z,w;z_i)=\dfrac{\braket{ -\frac{1}{2}\partial_z \varphi_k\partial_w \varphi_k^{*} \mathcal{T}_{a}(z_1)\tilde{\mathcal{T}}_{a}(z_2)\mathcal{T}_{a}(z_3)\tilde{\mathcal{T}}_{a}(z_4)}}{\braket{\mathcal{T}_{a}(z_1)\tilde{\mathcal{T}}_{a}(z_2)\mathcal{T}_{a}(z_3)\tilde{\mathcal{T}}_{a}(z_4)}}.
\end{equation} 
Imposing that for $z \to w $ we have $g(z,w;z_i) \sim (z-w)^{-2}$ and that for $z \to z_i$ we have $g(z,w;z_i) \sim (z-z_j)^{-a}$ for $j=1,3$ and $g(z,w;z_i) \sim (z-z_j)^{-(1-a)}$ for $j=2,4$, we can write (up to an additional constant independent of $z$ and $w$, $A(z_j,\bar{z}_j)$)
\begin{equation}
\label{eq:4twist1}
\begin{split}
g(z,w;z_i)=&\omega_a(z) \omega_{1-a}(w)\left[ a \frac{(z-z_1)(z-z_3)(w-z_2)(w-z_4)}{(z-w)^2}\right. \\
&\left. +(1-a) \frac{(z-z_2)(z-z_4)(w-z_1)(w-z_3)}{(z-w)^2} +A(z_j,\bar{z}_j) \right],
\end{split}
\end{equation}
where 
\begin{equation}
\label{eq:omega}
\omega_a(z)=[(z-z_1)(z-z_3)]^{-a}[(z-z_2)(z-z_4)]^{-(1-a)}.
\end{equation}
In the limit $ w \to z$
\begin{equation}
\label{eq:limit}
\lim_{w\to z} [g(z,w)-(z-w)^{-2}]=\frac{1}{2}a(1-a)\left(\frac{1}{z-z_1} +\frac{1}{z-z_2}+\frac{1}{z-z_3}+\frac{1}{z-z_4}\right)^2+\cdots
\end{equation}
This is exactly the expectation value of the insertion of the stress energy tensor of the field $\varphi_k$ in the four-point correlation function. From the comparison with the conformal Ward identity, we can understand that $\mathcal{T}_{a}$ and $\tilde{\mathcal{T}}_{a}$ are primary fields with scaling dimensions    
\begin{equation}
\label{eq:dim}
\Delta_a=\bar{\Delta}_a=\frac{1}{2}a(1-a)=\frac{1}{2}\left( \frac{k}{n}+\frac{|\alpha|}{2\pi n} \right)\left( 1-\frac{k}{n}-\frac{|\alpha|}{2\pi n}\right).
\end{equation} 
In order to obtain the conformal dimensions of the twist fields of the free Dirac field theory,
let us apply a similar procedure for the chiral or anti-chiral complex fermionic fields,  ($\psi_k,\psi^*_k$). The scaling dimension of $\mathcal{T}_{a} $ can be extracted from the Green's function in presence of two $\mathbb{Z}_n$ twist fields
\begin{equation}
g(z,w;z_i)=\dfrac{\braket{-\frac{1}{2}(\psi^*_k\partial_z\psi_k-\partial_w \psi^*_k \psi_k)\mathcal{T}_{a}(z_1)\tilde{\mathcal{T}}_{a}(z_2)}}{\braket{\mathcal{T}_{a}(z_1)\tilde{\mathcal{T}}_{a}(z_2)}}.
\end{equation}
Using the results in \cite{ising}, the previous expression can be explicitly written as
\begin{equation}
\label{eq:4twist12}
\begin{split}
g(z,w;z_i)=\omega_a(z) \omega_{-a}(w)&\left[ a \frac{(z_2-z_1)(w^2+z^2+2z_1z_2-(z_1+z_2)(w+z) )}{2(z-w)}+ \right. \\
&\left. -\frac{(w-z_1)(w-z_2)(z-z_1)(z-z_2)}{(z-w)^2} \right],
\end{split}
\end{equation}
where 
\begin{equation}
\label{eq:omega2}
\omega_a(z)=[(z-z_1)]^{-a-1}[(z-z_2)]^{a-1}.
\end{equation}
In the limit $ w \to z$
\begin{equation}
\label{eq:limit2}
\lim_{w\to z} [g(z,w)+(z-w)^{-2}]=\frac{1}{2}a^2\left(\frac{1}{z-z_1} -\frac{1}{z-z_2}\right)^2+\cdots
\end{equation}
This is the expectation value of the insertion of the stress energy tensor of the field $\psi_k$ in the two-point correlation function and, as before, the comparison with the conformal Ward identity gives the dimensions of the primary twist fields $\mathcal{T}_{a}$ and $\tilde{\mathcal{T}}_{a}$ as
\begin{equation}
\label{eq:dim2}
\Delta_a=\bar{\Delta}_a=\frac{1}{2}a^2=\frac{1}{2}\left( \frac{k}{n}+\frac{\alpha}{2\pi n} \right)^2.
\end{equation} 
Putting together the monodromy conditions (\ref{eq:rotation}) and the scaling dimension of the twist field in Eq.\,(\ref{eq:dim2}),  we deduce that the twist field of a fermionic field admits a bosonisation formula. We can write the complex fermionic field as $\psi_k \sim e^{i\varphi_k}$ and the twist field as $\mathcal{T}_{n,k,\alpha}(z)=e^{i\left( \frac{k}{n}+\frac{\alpha}{2\pi n} \right)\varphi_k}$. By introducing the vertex operators $V_{\beta}(z) = e^{i\beta\varphi(z)}$, the twist fields take the form $\mathcal{T}_{n,k,\alpha}(z)=V_{\frac{k}{n}+\frac{\alpha}{2\pi n}}(z)$ and  $\mathcal{\tilde{T}}_{n,k,\alpha}(z)=V_{-\frac{k}{n}-\frac{\alpha}{2\pi n}}(z)$ \cite{Belin-Myers-13-HolChargedEnt,ising}. Let us observe that at first sight this result could be misleading since the outcome for bosons in Eq.\,(\ref{eq:dim}) does not appear to agree with that of fermions in Eq.\,(\ref{eq:dim2}) given that they are related by bosonisation in 1+1 dimensions \cite{m-book,Gogolin-Tsvelik,Giamarchi2003}. However, via bosonisation of $U(1)$ complex fermions, the corresponding bosons transform by translation, and thus should instead satisfy the boundary condition $\varphi_k(e^{2\pi i} z) = \varphi_k(z) + a$. Therefore, our computation for charged bosons is not related to charged fermions by bosonisation. 

Before concluding this appendix, let us emphasise that while CFTs are well understood objects, $n$-copies of a CFT after modding out the $\mathbb{Z}_N$ symmetry among the replicas form a more complicated object known as orbifold \cite{k-87,dixon}. The operator product expansions of the twist fields with other fields have been extensively explored (e.g., see \cite{twist1,twist2,twist3,twist4,dei-18,cct-11,headrick,cz-13,cwz-16}), but, unless a bosonisation procedure for free theories can be used, as for the compact boson, they remain elusive in general and require a case-by-case study.

\section{Details for the analytic continuation for the Dirac field}\label{app:dirac}
In this appendix we provide some details about the analytic continuation of the quantities defined in Eq. \eqref{eq:kappa0}. 
First, we rewrite $\Omega(n,\alpha)$ (a similar result also holds for $\Lambda(n,\alpha)$) as
\begin{equation}
\begin{split}
&\Omega_n(\alpha) \equiv\frac{1}{n^2}\sum_{k=-(n-1)/2}^{(n-1)/2} \left(k+\frac{\alpha}{2\pi}\right)^2\left(\psi \left(\frac{k}{n}+\frac{\alpha}{2\pi n} \right)+\psi \left(-\frac{k}{n}-\frac{\alpha}{2\pi n} \right)\right)=\\
& \frac{1}{n^2}\sum_{k'=0}^{n-1} \left( k'+\frac{\alpha}{2\pi}-\frac{n-1}{2} \right)^2 \left( \psi \left( \frac{k'}{n}-\frac{n-1}{2n}+\frac{\alpha}{2\pi n} \right)+ \psi \left(-\frac{k'}{n}+\frac{n-1}{2n}-\frac{\alpha}{2\pi n} \right)\right),
\end{split}
\end{equation}
where we set $k'\equiv k+\frac{n-1}{2} $.
After this manipulation, the functions $\Omega(n,\alpha)$ and $\Lambda(n,\alpha)$ in (\ref{eq:kappa0}) can be split as 
\begin{equation}
\Omega(n,\alpha)=\sum_i B_i(n,\alpha), \qquad \Lambda(n,\alpha)=\sum_i C_i(n,\alpha), 
\end{equation}
where 
\begin{equation}\label{eq:app}
\begin{split}
&B_0(n,\alpha)=\frac{1}{n^2}\sum_{k=0}^{n-1} \left(\frac{\alpha}{2\pi}-\frac{n-1}{2}\right)^2 \left( \psi \left( \frac{k}{n}+\frac{1}{2}+\frac{1}{2n}+\frac{\alpha}{2\pi n} \right)+ \psi \left(-\frac{k}{n}+\frac{3}{2}-\frac{1}{2n}-\frac{\alpha}{2\pi n} \right)\right), \\
&B_1(n,\alpha)=\frac{2}{n^2}\sum_{k=0}^{n-1} \left(\frac{\alpha}{2\pi}-\frac{n-1}{2}\right)k \left( \psi \left( \frac{k}{n}+\frac{1}{2}+\frac{1}{2n}+\frac{\alpha}{2\pi n} \right)+ \psi \left(-\frac{k}{n}+\frac{3}{2}-\frac{1}{2n}-\frac{\alpha}{2\pi n} \right)\right),
\\
&B_2(n,\alpha)=\frac{1}{n^2}\sum_{k=0}^{n-1}k^2 \left( \psi \left( \frac{k}{n}+\frac{1}{2}+\frac{1}{2n}+\frac{\alpha}{2\pi n} \right)+ \psi \left(-\frac{k}{n}+\frac{3}{2}-\frac{1}{2n}-\frac{\alpha}{2\pi n} \right)\right),\\
&C_0(n,\alpha)=\frac{1}{n^2}\sum_{k=0}^{n-1} \left(\frac{\alpha}{2\pi}-\frac{n-1}{2}\right)^2 \left( \psi \left( \frac{k}{n}+\frac{1}{2}+\frac{1}{2n}+\frac{\alpha}{2\pi n} \right)+ \psi \left(-\frac{k}{n}+\frac{3}{2}-\frac{1}{2n}-\frac{\alpha}{2\pi n} \right)\right)^2, \\
&C_1(n,\alpha)=\frac{2}{n^2}\sum_{k=0}^{n-1} \left(\frac{\alpha}{2\pi}-\frac{n-1}{2}\right)k \left( \psi \left( \frac{k}{n}+\frac{1}{2}+\frac{1}{2n}+\frac{\alpha}{2\pi n} \right)+ \psi \left(-\frac{k}{n}+\frac{3}{2}-\frac{1}{2n}-\frac{\alpha}{2\pi n} \right)\right)^2,
\\
&C_2(n,\alpha)=\frac{1}{n^2}\sum_{k=0}^{n-1}k^2 \left( \psi \left( \frac{k}{n}+\frac{1}{2}+\frac{1}{2n}+\frac{\alpha}{2\pi n} \right)+ \psi \left(-\frac{k}{n}+\frac{3}{2}-\frac{1}{2n}-\frac{\alpha}{2\pi n} \right)\right)^2.
\end{split}
\end{equation}
We used the relation $\psi(x)=\psi(x+1)-\frac{1}{x}$ so that all the terms are in a suitable form for the use of the following integral representation of the digamma function, i.e.
\begin{equation} 
\psi(x)=-\gamma_E+\int_0^{1}dt \frac{1-t^{x-1}}{1-t}, \qquad x>0.
\end{equation}
Let us start from the analysis of the function $\Omega(n,\alpha)$. 
By inverting sums and integrals, we get 
\begin{equation}
\label{eq:kappa0b}
\begin{split}
&B_0(n,\alpha)= 
\left(\frac{\alpha}{2\pi n}-\frac{n-1}{2n}\right)^2\Big(-2n\gamma_E+\displaystyle \int_0^1dt \frac{2n}{1-t}+\frac{{t}^{\frac{1}{2n}-\frac12}}{t^{1/n}-1} 2\cosh\frac{\alpha\log t}{2 n \pi}  \Big),\\
&B_1(n,\alpha)=\left(\frac{\alpha}{\pi n}-\frac{n-1}{n}\right) \Big[(1-n) \gamma_E  +\\
&\displaystyle \int_0^1dt \frac{n-1}{1-t}-\frac{t^{-\frac{(-1 + n) \pi + \alpha}{
  2 n \pi} }[(1 -t) (-1 + t^{\frac{\pi + \alpha}{n \pi}}) \, + 
   n (-1 + t^{1/n}) (-1 + t^{1 +\frac{ \alpha}{n \pi}})]}{n(1-t)(-1+t^{1/n})^2}\Big],\\
 &B_2(n,\alpha)=\frac{1}{3n^2} \Big[ n (1 - n) (-1 + 2 n)\gamma_E +\displaystyle \int_0^1 dt \frac{n (1 - n) (1 - 2 n)}{1-t}-\frac{3t^{-\frac{(-1 + n) \pi + \alpha}{
  2 n \pi}}}{(1-t)(-1+t^{1/n})^3}\times
 \\& \Big ((-1 + t) (1 + t^{1/n}) (1 + t^{\frac{\pi+ \alpha}{
      n \pi})}) +
   2 n (1 - t^{1/n}) (1 + t^{\frac{\pi + n \pi + \alpha}{
      n \pi}}) - 
   n^2 (1 - t^{1/n})^2 (1 - t^{1 + \frac{\alpha}{n \pi}})\Big) \Big].
  \end{split}
\end{equation}

\begin{figure}
\centering
\subfigure
{\includegraphics[width=0.48\textwidth]{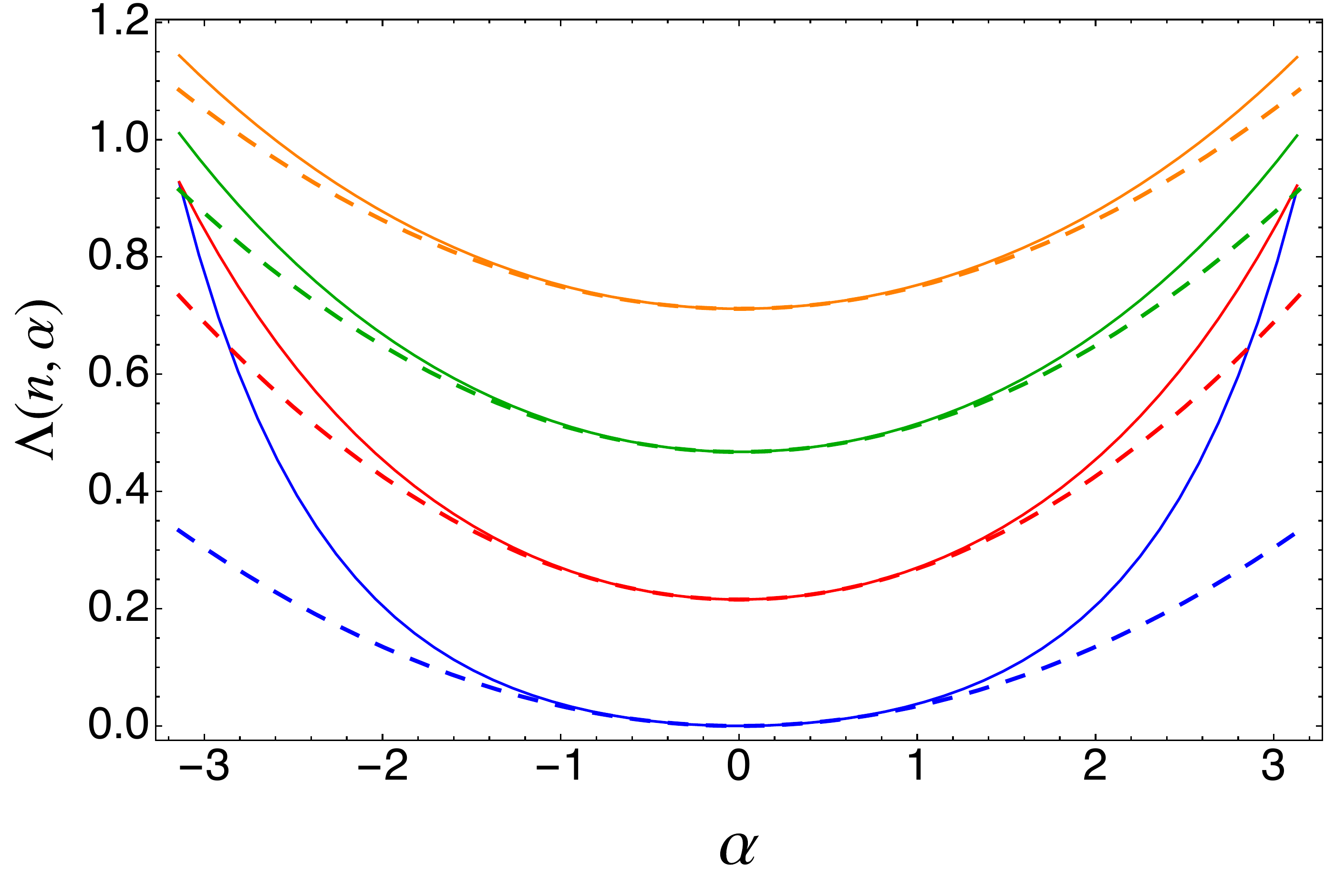}}
\subfigure
{\includegraphics[width=0.48\textwidth]{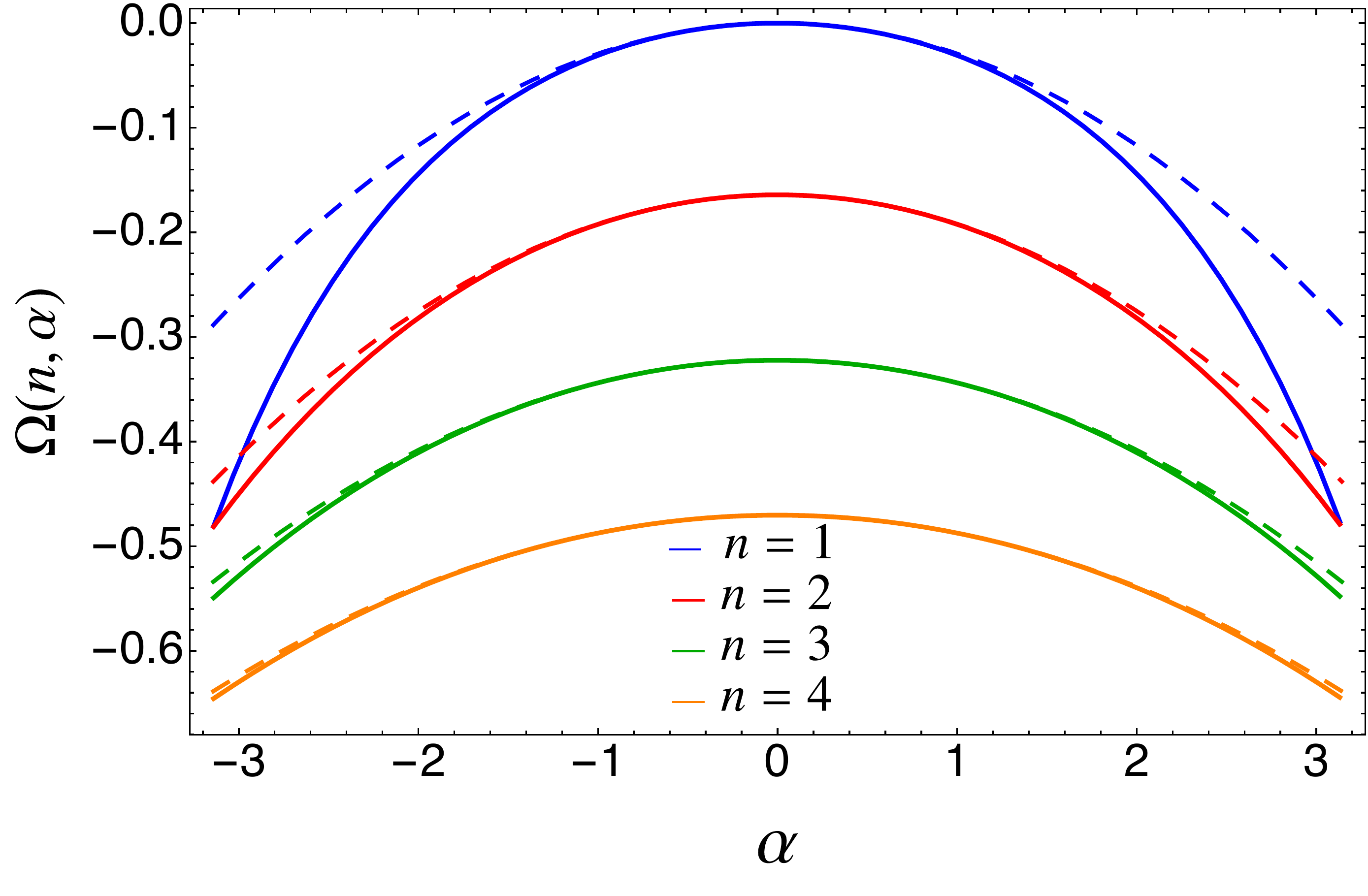}}
\subfigure
{\includegraphics[width=0.48\textwidth]{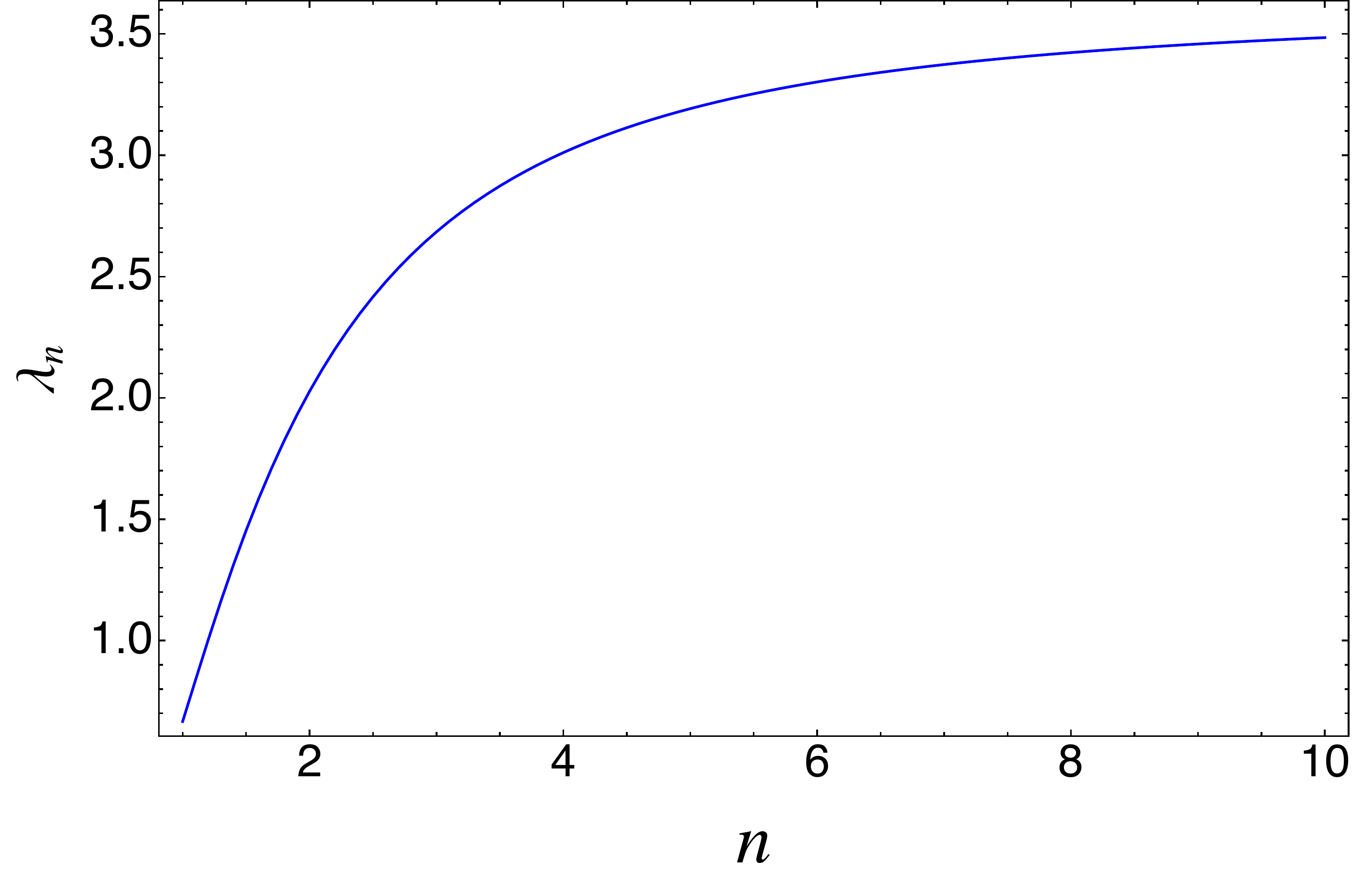}}
\subfigure
{\includegraphics[width=0.48\textwidth]{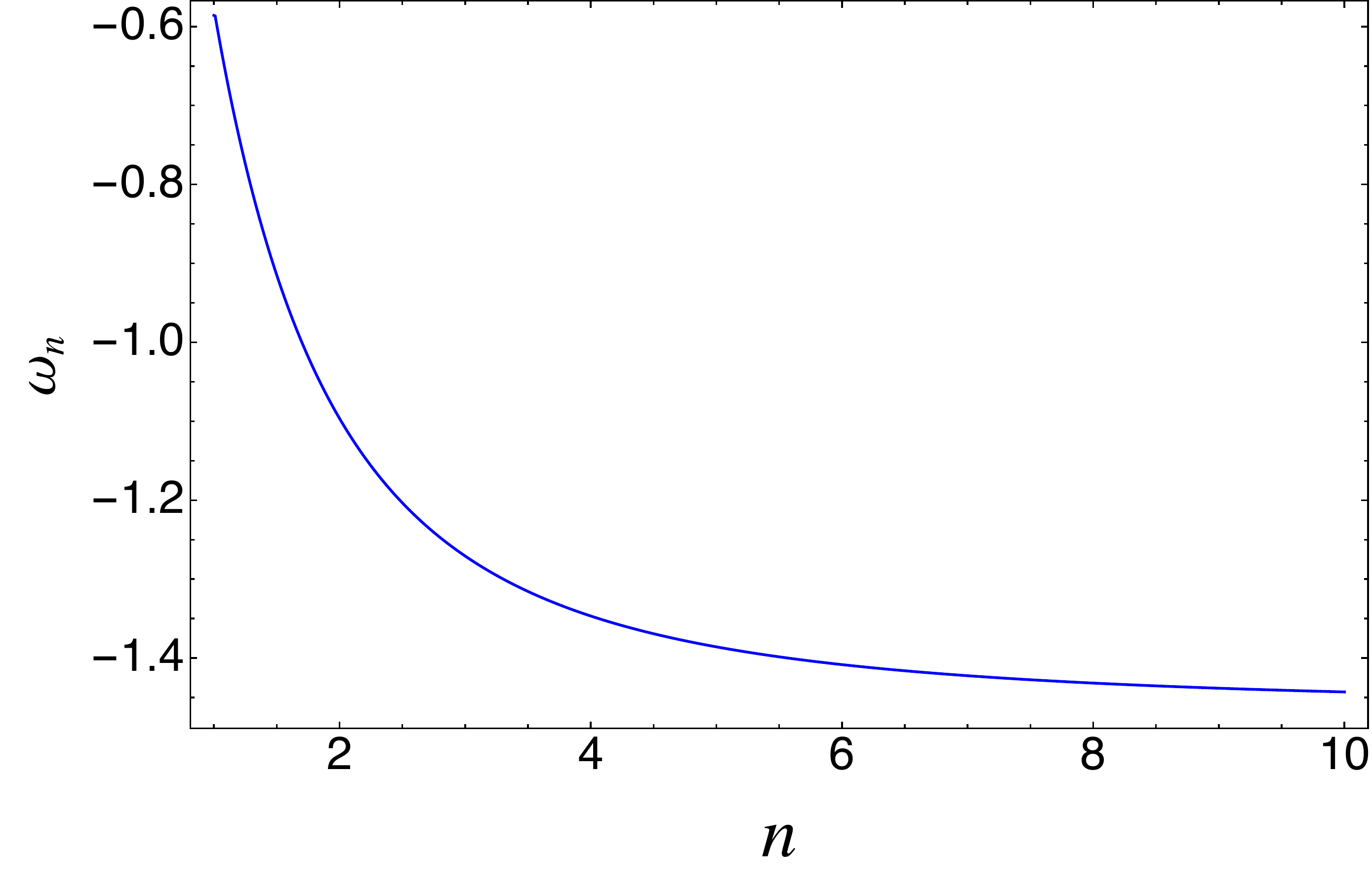}}
\caption{Top panels: $\Lambda(n,\alpha)$  (left) and $\Omega(n,\alpha)$ (right)  in Eq. (\ref{eq:kappa0}) as a function of $\alpha$ for $n = 1, 2, 3, 4$. The exact forms (full lines) are compared with the quadratic approximation (dashed lines) showing that, although very close, they are definitively different. Bottom panels: $\omega_n$ and $\lambda_n$ reported in Eq.\,(\ref{eq:kappa0}) are plotted as a function of $n$. Their non trivial dependence on $n$ gives rise to the subleading terms breaking the equipartition of entanglement in a massive Dirac field theory, as illustrated in Sec. \ref{sec:CHfermionsCL}.}
\label{fig:LO}
\end{figure}

The same strategy also works for $C_j(n,\alpha)$, $j=1,2,3$ in Eqs.\,(\ref{eq:app}), however this computation is longer and cumbersome and we do not report here.
In Figure \ref{fig:LO} we plot $\Lambda(n,\alpha)$ (left top panel) and $\Omega(n,\alpha)$ (right top panel)  as function of $\alpha$ for some $n$ and compare it with the quadratic approximation given in Eqs.\,(\ref{eq:kappa0}). The closeness of the two curves shows that the quadratic approximation is enough for most of the applications.
The quadratic approximation depends non trivially on $n$, as evident from the plot of $\omega_n$ and $\lambda_n$ in the bottom panels of Figure \ref{fig:LO}.

\section{The lattice models}
\label{app:lat}

For the numerical test of our field theory results we consider the following lattice discretisation of the Dirac fermion \cite{CFH}
\begin{equation}\label{eq:hamiltonianF}
H=-\frac{i}{2}\sum_{j=0}^{L-1}(c^{\dagger}_{j+1}c_j-c^{\dagger}_j c_{j+1})+m\sum_{j=0}^{L-1}(-1)^j c^{\dagger}_j c_j,
\end{equation}
where  $c_j$ satisfy the anti-commutation relations $\{ c_j,c^{\dagger}_k \}=\delta_{jk}$ and $L$ denotes the number of sites of the chain. 
The correlators in the thermodynamic limit, {\it i.e.} for $L \to \infty$, are 
\begin{equation}
\begin{split}
\braket{c^{\dagger}_jc_{k}}&=\frac{1}{2}\delta_{(j-k),0}+(-1)^j \displaystyle \int_0^{\frac{1}{2}}dx \frac{m \cos (2\pi x (j-k))}{\sqrt{m^2+\sin(2\pi x)^2}} \qquad \mathrm{for} \qquad |j-k| \quad \mathrm{even,} \\
\braket{c^{\dagger}_jc_{k}}&= i\displaystyle \int_0^{\frac{1}{2}}dx \frac{\sin (2\pi x )}{\sqrt{m^2+\sin(2\pi x)^2}}\sin (2\pi x (j-k)) \qquad \mathrm{for} \qquad |j-k| \quad \mathrm{odd.}
\end{split}
\end{equation}
 Denoting by $\varepsilon_{j}$ the eigenvalues of the correlation matrix restricted to the subsystem $A$ made by $\ell$ sites (with $j \in [1,\ell]$), simple algebra leads to the moments ${\rm Tr} \rho_A^n$ and to the R\'enyi entropies \cite{p-03,pe-09}.
The $\alpha$-dependent moments $Z_{n}(\alpha)$ can be
also easily written in terms of the eigenvalues of the correlation matrix $C_{jk}=\braket{c^{\dagger}_j c_k}$. For this purpose, it is necessary to write down the charge operator $Q_A$ in terms of $c_j$ and $c^{\dagger}_j$ operators in Eq.\,(\ref{eq:hamiltonianF}), i.e. \cite{grignani}
\begin{equation}
Q_A=\sum_{j=0}^{\ell-1} \left( c^{\dagger}_j c_j -\frac{1}{2}\right)
=\sum_{j=0}^{\ell-1}  c^{\dagger}_j c_j -\frac{\ell}{2}.
\end{equation}
Therefore, the charged moments read 
\begin{equation}\label{eq:Zalphanumerics}
Z_n(\alpha)=\displaystyle \prod_{j=1}^{\ell} [(\varepsilon_{j})^n e^{i\alpha/2}+(1-\varepsilon_{j} )^n e^{-i\alpha/2}],
\end{equation}
which provides a very simple formula for its numerical computation. The Fourier transform of $Z_n(\alpha)$ gives the symmetry resolved 
moments and entropies.

For the discretisation of the scalar field theory, we consider a chain of oscillators of mass $M = 1$ with frequency $\omega_0$, coupled together by springs with elastic 
constant $\kappa=1$. They are described by the Hamiltonian
\begin{equation}
\label{eq:hamHC1d}
H_{B}=\dfrac{1}{2}\sum_{i=0}^{L-1} p^2_i +\omega_0^2q^2_{i} +(q_{i+1}-q_{i})^2,
\end{equation}
where the $p_i$ and $q_i$ satisfy the canonical commutation relations $[q_i,q_j] = [p_i,p_j] = 0$ and $[q_i,p_j] = i\delta_{ij}$. In the thermodynamic limit, the correlators can be written in terms of the hypergeometric functions \cite{br-04}
\begin{eqnarray}
& & \hspace{-1.5cm}
\langle q_i q_j \rangle =
\frac{\zeta^{i-j+1/2}}{2}
\binom{i-j-1/2}{i-j}\;
_2F_1 \big(  1/2 , i-j+ 1/2 , i-j+1 ,  \zeta^2 \big),
\\
& & \hspace{-1.5cm}
\langle p_i p_j \rangle
=
\frac{\; \zeta^{i-j-1/2}}{2}
\binom{i-j-3/2}{i-j}\;
_2F_1 \big( - 1/2 , i-j- 1/2 , i-j+1 ,  \zeta^2 \big),
\end{eqnarray}
where the parameter $\zeta$ is defined by
\begin{equation}
\zeta \equiv \frac{\big(\omega - \sqrt{\omega^2+4} \big)^2}{4}.
\end{equation}
Let us denote as $X$ and $P$ the matrices of the correlators of positions and momenta  (i.e. $X_{ij}=\braket{q_{i}q_{j}}$ and $P_{ij}=\braket{p_{i}p_{j}}$). 
The moments of the reduced density matrix of $A$ can be written in terms of the eigenvalues of $\sqrt{X P}$ that we call $\sigma_{i}$ (with $i \in [1, \ell]$).

To have a continuous symmetry, we consider a complex bosonic theory which on the 
lattice is a chain of complex oscillators. 
It is the sum of two real harmonic chains in the variables $(p^{(1)},q^{(1)})$ and $(p^{(2)},q^{(2)})$, i.e.
\begin{equation}
\label{eq:complex}
H_{CB}(p^{(1)}+ip^{(2)},q^{(1)}+iq^{(2)})=H_{B}(p^{(1)},q^{(1)})+H_{B}(p^{(2)},q^{(2)}).
\end{equation}
The Hamiltonian (\ref{eq:complex}) can be also written
in terms of particles and antiparticles mode operators, $a_k$ and $b_k$, satisfying $[a_k,a^{\dagger}_j]=\delta_{j,k}$, $[b_k,b^{\dagger}_j]=\delta_{j,k}$. In terms of these operators, the Hamiltonian (\ref{eq:complex}) is 
\begin{equation}
H_{CB}=\sum_{k=0}^{L-1} \omega_k(a^{\dagger}_ka_k+b_{k}^{\dagger}b_k), \qquad \omega_k=\sqrt{\omega_0^2+4\sin^2 \left(\frac{\pi k}{L} \right)},
\end{equation}
while the charge operator reads
\begin{equation}
Q=\sum_{k=0}^{L-1}(a^{\dagger}_ka_k- b_{k}^{\dagger}b_k),
\end{equation}
i.e. the total number of particles {\it minus} the number of antiparticles. 
The conserved charge can be as well written in real space and its value in a given subsystem $A$ is the same sum restricted to $A$, i.e. 
 \begin{equation}
\label{eq:chargeanti}
Q_A=\displaystyle \sum_{j\in A} a^{\dagger}_ja_j-b^{\dagger}_jb_j.
\end{equation}
For the charged moments, we need to compute ${\rm Tr} ( \rho_A^n e^{i {Q}_A\alpha} )$, but using the form in Eq.~\eqref{eq:chargeanti} for ${Q}_A$, the trace factorises as 
\begin{equation}
Z_n(\alpha)={\rm Tr}\rho_A^n e^{i {Q}_A\alpha}= {\rm Tr}[(\rho_A^{a})^n e^{iN_A^{a}\alpha}]\times [{\rm Tr}(\rho_A^{b})^n e^{-iN_A^{b}\alpha}],
\end{equation}
where $N_A^{a}=\sum_{j\in A} a^{\dagger}_{j}a_{j}$ and 
$N_A^{b}=\sum_{j\in A} b^{\dagger}_{j}b_{j}$. Using the relations between the number operator $N_A$ and the eigenvalues of the correlation matrix $\sigma_i$, one finds \cite{MDC-19-CTM}
\begin{equation}
\label{eq:logZ n alfa lattice} 
Z_{n}(\alpha) 
=
\prod_{j=1}^\ell
\frac{1}{
\left(
\sigma_{j}+\frac{1}{2}
\right)^n
-
e^{i\alpha}
\left(
\sigma_{j}-\frac{1}{2}
\right)^n
}
\frac{1}{
\left(
\sigma_{j}+\frac{1}{2}
\right)^n
-
e^{-i\alpha}
\left(
\sigma_{j}-\frac{1}{2}
\right)^n
}.
\end{equation}
Again, this is the starting point for  the computation of the symmetry resolved 
moments and entropies by a Fourier transform.


\end{document}